\documentclass[preprint]{JHEP3} 

\usepackage{epsfig}
\usepackage{amsfonts}
\usepackage{amssymb,amsmath}
\usepackage[ps,matrix,arrow]{xy}

\newcommand{\Tr}[1]{\:{\rm Tr}\,#1}

\newcommand{\mbf}[1]{{\boldsymbol {#1} }}
\newcommand{\complex}{{\mathbb C}} 
\newcommand{\zed}{{\mathbb Z}} 
\newcommand{\nat}{{\mathbb N}} 
\newcommand{\real}{{\mathbb R}} 
\newcommand{\rat}{{\mathbb Q}} 
\newcommand{\torus}{{\mathbb T}}
\newcommand{\sphere}{{\mathbb S}}
\newcommand{\disk}{{\mathbb D}}
\newcommand{\interval}{{\mathbb I}}
\def\e{{\,\rm e}\,}

\newcommand{\id}{{1\!\!1}}

\newcommand{\vm}{{\mbf m}}
\newcommand{\vnu}{{\mbf\nu}}

\def\ii{{\,{\rm i}\,}}
\def\dd{{\rm d}}
\def\DD{{\rm D}}

\def\beq{\begin{equation}}
\def\bee{\begin{equation}}
\def\eeq{\end{equation}}
\def\bea{\begin{eqnarray}}
\def\eea{\end{eqnarray}}
\def\bd{\begin{displaymath}}
\def\ed{\end{displaymath}}

\newcommand{\Cint}{\int\kern-10.5pt-\kern7pt}

\newcommand{\PP}{{\mathbb{P}}}

\newcommand{\be}{\begin{equation}}
\newcommand{\ee}{\end{equation}}

\newcommand\fverb{\setbox\pippobox=\hbox\bgroup\verb}
\newcommand\fverbdo{\egroup\medskip\noindent%
                        \fbox{\unhbox\pippobox}\ }
\newcommand\fverbit{\egroup\item[\fbox{\unhbox\pippobox}]}
\newbox\pippobox

\allowdisplaybreaks

\title{Topological Strings, Two-Dimensional Yang-Mills Theory and
  Chern-Simons Theory on Torus Bundles}
\author{Nicola Caporaso$^{(a)}$, Michele Cirafici$^{(b)}$, Luca
Griguolo$^{(c)}$,
 Sara Pasquetti$^{(c)}$,  Domenico~Seminara$^{(d)}$ and
 Richard~J.~Szabo$^{(b)}$ \\
$^{(a)}$ Center for Theoretical Physics, Massachusetts Institute of
Technology, Cambridge, MA 02139, USA\\
$^{(b)}$ Department of Mathematics, Heriot-Watt University and\\
Maxwell Institute for Mathematical Sciences\\
Colin Maclaurin Building, Riccarton, Edinburgh EH14 4AS, UK\\
$^{(c)}$ Dipartimento di  Fisica, Universit\`a  di Parma,
INFN Gruppo Collegato di Parma\\
Parco Area delle Scienze 7/A, 43100 Parma, Italy\\
$^{(d)}$ Dipartimento di Fisica, Polo Scientifico Universit\`a di
Firenze,\\ INFN Sezione di Firenze
Via  G. Sansone 1, 50019 Sesto Fiorentino, Italy\\
\email{caporaso@fi.infn.it , M.Cirafici@ma.hw.ac.uk ,
griguolo@fis.unipr.it , pasquetti@fis.unipr.it ,
seminara@fi.infn.it , R.J.Szabo@ma.hw.ac.uk } }

\received{\today}               

\accepted{\today}               
\preprint{ {\tt HWM-06-34} \ \ {\tt EMPG-06-07}
\\ \hepth{0609129}} 

\date{data}
\abstract{We study the relations between two-dimensional
Yang-Mills theory on the torus, topological string theory on a
Calabi-Yau threefold whose local geometry is the sum of two line
bundles over the torus, and Chern-Simons theory on torus bundles.
The chiral partition function of the Yang-Mills gauge theory in the
large $N$ limit is shown to coincide with the topological string
amplitude computed by topological vertex techniques. We use Yang-Mills
theory as an efficient tool for the computation of Gromov-Witten
invariants and derive explicitly their relation with Hurwitz
numbers of the torus. We calculate the Gopakumar-Vafa invariants,
whose integrality gives a non-trivial confirmation of the conjectured
nonperturbative relation between two-dimensional Yang-Mills theory and
topological string theory. We also demonstrate how the gauge theory
leads to a simple combinatorial solution for the Donaldson-Thomas
theory of the Calabi-Yau background. We match the instanton representation
of Yang-Mills theory on the torus with the nonabelian localization of
Chern-Simons gauge theory on torus bundles over the circle. We also
comment on how these results can be applied to the computation of
exact degeneracies of BPS black holes in the local Calabi-Yau
background.}


\begin{document}

\section{Introduction and Summary}

The OSV conjecture~\cite{Ooguri:2004zv}--\cite{Aganagic:2004js}
proposes an intriguing relation between topological string theory
and the physics of black holes. It asserts that the partition function
of an extremal black hole arising in a Calabi-Yau compactification of
Type~II string theory is related to the modulus squared of the
partition function of topological string theory whose target space is
the same Calabi-Yau manifold. This insight
relies on the well known fact that topological string amplitudes compute
F-terms in the four-dimensional supersymmetric theory in a
graviphoton background~\cite{Antoniadis:1993ze,Bershadsky:1993cx}.
This provides a powerful tool for calculating higher derivative
corrections to black hole entropy, extending previous results
based on supergravity
calculus~\cite{LopesCardoso:1998wt}--\cite{LopesCardoso:1999ur}.

A concrete realization of the OSV conjecture is modelled on a specific
class of Calabi-Yau threefolds whose local geometry is given by the
sum of two line bundles over a compact Riemann
surface~\cite{Vafa:2004qa,Aganagic:2004js}. A four-dimensional BPS
black hole\footnote{Strictly speaking, these are not black holes
  because the non-compactness of the threefold prevents them from
  being particle-like. In what follows we refer only to the BPS state
  counting.} is engineered by wrapping a system of D4--D2--D0 branes on
holomorphic cycles of the Calabi-Yau space,
realizing an explicit construction of the black hole microstates of
charge $N$. The counting of these microstates is thus given by the
number of bound states that the D2 and D0 branes can form with $N$ D4
branes, which in turn corresponds to excited gauge field
configurations of the $\mathcal{N}=4$ (topologically twisted)
$U(N)$ gauge theory living on the worldvolume of the D4 branes. These
configurations are then argued~\cite{Vafa:2004qa} to localize in a two-dimensional
$q$-deformed $U(N)$ Yang-Mills theory living on the Riemann surface
over which the Calabi-Yau manifold is fibred. This chain of arguments
closes with the identification between the partition function of the
two-dimensional Yang-Mills theory in the large $N$ limit, which admits
a chiral/antichiral factorization, and the modulus squared of the
topological string partition function. In this way the $q$-deformed
gauge theory is conjectured to provide a nonperturbative completion of
topological string theory in the local Calabi-Yau background.

This conjecture has been studied in a variety
of different contexts. The dynamics of the $q$-deformed theory
on the sphere has been analyzed in detail
in~\cite{Arsiwalla:2005jb}--\cite{Caporaso:2005np}.
While~\cite{Arsiwalla:2005jb}--\cite{Caporaso:2005ta}
are focused mostly on the gauge theoretical aspects of the
correspondence and deal primarily with the phase transition of $q$-deformed
Yang-Mills theory on the sphere,\footnote{See~\cite{Caporaso:2006gk}
  for a comparison of this phase transition with that of perturbative
  topological string theory on the corresponding local Calabi-Yau
  threefold.} in~\cite{Caporaso:2005fp} we have
exhibited a precise link between the two-dimensional theory and the
underlying toric Calabi-Yau geometry through an explicit topological
string computation. The interplay between the various geometrical
invariants that arise in topological string theory is crucial for this
correspondence. A different perspective has been proposed
in~\cite{Caporaso:2005ta,deHaro1}--\cite{Blau:2006gh},
where it is shown that the $q$-deformation is very natural from the
point of view of three-dimensional Chern-Simons gauge theory defined
on a Seifert manifold fibred over the Riemann surface. In this case the
three-dimensional gauge theory, which describes open topological
strings on the cotangent bundle of the total space of the Seifert
fibration, localizes to a $q$-deformed gauge theory on the base
space. The two-dimensional gauge theory has been used in this way as
an efficient tool for the computation of various three-manifold and
knot invariants~\cite{Boulatov}--\cite{deHaro3}. The $q$-deformed
theory has also been studied from the perspective of its large $N$
Gross-Taylor string expansion in terms of central elements of Hecke
algebras in~\cite{deHaro}, while its use in more general BPS state
counting is described in~\cite{Aganagic2:2005,Jafferis}. Moreover, the
relevant topological string theory has been studied in great detail
in~\cite{Bryan:2004iq} where it was shown that the isometries of these
backgrounds lift to the moduli space of curves and thereby
provide a powerful tool for the computation of equivariant
Gromov-Witten invariants.

The aim of this paper is to further connect the various points of
view that lead to two-dimensional Yang-Mills theory in the case where
the base Riemann surface is a torus and the two-dimensional gauge
theory is not deformed. Our analysis provides, amongst other things, a
detailed elucidation of some of the constructs proposed
in~\cite{Vafa:2004qa}. At the same time the local elliptic threefolds
we study provide a class of {\it non-toric} examples in which we can
explicitly verify a number of conjectured physical and mathematical
equivalences in (non-perturbative) topological string theory,
extending previous correspondences for local theories of rational
curves~\cite{Caporaso:2005fp} and other toric Calabi-Yau
backgrounds~\cite{Aganagic2:2005}.

We compute explicitly the large $N$ limit of
the two-dimensional gauge theory and extract the chiral contribution
following the standard Gross-Taylor expansion~\cite{GrossTaylor}. In
particular, we show explicitly that the role played by the $U(1)$ charges is
intimately related to the appearence of open string degrees of
freedom. The precise matching with topological string theory on the
elliptic threefold would provide an explicit realization of
Gross-Taylor string theory on the elliptic curve in terms of an
equivariant topological sigma-model in six dimensions. This is to be
compared with its relation~\cite{Bershadsky:1993cx} to a
two-dimensional topological sigma-model coupled to topological gravity
and perturbed by the K\"ahler class of the underlying elliptic
curve. We will see very directly that these two topological string
theories are in fact the same.

To compare with the conjecture of~\cite{Vafa:2004qa}, we explicitly
construct the pertinent Calabi-Yau geometry by using blowup techniques
to relate it to a formal toric variety. Topological string amplitudes
on formal toric varieties can be computed by means of the topological
vertex~\cite{Aganagic:2003db,Li:2004uf}. After blowing down and a
change of framing in the topological vertex gluing rules to recover
the non-toric geometry of~\cite{Vafa:2004qa}, the resulting
topological string amplitude matches the two-dimensional
prediction. This allows us to exploit the two-dimensional gauge theory
to compute geometrical invariants of the elliptic Calabi-Yau
threefold. In particular, we find that the computation
of Gromov-Witten invariants can be reduced to the evaluation
of Hurwitz numbers that count (connected) coverings of the
torus. In this way we deduce a closed, explicit combinatorial
formula for the Gromov-Witten invariants. These ideas easily
generalize to the computation of Gopakumar-Vafa invariants, and their
integrality confirms our computations. The two-dimensional gauge
theory also leads to a simple combinatorial solution for the
Donaldson-Thomas theory of the local elliptic threefold.

Finally, we describe the point of view from Chern-Simons gauge
theory, a relation anticipated from open-closed string duality in the
A-model with D-branes. The pertinent three-dimensional geometry is that
of a torus bundle fibred over the circle. The partition function
localizes onto a sum over flat connections, the critical points of the
Chern-Simons action functional. We classify the flat connections and
provide an explicit matching with two-dimensional instantons. The
precise matching between Chern-Simons gauge theory and two-dimensional
Yang-Mills theory relies on performing an analytic continuation of the
Chern-Simons coupling $k$ to imaginary non-integer levels, leading to
some subtleties in the correspondence. The link to two-dimensional
conformal field theory plays a crucial role here. This provides a very
explicit demonstration of the formal
equivalence~\cite{Beasley:2005vf,Blau:2006gh} between Chern-Simons
theory on a Seifert manifold and two-dimensional Yang-Mills theory on
the base of the Seifert fibration.

The organization of this paper is as follows. In Sect.~2 we
briefly review the proposed relationship between the OSV conjecture
and the role played by two-dimensional Yang-Mills theory. In
particular, we use nonabelian localization techniques to express the
partition function of the gauge theory as a sum over instanton
contributions and describe their role in reproducing the counting of
microstates through the $\mathcal{N}=4$ gauge theory in four
dimensions. The factorization into chiral and antichiral
components in the large $N$ limit is the subject of Sect.~3. In
Sect.~4 we construct the relevant Calabi-Yau geometry and compute the
topological string amplitude using the topological vertex gluing
rules. In Sect.~5 we examine the various geometric invariants that
characterize the background. We use two-dimensional Yang-Mills theory
to compute the Gromov-Witten invariants and study their relation
with Hurwitz numbers and Gopakumar-Vafa invariants, the
integrality of the latter giving strong support to the
validity of the conjecture. We also study their relation with the
Donaldson-Thomas invariants of the background, for which we derive a
closed combinatorial formula. Finally, in Sect.~6 we analyze in
detail the relation between the Yang-Mills theory and
Chern-Simons gauge theory on a torus bundle. Some technical details
are collected in three appendices at the end of the paper.

\section{Yang-Mills Theory and the D-Brane Partition
  Function\label{DPartFn}}

In this section we will state more precisely the conjecture
of~\cite{Vafa:2004qa,Aganagic:2004js} in the case of the local
elliptic curve.\footnote{This case has also been studied from a different
perspective in~\cite{Rangoolam,BabyUniverse}, in an attempt to evaluate and
give a black hole interpretation to the nonperturbative $\e^{-N}$
corrections at large $N$.} This will lead to the introduction of
two-dimensional Yang-Mills theory on the torus. We will review its
instanton representation and modular properties. We also describe its
relationship to the computation of the Euler characteristic of
instanton moduli space in $\mathcal{N}=4$ gauge theory, clarifying
some points which were missed in the analysis
of~\cite{Aganagic:2004js}.

\subsection{Black Hole Entropy}

Consider Type~IIA superstring theory on $\mathbb{R}^{3,1} \times
X_p$ where the local Calabi-Yau threefold $X_p$ is the total space
of the holomorphic rank~2 vector bundle
\begin{equation} \label{threefold}
\mathcal{O}_{\torus^2} (-p) \oplus \mathcal{O}_{\torus^2}(p) ~\longrightarrow~
\mathbb{T}^2 \ ,
\end{equation}
with $\mathcal{O}_{\torus^2}(p)$ a holomorphic line bundle of degree
$p\geq0$ over the torus $\mathbb{T}^2$ (i.e. a holomorphic section of
this bundle has exactly $p$ zeroes) and $\mathcal{O}_{\torus^2}(-p)$
its dual. The restriction to non-negative $p$ is possible without
losing generality because of the reflection symmetry $p\to-p$ which
interchanges the two summands of (\ref{threefold}). The
non-compact space $X_p$ may be regarded as the local neighbourhood
of an elliptic curve embedded in a compact Calabi-Yau threefold, with
the local Calabi-Yau condition in this case being $c_1(TX_p)=0$. We
are free to twist the fibration (\ref{threefold}) by any torsion-free
holomorphic line bundle over $\torus^2$ without affecting the gauge
and string theory physics.

The low energy effective four-dimensional theory on $\mathbb{R}^{3,1}$
is $\mathcal{N}=2$ supergravity. Its charged extremal black hole
solutions are uniquely characterized through the attractor
mechanism~\cite{LopesCardoso:1998wt}--\cite{LopesCardoso:1999ur,Ferrara:1995ih,Strominger:1996kf}
by the magnetic and electric charges $(m^l,e_l)$,
where $l=0,1,\dots,n_V$ with $n_V=h^{1,1}(X_p)=\dim H^{1,1}(X_p)$ the
number of vector multiplets in the low energy supergravity
theory. This fixes the K\"ahler moduli $k\in H^{1,1}(X_p)$ to be
proportional to the charge vector $m^l$. Since the only compact
holomorphic two-cycle of $X_p$ is the base $\mathbb{T}^2$, embedded in
$X_p$ as the zero section of (\ref{threefold}), four-dimensional BPS
black holes are uniquely characterized by the sets of charges
$(m^0,m^1,e_0,e_1)$. Motivated by insight from topological string
theory~\cite{Vafa:2004qa,Aganagic:2004js}, in the following we will
choose the configuration $(m^0=0, m^1= p \,N, e_0,e_1)$ with arbitrary
$e_0,e_1\in\nat_0$.

After taking into account the effects of supergravity backreaction,
the black hole solution can be realized by wrapping $N$ D$4$ branes on
the non-compact four-cycle $C_4$ in $X_p$ which is the total space of
the holomorphic line bundle $\mathcal{O}_{\torus^2}(-p) \rightarrow
\mathbb{T}^2$, along with D2 branes wrapping the base torus
$\mathbb{T}^2$ and D0 branes. This brane configuration breaks the
symmetry $p\to-p$. Note that the number of D2 and D0 branes
is not fixed, and that no D6 brane charge is turned on. The number of
microstates of the four-dimensional black hole is given by the number
of bound states that the D2 and D0 branes can form with the D4
branes. This number can be counted by studying the $\mathcal{N}=4$
topologically twisted $U(N)$ gauge theory living on the worldvolume of
the D4 branes{\footnote{The maximally supersymmetric gauge theory is
    topologically twisted due to the non-triviality of the D4 brane
    geometry~\cite{Sadov}. The topological twist is the only way to
    realize covariantly constant spinors, since they are transformed
    into scalars. }} with certain interactions that correspond to
turning on chemical potentials for the D2 and D0 branes.

D$p$ branes wrapping a $(p+1)$-manifold $M_{p+1}$ couple to all
Ramond-Ramond fields $C_{(q)}$ through anomalous couplings of the
form~\cite{Douglas:1995bn} $\int_{{M}_{p+1}}\, \sum_q\, C_{(q)}\wedge\Tr
\e^{2\pi\,\alpha'\,F}$, where $F$ is the gauge field strength living on the brane
worldvolume. In particular, there is a coupling quadratic in $F$ given
by $\int_{{M}_{p+1}}\, C_{(p-3)} \wedge \Tr F \wedge F$. This
means that an instanton configuration excited on a four-dimensional
submanifold of the worldvolume is equivalent to a D$(p-4)$ brane
charge. Thus counting D4--D0 brane bound states is equivalent to counting
instantons on the four-dimensional part of the D4 brane
worldvolume that lies inside the local Calabi-Yau threefold. A
similar reasoning applies to the chemical potential for the D2
branes which is related to the anomalous couplings
$\int_{C_4}\,C_{(3)}\wedge\Tr F$. To obtain the correct charges one
then takes the Ramond-Ramond three-form field $C_{(3)}$ to be
constant, i.e. proportional to the volume form of the wrapped
submanifold $\torus^2$. One thus concludes that the counting of
BPS bound states is equivalent to introducing the observables
\begin{equation} \label{4dobservables}
\mathcal{S}=\frac{1}{2 g_s}\, \int_{C_4}\, 
\Tr F \wedge F + \frac{\theta}{g_s}
\,\int_{C_4}\, \Tr F \wedge \omega_{\torus^2} \ ,
\end{equation}
where $g_s$ is the string coupling constant and $\omega_{\torus^2}$ is
the unit volume form on $\mathbb{T}^2$
(i.e. $\int_{\torus^2}\,\omega_{\torus^2}=1$). The chemical potentials
for the D0 and D2 branes are then respectively given by $\phi^0 = {4
  \pi^2}/{g_s}$ and $\phi^1 = {2\pi\,\theta}/{g_s}$.

\subsection{Two-Dimensional Yang-Mills Theory\label{2DYM}}

According to the conjecture of~\cite{Vafa:2004qa,Aganagic:2004js}, the
counting of the number of black hole microstates, or equivalently the
computation of the ${\cal N}=4$ gauge theory vacuum expectation value
of the observables (\ref{4dobservables}),
reduces to the evaluation of the partition function of $U(N)$
Yang-Mills theory on the base torus. In the more general case where
the local threefold is a fibration over a genus $g$ Riemann surface,
the gauge theory carries a deformation that reflects the non-triviality
of the holomorphic vector bundle. This deformation is tantamount to
replacing, in the usual Migdal heat kernel
expansion~\cite{Migdal:1975zg,Rusakov:1990rs}, the dimension of $U(N)$
representations with their {\it quantum} dimension. However, the case
of the torus is special. Since in general the dimension appears to the
power of the Euler characteristic $2-2g$ of the surface, the gauge
theory at $g=1$ is insensitive to the deformation.
The only remnant of the non-triviality of the fibration is in the
insertion of the four-dimensional mass deformation $ p \,\Tr(\Phi^2) $,
where the scalar field $\Phi$ is the holonomy of the gauge field
along the fibre at infinity in $C_4$. The four-dimensional gauge
theory localizes onto field configurations which are invariant under
the natural $U(1)$ scaling action along the fibre, leading to the
action
\begin{equation} \label{qYMaction}
S =\frac{1}{g_s}\,\int_{\mathbb{T}^2}\, \mathrm{Tr}(\Phi \,F)-
\frac{p}{2 g_s}\,\int_{\mathbb{T}^2}\,
\mathrm{Tr}\left(\Phi^2\right)~\omega_{\torus^2}+\frac{\theta}{g_s}\,
\int_{\torus^2}\,\mathrm{Tr}(\Phi)~\omega_{\torus^2} \ .
\end{equation}
By integrating out $\Phi$, one can recast (\ref{qYMaction}) in the
standard form where the role of the dimensionless Yang-Mills coupling
is played by the combination $g_s \, p$.

The partition function of $U(N)$ Yang-Mills theory on a torus can be
expressed as the heat kernel
expansion~\cite{Migdal:1975zg,Rusakov:1990rs}
\begin{equation} \label{Z2dtorusgen}
{\cal Z}^{\rm YM} = \sum_{R}\, \e^{ - \frac{g_s \, p}{2}\, C_2
(R)+\ii\theta\, C_1(R)} \ ,
\end{equation}
where the sum over $R$ runs through all irreducible representations of
the gauge group $U(N)$ which can be labelled by sets of $N$ increasing
integers $+\infty > n_1(R) > n_2(R) > \cdots
> n_N(R) > - \infty $ giving the lengths of the rows of the Young
tableau corresponding to $R$. The first and second Casimir
invariants of $R$ can be expressed in term of these integers as
\begin{equation}
C_1(R)=\sum_{i=1}^N\, n_i(R) \ ,\,\,\,\,\,\,\, C_2 (R)=-\frac{N}{12}\,
\left( N^2 -1 \right) + \sum_{i=1}^N \,\left( n_i(R) - \frac{N-1}{2}
\right)^2 \ .
\end{equation}
Note that when the degree $p$ is positive, it can be absorbed into a
redefinition of the string coupling constant $g_s$. This fact is
peculiar to the torus since the heat kernel expansion is not
weighted by the (quantum) dimension of the representation. On any
other Riemann surface, this redefinition would not be possible. We
will nevertheless keep the explicit dependence on $p$ in order to
better describe later on the dependence of topological invariants and
the relation with Chern-Simons gauge theory on Seifert fibrations.

\subsection{The Nonabelian Localization Formula\label{YMNLF}}

Two-dimensional Yang-Mills theory has a dual description in terms of
instantons~\cite{Witten:1992xu} in which its modular properties are
manifest. The sum over $U(N)$ weights ${\mbf
  n}(R)=\{n_i(R)\}$ in eq.~(\ref{Z2dtorusgen}) can be traded for a sum
over conjugacy classes of the symmetric group ${S}_N$ on $N$
elements~\cite{Griguolo,Paniak:2002fi}. The conjugacy
classes are labelled by sets of $N$ integers $0 \le \nu_a \le
\lfloor N/a\rfloor$ which define a partition of $N$, $\nu_1 + 2\,
\nu_2 + \cdots+ N\, \nu_N = N$. A conjugacy class contains
${N!}/{\prod_{a=1}^{N} \,a^{\nu_a}\, \nu_a !}$ elements, each of which
gives the same contribution to the partition function and comes with
parity factors $(-1)^{\nu_{a}}$. The heat kernel expansion can be recast
in the form of a sum over instanton solutions by means of the
Poisson resummation formula
\begin{equation}
\sum_{n= -\infty}^{\infty}\, f(n) = \sum_{m=-\infty}^{\infty}\,
\int_{-\infty}^{\infty} \,\mathrm{d} s ~ f(s) ~ \e^{2 \pi \ii m \,
s} \ .
\end{equation}

In this way one finds that the $U(N)$ Yang-Mills partition function
on the torus in the instanton representation is given up to
normalization by~\cite{Paniak:2002fi,Griguolo:2004jp}
\begin{eqnarray} \label{instantonrep}
{\cal Z}^{\rm YM} &=&  \sum_{\stackrel{\scriptstyle
\vnu\in\nat_0^{N}}{\scriptstyle\sum_a\,a\,\nu_a=N}}~
\prod_{a=1}^{N}\,\frac{(-1)^{\nu_a}}{\nu_a!}\,\left(
\frac{2\pi}{a^3\, g_s \, p}\right)^{\nu_a/2} \cr &&\times\,
\sum_{\vm\in\zed^{|\vnu|}}\,(-1)^{(N-1)\sum_am_a}\,\exp\left[-
\frac{2\pi^2}{g_s \,
p}\,\sum_{l=1}^N\,\frac1l\,\sum_{j=1+\nu_1+\dots+
\nu_{l-1}}^{\nu_1+\dots+\nu_l}\,m_j^2\right] \ .
\end{eqnarray}
(For simplicity we set the $\theta$-angle to zero in the following
unless explicitly stated otherwise.) We can simplify the form of this
expression somewhat by introducing the elliptic Jacobi theta-functions 
\begin{equation}
\vartheta_3(z,\tau):= \sum_{m= -\infty}^{\infty} \, \e^{2\pi\ii m\,z+
\pi\ii m^2\,\tau} \ .
\label{Jacobitheta}\end{equation}
Then (\ref{instantonrep}) can be written as
\begin{equation} \label{YMinstanton}
{\cal Z}^{\rm YM}  =  \sum_{\stackrel{\scriptstyle
\vnu\in\nat_0^{N}}{\scriptstyle\sum_a\,a\,\nu_a=N}}~
\prod_{a=1}^{N}\,\frac{(-1)^{\nu_a}}{\nu_a!}\,\left(
\frac{2\pi}{a^3\, g_s\, p}\right)^{\nu_a/2}\, \left[ \vartheta_3 \bigl(
\mbox{$\frac{1}{2}\, (N-1) \,,\, \frac{2 \pi \ii}{a\, g_s\,  p}$} \bigr)
\right]^{\nu_a} \ .
\end{equation}

In this form the localization of the path integral onto critical
points of the Yang-Mills action is manifest, and these formulas
have a natural interpretation in terms of instantons. For this, let us
recall how to construct general solutions of the $U(N)$
Yang-Mills equations on a compact Riemann surface $\Sigma$ of genus
$g$~\cite{AtiyahBott}. The Yang-Mills equations on $\Sigma$ can be
written as $\dd_AX=0$ where $\dd_A$ is the covariant derivative with
respect to a gauge connection $A$ on a principal $U(N)$-bundle
$\mathcal{P}_N\to\Sigma$, and $X={}^*F$ with $F=F_A$ the
curvature of $A$. This equation means that $X$ is a covariantly constant
section of the adjoint bundle ${\rm ad}(\mathcal{P}_N)$, and
thus it yields a reduction of the $U(N)$ structure group to the
centralizer subgroup $U_X\subset U(N)$ which commutes with $X$. As a
consequence, any $U(N)$ Yang-Mills solution can be described as a {\it flat}
connection of an associated $U_X$-bundle which is twisted by a
constant curvature line bundle associated to the $U(1)$ subgroup of
$U(N)$ generated by $X$. Flat connections in turn are uniquely
characterized by their holonomies around closed curves on $\Sigma$,
and can be concretely described in terms of group homomorphisms from
the fundamental group of $\Sigma$ to the structure group of the gauge
bundle.

The moduli space of gauge inequivalent Yang-Mills
connections can be conveniently described by introducing
the universal central extension
$\Gamma_\real=\pi_1(\Sigma)\times_\zed\real$ of the fundamental group
$\pi_1(\Sigma)$ of $\Sigma$, with the center of the $U(N)$ gauge group
extended to $\real$. Then according to the above discussion, there is
a one-to-one correspondence between gauge equivalence classes of
connections solving the Yang-Mills equations on $\Sigma$ and
conjugacy classes of homomorphisms $\rho:\Gamma_\real\rightarrow U(N)$
with $\rho(\pi_1(\Sigma))\subset SU(N)$. Explicitly, the
gauge connection $A^{(\rho)}$ associated to $\rho$ has curvature
$F^{(\rho)}=X^{(\rho)}\otimes\omega_\Sigma$, where $\omega_\Sigma$ is
the unit volume form of $\Sigma$ and $X^{(\rho)}$ is an element of the
Lie algebra of $U(N)$ defined by the map $\dd\rho:\real\rightarrow
\mathfrak{u}(N)$. For the purpose of evaluating the Yang-Mills action
on classical solutions, we have only to find all possible
$X^{(\rho)}$. This is straightforward to do, since the homomorphism
$\rho$ furnishes an $N$-dimensional unitary representation of
$\Gamma_\real$.

When the representation $\rho$ is {\it irreducible},
$X^{(\rho)}$ is central with respect to the adjoint action of $U(N)$ and
therefore its eigenvalues are all equal to a certain real number
$\lambda$. The Chern class of a principal $U(N)$-bundle
$\mathcal{P}_N$ over $\Sigma$ is always an
integer, $\frac{1}{2\pi}\,\int_\Sigma\, {\rm Tr}\,F^{(\rho)}=m\in\zed$,
and one has $\frac1{2\pi}\,{\rm Tr}\,X^{(\rho)}=m$
which completely determines $\lambda$ to be
$\lambda=\frac{2\pi\,m}{N}$. On the other hand, if $\rho$ is {\it
  reducible} then $X^{(\rho)}$ is generically central with respect to
a subgroup of $U(N)$ of the form
\begin{equation}
U_{X^{(\rho)}}=U(N_1)\times U(N_2)\times\cdots\times U(N_r)
\label{centralsubgp}\end{equation}
with $N_1+N_2+\cdots+N_r=N$. Because $X^{(\rho)}$ is constant, the
adjoint action ${\rm ad}_{X^{(\rho)}}=[X^{(\rho)},-]$ is a bundle map
${\rm ad}_{X^{(\rho)}}:{\rm ad}(\mathcal{P}_N)\to{\rm
  ad}(\mathcal{P}_N)$, and we can decompose the bundle ${\rm
  ad}(\mathcal{P}_N)$ under this action into a direct sum of
sub-bundles each associated to a distinct eigenvalue of ${\rm
  ad}_{X^{(\rho)}}$. This effectively manifests itself as a
decomposition of the original $U(N)$ gauge bundle into eigenbundles
associated to the distinct eigenvalues of $X^{(\rho)}$ itself as
\begin{equation}
\mathcal{P}_N=\bigoplus_{i=1}^r\,
\mathcal{P}_{N_i} \ . \label{riduco}
\end{equation}
Since the individual Chern classes in this decomposition are
necessarily integers, $X^{(\rho)}$ has eigenvalues
$\lambda_1,\lambda_2,\dots,\lambda_r$ with multiplicities
$N_1,N_2,\dots,N_r$ and
\begin{equation}
\lambda_i=\frac{2\pi\,m_i}{N_i} \ ,
\label{lambdaiexpl}\end{equation}
where $m_i\in\zed$ and $\sum_{i}\,m_i=m$ is the total Chern class. 

For fixed Chern class $m$, the absolute minimum value of
the Yang-Mills functional
$\frac1{2g_s\,p}\,\int_{\Sigma}\,\Tr(F\wedge{}^*F)$ is reached when
all the eigenvalues are equal to $\lambda={\frac{2\pi\,m}{N}}$. The
action at this minimum is
\begin{equation}
S_{{\rm min}}=\frac{2\pi^2}{g_s\,p}\,\frac{m^2}{N} \ .
\end{equation}
The generic $U(N)$ Yang-Mills connection is a direct sum of Yang-Mills
minima for the sub-bundles in the decomposition
(\ref{riduco}). The eigenvalues of $X^{(\rho)}$ can be
rational-valued with denominator $N$ when the representation $\rho$ is
irreducible. At the opposite extreme, if a representation $\rho$
which is completely reducible to the maximal torus $U(1)^N$ exists,
then the eigenvalues are all integers. All intermediate possibilities
according to eq.~(\ref{riduco}) can also appear.

It is now easy to understand eq.~(\ref{instantonrep}) in light
of this discussion. The exponential factors are the
Boltzmann weights of the Yang-Mills action on a solution determined by
$\nu_a$ eigenvalues $\lambda_{i_a}=\displaystyle{2\pi\, m_{i_a}/a}$ with
multiplicity $a$, for $a=1,\dots,N$, corresponding to the reduction in
structure group $U(N)\to U(1)^{\nu_1}\times
U(2)^{\nu_2}\times\cdots\times U(N)^{\nu_N}$. The full partition
function sums over all classical solutions, with the Boltzmann factor
of the Yang-Mills action accompanied by a fluctuation determinant
according to the nonabelian localization
principle~\cite{Witten:1992xu,Beasley:2005vf}. The singular terms in
eq.~(\ref{instantonrep}) as $g_s\,p\to0$ arise precisely from these
determinants and are in principle recovered by an integration over the moduli
space of classical solutions. The only subtle point here is that
distinct partitions in eq.~(\ref{instantonrep}) are not necessarily
related to gauge inequivalent solutions. One has to identify,
as coming from the same instanton, some contributions related to
different conjugacy classes $\{\nu_a\}$ of $S_N$. Such a
reorganisation of the instanton sum provides the effective form of the
fluctuation determinants.

\subsection{S-Duality and Instanton Moduli Spaces\label{SDuality}}

To further understand the structure of the torus partition
function and its relation to the $\mathcal{N}=4$ topological gauge
theory in four dimensions, we have to describe the moduli space of
classical instantons in order to be able to single out the
contribution of the fluctuations to the nonabelian localization
formula. For this, we recall the classification of unitary
representations of the group $\Gamma_\real$. According to the
Narasimhan-Seshadri theorem~\cite{indi}, for genus $g>1$ there is
associated to any pair of integers $(N,m)$ a $d$-dimensional
irreducible representation of $\Gamma_\real$, where $d={\rm
  gcd}(N,m)$, giving the Yang-Mills
ground state for gauge group $U(N)$ and with Chern class $m$. The
genus one case is different, because the fundamental group
$\pi_1(\torus^2)=\zed^2$ is abelian and so it has no irreducible
representations for $d>1$. Thus for $m=0$ and $N>1$ there are no
irreducible representations of $\Gamma_\real$, corresponding
to the well known fact that every flat connection on $\torus^2$ is
reducible (recall that flat $U(N)$ gauge connections on $\Sigma$ are
described by group homomorphisms $\gamma:\pi_1(\Sigma)\to U(N)$). Only
for $(N,m)$ coprime do irreducible
representations of the universal central extension $\Gamma_\real$
exist. In this case the moduli space of Yang-Mills solutions is a
smooth manifold. When reducible representations of $\Gamma_\real$
occur, orbifold singularities appear in the instanton moduli
space~\cite{Paniak:2002fi}. The result can be summarized as follows.

Again we let $N = \sum_{a=1}^{N}\,{a \,\nu_a}$ be a partition of the
rank $N$ corresponding to the gauge symmetry breaking $U(N)
\rightarrow\prod_{a = 1}^{N}\, U(a)^{\nu_a}$. Consider a classical
Yang-Mills solution with component Chern numbers $m_a$ on a $U(N)$
gauge bundle of Chern class $m=\sum_{a=1}^N\,m_a$. Then the moduli
space of instantons is given by the symmetric products
\begin{equation}
{\cal M}_{N,m} \left( \vnu,\vm\right) = \prod_{a=1}^{N}\,
\mathrm{Sym}^{\nu_a}\, \tilde{\torus}^2 \ ,
\label{modspinstNq}\end{equation}
where $\tilde{\torus}^2$ is a dual torus. This result generalizes the
moduli space of flat connections, with $\nu_N=1$, given by ${\cal
  M}_{N,0}(1,0)\cong{\rm
  Hom}(\pi_1(\torus^2),U(N))/U(N)\cong(\tilde\torus{}^2)^N/S_N$, where
the Weyl subgroup $S_N\subset U(N)$ is the residual gauge symmetry acting
by permuting the components of the maximal torus $U(1)^N$. The
symmetric orbifold structure in (\ref{modspinstNq}) arises from the
non-trivial fixed points of the action of the symmetric group
${S}_{\nu_a}$ on $( \tilde{\torus}^2)^{\nu_a}$, which is the subgroup
of the stabilizer group of the conjugacy class $\vnu$ of $S_N$ which
permutes the $\nu_a$ cycles of length $a$. The orbifold
singularities for coincident instantons are directly related to
the singular behaviour of the Yang-Mills partition function in the
instanton representation in the weak coupling
limit~\cite{Paniak:2002fi}.

However, as mentioned already above, one needs to carefully
reorganise the sum over partitions in the nonabelian localization
formula to ``instanton partitions'', reflecting a bundle splitting of the type
  (\ref{riduco}) into gauge inequivalent eigenbundles, in order to
  prevent the overcounting of distinct Yang-Mills stationary
  points. This consists in writing the integer pairs corresponding to
  (\ref{riduco}) as $(N_i,m_i)=d_i\,(N_i',m_i')$ , where $d_i={\rm
    gcd}(N_i,m_i)$ and $(N_i',m_i')$ are coprime integers. As the eigenvalues
  (\ref{lambdaiexpl}) are independent of the ranks $d_i$, one restricts
  the counting of Yang-Mills critical points to those labelled by
  triples of integers $(\nu_a,N_a,m_a)$ which satisfy, in addition to
  the partition constraints above, the requirement that $(N_a,m_a)$ be
  {\it distinct} coprime integers. These additional constraints are
  implicitly understood in (\ref{modspinstNq}) and they ensure
  that one does not count as distinct those bundle splittings
  (\ref{riduco}) which contain some sub-bundles that can themselves be
  decomposed into irreducible components. (See
  Sect.~9 of~\cite{Paniak:2002fi} for a more detailed treatment.)

To see how this works in practice, let us consider the simple example
of $U(2)$ gauge theory on $\torus^2$. The partition function may then
be written out explicitly as
\beq
{\cal Z}^{\rm YM}=
\sum_{m_1,m_2=-\infty}^{\infty}\,(-1)^{m_1+m_2}\,\frac{2\pi}{g_s\, p}~
\e^{-\frac{2\pi^2}{g_s\,p}\,(m_1^2+m_2^2)}
+\sum_{m_0=-\infty}^{\infty}\,\frac{(-1)^{m_0}}{\sqrt{2}}\,
\sqrt{\frac{2\pi}{g_s\,p}}~\e^{-\frac{\pi^2}{g_s\,
p}\,m_0^2} \ . \label{su2}
\eeq
The first term in (\ref{su2}) corresponds to solutions coming from the
reduction $U(2)\rightarrow U(1)\times U(1)$, while the second term receives
contributions only from global minima with all eigenvalues equal and
proportional to ${\frac{m_0}{2}}$. In both sums there are connections
contributing to the same value of the Yang-Mills action. For
instance, by taking $m_1=m_2=m'$ in the first sum and $m_0=2{m'}$ in the
second sum, both contributions represent the minimum value of the
action for the Chern class $m=2{m'}$. Are these connections gauge
inequivalent? The answer is negative because, as discussed above, for
$g=1$ the solutions are irreducible only when the integers $N$ and $m$ are
coprime. When the Chern class is an even integer, the minima therefore
originate from completely reducible connections (of the type
$U(1)\times U(1)$). It is better to rewrite the partition function in
the form
\bea
{\cal Z}^{\rm YM}&=&
\sum_{m_1,m_2=-\infty}^{\infty}\, \left((-1)^{m_1+m_2}\frac{2\pi}{g_s\,
p}+\delta_{m_1,m_2}\,\frac{1}{\sqrt{2}}\,\sqrt{\frac{2\pi}{g_s\, p}}~
\right)~\e^{-\frac{2\pi^2}{g_s\,
p}\,(m_1^2+m_2^2)}\nonumber\\ &&-\,
\sum_{m_0=-\infty}^{\infty}\,\frac{1}{\sqrt{2}}\,
\sqrt{\frac{2\pi}{g_s\,p}}~\e^{-\frac{\pi^2}{g_s\,
p}\,(2m_0+1)^2} \ , \label{su22}
\eea
where now every classical contribution appears together with the part
coming from the quantum fluctuations. Note that reducible and
irreducible connections are completely disentangled in this
rewriting.

When the Chern number $m=2m_0+1$ is odd the two terms in (\ref{su22})
respectively come from instantons in the smooth moduli spaces
\bea
{\cal M}_{2,m=2m_0+1}(2,m)&=&\tilde{\torus}{}^2 \ ,
\label{calM2qodd}\\[4pt] {\cal
M}_{2,m} \big((1,1)\,,\,(m_1,m_2)\big)&=&\tilde{\torus}{}^2\times
\tilde{\torus}{}^2 \ .
\label{calMqodd}\eea
Heuristically, with the appropriate symmetry factors, each factor
$\tilde{\torus}{}^2$, representing the one-instanton moduli space,
contributes a mode with fluctuation determinant $-1/\sqrt{4g_s\,
  p}$. On the other hand, when $m=2m'$ is even only the first term
contributes, and we find again instantons in the smooth moduli space
(\ref{calMqodd}) with $m=2m'$ and $m_1\neq m_2$. The novelty now comes
from the two gauge equivalent instanton contributions for $m_1=m_2=m'$
in the singular moduli space
\beq
{\cal M}_{2,m=2m'}(2,2m'\,)={\rm Sym}^2\,\tilde{\torus}{}^2 \ .
\label{calMqextra}\eeq
The singular locus of the symmetric orbifold (\ref{calMqextra}) is the
disjoint union ${\rm Sym}^2\,\tilde{\mathbb
T}{}^2\,\amalg\,\tilde{\mathbb T}{}^2$ with the disjoint sets
corresponding to the identity and order two elements of the cyclic
group $S_2=\zed_2$, respectively. In principle the same analysis can
be repeated for any $N$ (see~\cite{Paniak:2002fi} for the case $N=3$),
with increasing complexity in deriving the explicit forms of the
singular fluctuations.

This partition function should reproduce instanton calculus for the
${\cal N}=4$ topologically twisted Yang-Mills theory on the
non-compact four-cycle $C_4$. In particular, it should compute the
generating functional (at $\theta=0$) for Euler characteristics of the
moduli spaces of self-dual $U(N)$ gauge connections on $C_4$ given
by~\cite{VafaWitten}
\beq
{\cal Z}^{{\cal N}=4}=\sum_{e_0,e_1\geq0}\,\Omega_N(e_0,e_1)~
\exp\left(-\frac{4\pi^2}{g_s}\,e_0-\frac{2\pi\,\theta}{g_s}\,e_1
\right) \ ,
\label{Eulergenfn}\eeq
where $e_0$ and $e_1$ are respectively the D4--D0 and D4--D2 electric
charges, obtained from summing over different topological classes of
gauge bundles in the ${\cal N}=4$ Yang-Mills amplitude with the observables
(\ref{4dobservables}) inserted, and $\Omega_N(e_0,e_1)$ is the index
of BPS bound states with the given charges.
One would thus expect that the final result respects the well known
S-duality properties of ${\cal N}=4$ supersymmetric Yang-Mills theory,
so that ${\cal Z}^{\rm YM}$ should be a modular form. Moreover,
because the Euler characteristic is an indexed counting of classical
solutions, we would not expect any contributions from quantum
fluctuations. We see that apparently Yang-Mills theory on the torus
disagrees with both expectations.

Let us analyse these discrepencies in more detail, beginning with the
modular structure. The explicit form of the instanton partition
function in (\ref{YMinstanton}) is a sum of
theta-functions each with a coupling dependent factor. This is a
(finite) sum of modular forms, each transforming with
different weights, according to the eigenbundle decomposition
(\ref{riduco}). As a result it is not immediate how to extract the
information on the number of black hole microstates. In the dual
black hole language the number of microstates with a given charge
configuration arises as a combinatorial problem where one has to
sum over all prefactors whose Boltzmann weights correspond to
the same set of global charges in (\ref{Eulergenfn}). In light of the
above discussion, this is equivalent to organizing the sum over
two-dimensional instantons into a sum over gauge inequivalent
configurations. In doing so one has to face the effect of the singular
fluctuations, depending explicitly on the Yang-Mills coupling, and
therefore its interpretation as the Witten index of black hole states
appears difficult.

On the other hand, the occurence of the bundle reduction
(\ref{riduco}) in the description of the two-dimensional gauge theory
partition function is at the very heart of the interpretation
problem. The same phenomenon has been found in~\cite{Aganagic2:2005}
in studying the counting of black hole microstates on local
$\mathbb{P}^2$. There it was suggested that the sectors corresponding
to a non-trivial reduction in structure group should come from
marginally bound states of D-branes, and only the instantons associated to the
trivial partition with $\nu_N=1$ should be considered. However, a
direct comparison between four-dimensional gauge field configurations
and two-dimensional instantons is presently out of reach because of
our poor understanding of four-dimensional gauge theories on curved
backgrounds and non-compact spaces.

Some of these unexpected features could also be related to the
non-compactness of the four-cycle $C_4$ on which the supersymmetric
gauge theory is defined. Recall that the two-dimensional gauge
theory is the localization under an appropriate $U(1)$-action of
the full $\mathcal{N}=4$ topologically twisted gauge theory whose
topological sectors are given by the black hole charges, as in
(\ref{Eulergenfn}). In this setup it is tempting to suggest that the
$g_s$-dependent factors in (\ref{YMinstanton}) arise as a consequence
of imposing boundary conditions on the four-dimensional gauge field at
infinity within each superselection sector separately. As we will see
in Sect.~6, it is possible to give a geometrical interpretation of the
$g_s$-dependent factors as the fluctuations around a flat connection
in Chern-Simons gauge theory on the boundary of $C_4$, in complete
analogy with what was found in~\cite{Caporaso:2005ta} in the case of
local $\mathbb{P}^1$.

\section{Gauge String Theory on the Elliptic Curve\label{YMLargeN}}

In preparation for our description in terms of closed topological
strings on the local elliptic curve, in this section we will study the
large $N$ limit of $U(N)$ Yang-Mills theory on the base torus
$\torus^2$. We will pay particular attention to the role played by the
$U(1)$ factor of the gauge group, in light of the observations
of~\cite{Vafa:2004qa}. We will find rather explicitly that the
partition function has a natural interpretation in terms of {\it open}
string degrees of freedom in the non-trivial $U(1)$ charge sectors.

\subsection{Large~$N$ Expansion\label{LargeNExp}}

\EPSFIGURE{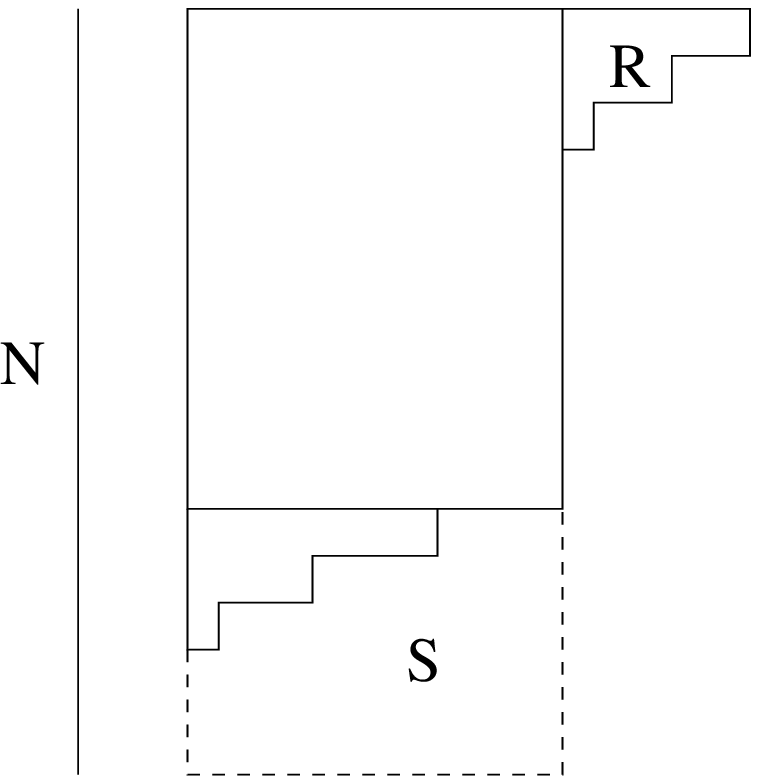,width=2in}{Coupled representations with $S=R_+$
  and $R=R_-$.\label{coupledreps}}

It is convenient to use the relation $U(N) = SU(N) \times U(1) /
\mathbb{Z}_N$ to decompose irreducible representations $R$ of
$U(N)$ in terms of $SU(N)$ representations $\hat{R}$ and $U(1)$
charges $m = N\, r + |\hat R|$, where $|\hat R|$ is the total number of
boxes in the Young tableau associated with $\hat{R}$ and $r \in
\mathbb{Z}$. The quadratic Casimir invariant of $R$ can be related to
$SU(N)$ and $U(1)$ invariants as
\begin{equation}
C_2 (R) = C_2 ( \hat{R} , m ) = C_2 ( \hat{R} ) + \frac{m^2}{N}
\label{C2RC2hatR}\end{equation}
where $C_2 ( \hat{R} ) = |\hat R| \, N + \kappa_{\hat{R}} -
\frac{|\hat R|^2}{N}$. If we label by $n_i(\hat R)$ the length of the
$i$-th row of the Young tableau for $\hat R$, with $n_1(\hat R) \ge
n_2(\hat R) \ge\cdots \ge n_{N-1}(\hat R)\ge 0$ and $\sum_{i=1}^{N-1}\,
n_i(\hat R) = |\hat R|$, then we can write
\beq
\kappa_{\hat{R}}=
\sum_{i=1}^{N-1}\, n_i(\hat R)\, \big(n_i(\hat R) +1 -2i\big) \ .
\label{kappahatRdef}\eeq

In the large $N$ limit, any representation of $SU(N)$ can be
expressed uniquely~\cite{GrossTaylor} in terms of coupled
representations $ \hat{R} = \overline{\hat{R}_+} \, \hat{R}_-$ such that
the Young tableau for $R$ is given by joining a chiral tableau $R_+$ to an
antichiral tableau $R_-$ as depicted in Fig.~\ref{coupledreps}.
The quadratic Casimir invariant of the
representation $\hat{R}$ is given by
\begin{equation}
C_2 ( \hat R ) = C_2 (\hat{R}_+) + C_2 (\hat{R}_-) + \frac{2\, |\hat R_+| \,
|\hat R_-|}{N}
\end{equation}
and the total number of boxes in $\hat{R}$ is
\begin{equation}
|\hat R| = |\hat R_-| + N\, n_{1}(\hat{R}_+) - |\hat R_+|
\end{equation}
where $n_{1}(\hat R_+)$ is the number of boxes in the first row of
$\overline{\hat{R}_+}$. The $U(1)$ charge associated to
$\hat{R}$ can be written as
\begin{equation}
m = |\hat R| + r \,N = |\hat R_-| - |\hat R_+| + \ell\, N
\end{equation}
where $\ell = n_{1}(\hat{R}_+) + r$. In the large $N$ limit the
quantities $|\hat R_\pm|$ and $\ell$ are all understood as being
small compared to $N$.

Given a coupled $SU(N)$ representation $\hat R$, the quadratic Casimir
invariant of the corresponding $U(N)$ representation $R$
is given by
\bea
C_2 (R) &=& |\hat R_-| \, N + \kappa_{\hat{R}_-} - \frac{|\hat
  R_-|^2}{N} +|\hat R_+|
\, N + \kappa_{\hat{R}_+} - \frac{|\hat R_+|^2}{N} \nonumber\\
&& +\, \frac{2\,|\hat R_+| \,
|\hat R_-|}{N} + \frac{\big(|\hat R_-| - |\hat R_+| + \ell\, N
\big)^2}{N} \ .
\label{TCasimir}\eea
Note that changing the sign of $\ell$ in (\ref{TCasimir}) is
equivalent to exchanging $|\hat R_+|$ with $|\hat R_-|$. We will
momentarily set $\theta=0$. By using
(\ref{TCasimir}) and introducing the 't~Hooft coupling
\begin{equation}
\lambda = g_s \, p \, N \ ,
\label{tHooftdef}\end{equation}
we can recast the torus partition function (\ref{Z2dtorusgen}) in the
form
\begin{equation} \label{Z2dtorus}
{\cal Z}^{\rm YM} = \sum_{\ell= - \infty}^{ \infty}\, \e^{-{\lambda\,
\ell^2}/{2}}~ \sum_{\hat{R}_+}\, \e^{- \frac{\lambda}{2N}\,
 (|\hat R_+|\,N + \kappa_{\hat{R}_+ } +2\,|\hat R_+|\, \ell)}~
\sum_{\hat{R}_-}\,\e^{- \frac{\lambda}{2N}\, (|\hat R_-|\,N + \kappa_{
\hat{R}_- } - 2\,|\hat R_-|\, \ell)} \ .
\end{equation}
The partition function can be naturally written as a sum over
$U(1)$ charge sectors, each sector $\ell\in\zed$ being weighted by the
factor $\e^{-{\lambda\, \ell^2}/{2}}$ determined by the energy of the
charge. We are thus led to define
\begin{equation}
{\cal Z}^{\rm YM} = \sum_{\ell = - \infty}^{\infty}\, {\cal Z}(\ell) :=
\sum_{\ell = - \infty}^{\infty}\, \e^{-{\lambda\, \ell^2}/{2}}~ {\cal
Z}_+ (\ell) \,{\cal Z}_- (\ell)
\end{equation}
where the expressions for the partition functions ${\cal Z}(\ell)$ and
${\cal Z}_{\pm}(\ell)$ can be read off from (\ref{Z2dtorus}).

To analyse the contributions from a given charge sector, we follow the
trick employed in~\cite{Rudd:1994ta}. Consider the function
 \begin{equation}
{\cal Z}_+ (\lambda, \lambda'\,) := \sum_{\hat{R}_+}\, \e^{-
{\lambda\, \kappa_{\hat{R}_+} }/{2N}}~ \e^{- {\lambda'\,
|\hat R_+|}/{2}}
\label{calZlambdaprime}\end{equation}
with $\lambda$ and $\lambda'$ regarded formally as independent
variables. Then the $\ell$ dependence of the chiral partition function
${\cal Z}_+ (\ell)$ can be generated by differentiating
(\ref{calZlambdaprime}) with respect to $\lambda'$ to get
\begin{equation} \label{rudd}
{\cal Z}_+ (\ell) = \left.\exp\left(\mbox{$\frac{2 \lambda\,
          \ell}{N}\,\frac\partial{\partial{\lambda'}}$}\right)
{\cal Z}_+ (\lambda, \lambda'\,) \right|_{\lambda' = \lambda} \ .
\end{equation}
The action of the translation operator in
(\ref{rudd}) implies that the free energies associated to ${\cal
Z}_+ (\ell)=\e^{{\cal F}_+(\ell)}$ and ${\cal Z}_+ (\lambda,
\lambda'\,)=\e^{{\cal F}_+(\lambda,\lambda'\,)}$ are related by ${\cal
  F}_+(\ell) = {\cal F}_+ (\lambda, \lambda + \frac{2 \lambda\, \ell}{N})$. In
\cite{Rudd:1994ta} it was shown that the free energy ${\cal F}_+
(\lambda, \lambda'\,)$ admits the large $N$ genus expansion
\begin{equation}
{\cal F}_+ (\lambda, \lambda'\,) = \sum_{g=1}^{\infty}\, \left(
\frac{\lambda}{2 N} \right)^{2g-2} ~{F}_g (\lambda'\,)
\end{equation}
where each coefficient function ${F}_g (\lambda'\,)$ can be expressed
in terms of quasi-modular forms of weight $6g-6$. We
conclude that the chiral free energy is simply given by
\begin{equation} \label{chiralFE}
{\cal F}_+ (\ell) = \sum_{g=1}^{\infty} \,\left( \frac{\lambda}{2 N}
\right)^{2g-2}~ {F}_g  \big( \mbox{$\lambda+ \frac{2 \lambda\, \ell}{N}$}
\big) \ .
\end{equation}

\subsection{String Interpretation\label{StringInterpret}}

The effect of the $U(1)$ charge in (\ref{chiralFE}) appears only as a
shift in the argument of the forms ${F}_g (\lambda'\,)$ weighted with
a factor of $\frac{1}{N}$. The occurence of {\it odd} powers of
$\frac{1}{N}$ in their large $N$ expansion suggests an interpretation
in terms of {\it open} string degrees of freedom. For this, we expand
the quasi-modular forms in a Taylor series at $N\to\infty$ as
\begin{equation} \label{quasimod}
{F}_g \big(\mbox{$\lambda+ \frac{2 \lambda \,\ell}{N}$}\big) =
\sum_{b=0}^{\infty}\,\frac{1}{b!} \,\frac{\mathrm{d}^b{F}_g
(\lambda) }{\mathrm{d}
\lambda^b} ~\left( \frac{2 \lambda \,\ell}{N}  \right)^b \ .
\end{equation}
Each form ${F}_g (\lambda)$ in (\ref{quasimod}) can in turn be
expanded for $\lambda\to\infty$ as~\cite{Griguolo:2004jp}
\begin{equation} \label{hurwitz}
{F}_g (\lambda) = \frac{1}{(2g-2)!}\, \sum_{d=0}^{\infty}\,
H_{g,d}~\e^{-{\lambda\,d}/{2}}
\end{equation}
where $H_{g,d}=H_{g,d}^{\torus^2}(1^d)$ are the simple
Hurwitz numbers that count the \textit{connected}, simple branched
covering maps of degree $d$ and genus $g$ to the torus
$\mathbb{T}^2$. Collecting together
eqs.~(\ref{chiralFE})--(\ref{hurwitz}) we arrive finally at
\begin{equation} \label{chiralFE2}
{\cal F}_+ (\ell) = \sum_{g=1}^{\infty} \,\frac{1}{(2g-2)!}\, \left(
\frac{\lambda}{2 N} \right)^{2g-2}~ \sum_{b=0}^{\infty}~
\sum_{d=0}^{\infty}\, \frac{H_{g,d} \, d^b}{b!} \left( -
\frac{\lambda \,\ell}{N} \right)^b~ \e^{-{\lambda\,d}/{2}} \ .
\end{equation}

The expansion (\ref{chiralFE2}) can be given a very suggestive
interpretation. If we interpret the integer $b$ as the number of
boundaries of a Gross-Taylor string worldsheet $\Sigma$, then we can
define the total Euler characteristic $\chi=\chi(\Sigma) = 2g - 2 + b$
at genus $g$ and recast (\ref{chiralFE2}) in the form
\begin{equation} \label{chiralFE3}
{\cal F}_+ (\ell) = \sum_{\chi=0}^{\infty}\, \left(\frac{1}{N}
\right)^\chi~{F}_{\chi}(\lambda, \ell )
\end{equation}
where
\begin{equation}
{F}_{\chi} (\lambda, \ell ) = \left( \frac{\lambda}{2}
\right)^{\chi}\, \sum_{\stackrel{\scriptstyle b=0}
{\scriptstyle\chi-b~{\rm even}}}^{\chi}\, \frac{(-2\ell)^b}{b!\, (\chi - b)!}~
\sum_{d=0}^{\infty}\,H_{\chi,d,b} ~ \e^{-
{\lambda\,d}/{2}} \ .
\end{equation}
We can interpret (\ref{chiralFE3}) as a sum over simple branched
covering maps of winding number $d$ from a Riemann surface of
Euler characteristic $\chi$ to the torus, where each boundary is
mapped to a fixed point of $\torus^2$. By the Riemann-Hurwitz
theorem, these covering maps have $2g-2$ simple branch points which
contribute a factor $\big(\frac{\lambda}{2}\big)^{2g-2}$ from the
moduli space integration over the positions of the branch points
and with the combinatorial factor $\frac{1}{(\chi -b)!} =
\frac{1}{(2g-2)!} $ since the branch points are indistinguishable.
Similarly, the boundaries contribute $( \frac
{\lambda}{2})^b$ with the combinatorial factor $\frac{1}{b!}$
and with each weighted by the $U(1)$ charge $\ell$ which corresponds to a
boundary holonomy. The exponential factor
$\e^{-{\lambda\,d}/{2}}$ plays the role of the Nambu-Goto
action. Finally, the relative Hurwitz numbers $H_{\chi,d,
b} = H_{g,d} \, d^b$ count the number of inequivalent
$d$-sheeted coverings of the torus by a Riemann surface with $b$
boundaries and genus $g$, and it is given by the number of simple
covering maps $H_{g,d}$ times the total number of ways $d^b$
in which one can map any of the $b$ boundaries to the given fixed
point of $\torus^2$. Note that the open string degrees of freedom
disappear in the $\ell=0$ sector (as then only the $b=0$ term
contributes to the sum) corresponding to turning off the boundary
holonomies.

In order to write the total free energy, we have to sum the chiral
and the antichiral contributions. From (\ref{Z2dtorus}) we see that
the antichiral free energy ${\cal F}_-(\ell)$ is obtained from the
chiral free energy ${\cal F}_+(\ell)$ by reversing the sign of the
$U(1)$ charge $\ell$. But from the form of eq.~(\ref{chiralFE2}) it
follows that, in the sum of the chiral and antichiral free energies,
only the {\it even} powers of $\ell$ survive since they are
the only ones which are insensitive to the sign of $\ell$. This fixes
the number of boundaries $n$ to be an {\it even} integer. Therefore,
the total free energy is given by
\begin{equation} \label{totalFE}
{\cal F}(\ell) = 2\, \sum_{\stackrel{\scriptstyle\chi=0 }{\scriptstyle
    \chi ~\mathrm{even}}}^{\infty}\, \left( \frac{1}{N} \right)^\chi~
{F}_{\chi} (\lambda, \ell)
\end{equation}
and the full partition function (\ref{Z2dtorus}) can then be rewritten
as
\begin{equation}
{\cal Z}^{\rm YM} = \sum_{\ell= -\infty}^{\infty}\, \e^{-\frac{\lambda\,
\ell^2}{2} + {\cal F}(l)} \ .
\end{equation}

Finally, the effect of the $\theta$-angle can be easily taken into
account. It amounts to including in (\ref{Z2dtorus}) the holonomy
factor $\e^{\ii \theta \, C_1 (\hat{R} \, , \, q)} =
\e^{\ii\theta\,q}=\e^{\ii \theta \, (|\hat R_-|-|\hat R_+| + \ell \,
  N)}$ determined by the $U(1)$ flux. This leads to the final form
\begin{equation}\label{Z2dtorus2}
{\cal Z}^{\rm YM}= \sum_{\ell= -\infty}^{\infty}\, \e^{-\frac{\lambda\,
\ell^2}{2} + \ii \ell\, N\, \theta}~ \e^{{\cal F}_+ (\lambda \, , \, t + \frac{2
\lambda \,\ell}{N}) + {\cal F}_- (\lambda \, , \, \overline{t} - \frac{2
\lambda \,\ell}{N})}
\end{equation}
where
\beq
t= \frac{\lambda}{2}+ \ii \theta
\label{Kahlerpar}\eeq
is the complex coupling constant, with $t$ and $\overline{t}$ formally
regarded as independent variables, and
\begin{eqnarray}
{\cal F}_+ \big(\mbox{$\lambda\,,\, t+ \frac{2 \lambda \,\ell}{N}$}
\big)& =& \sum_{\chi=0}^{\infty}\,\left(
\frac{1}{N}\right)^{\chi}~{F}_{\chi} ( \lambda, t , \ell) \ , \\[4pt]
{\cal F}_- \big(\mbox{$\lambda\,,\, \bar{t} - \frac{2 \lambda
    \,\ell}{N}$}\big) &=&
\sum_{\chi=0}^{\infty}\,\left( \frac{1}{N}\right)^{\chi}~{F}_{\chi} (
\lambda, \bar{t} , -\ell)
\end{eqnarray}
with
\begin{equation}
{F}_{\chi}  (\lambda, t, \ell) = \left( \frac{\lambda}{2}
\right)^{\chi}\, \sum_{\stackrel{\scriptstyle b=0}
{\scriptstyle\chi-b~{\rm even}}}^{\chi}\, \frac{(-2\ell)^b}{b!\, (\chi - b)!}~
\sum_{d=0}^{\infty}~H_{\chi,d,b}~ \e^{- t \, d} \ .
\label{Fchilambdatell}\end{equation}
Note that both free energy functions ${\cal F}_\pm$ have an explicit
dependence on the real 't~Hooft coupling $\lambda$ through the factor
$\lambda^{\chi}$ in (\ref{Fchilambdatell}).

In the next section we will reconcile this string expansion with the
A-model topological string theory on the geometry
(\ref{threefold}). In that instance the complex coupling
(\ref{Kahlerpar}) is the cohomology class $t\in H^{1,1}(\torus^2)$ of
the complexified K\"ahler form $\omega_{\torus^2}$ on $\torus^2$, as follows from
the identification of closed and open string moduli required to match
the black hole and closed topological string partition functions in
the OSV conjecture~\cite{Vafa:2004qa,Aganagic:2004js}. Indeed,
eq.~(\ref{Kahlerpar}) is just the attractor equation in this
instance. The sum over $U(1)$ charges $\ell$ corresponds to a sum over
Ramond-Ramond fluxes coupled to the D2 branes wrapping
$\torus^2$~\cite{Vafa:2004qa,Aganagic:2004js}, consistent with the
observation above that the $U(1)$ sector of the $U(N)$ gauge theory is
a source of open string moduli.

\section{Topological String Theory on the Local Elliptic
  Threefold\label{TopStringElliptic}}

The chiral partition function of two-dimensional Yang-Mills theory on
the torus can be interpreted as the topological string amplitude
in the Calabi-Yau background
(\ref{threefold})~\cite{Vafa:2004qa,Aganagic:2004js}. The aim of this
section is to make this correspondence more precise and exhibit a
rigorous construction of the relevant toric geometry. The perturbative
topological string amplitude can be computed by means of topological vertex
techniques and it is shown to agree with the chiral two-dimensional 
gauge theory partition function found in Sect.~\ref{StringInterpret}.
As we discuss
in detail below, the calculation is somewhat subtle and must be
handled with care because, unlike the case of local $\PP^1$ studied
in~\cite{Caporaso:2005fp}, the local elliptic threefold is {\it not} a
toric manifold and so the standard topological vertex methods do not
immediately apply. We will also describe, for completeness, an alternative
derivation that follows the techniques of~\cite{Bryan:2004iq} where
the topological string theory is interpreted as a two-dimensional
topological quantum field theory.

\subsection{The Formal Toric Geometry\label{FTG}}

We begin with a detailed construction of the local elliptic curve
$X_p$ as a toric Calabi-Yau manifold. The idea is to start from the
familiar toric description of the trivial K\"ahler geometry of
$\complex^3$ and then compactify one of the complex coordinates to get
the geometry $\complex^2\times \torus^2$, which is the total space
$X_0$ of the trivial rank~2 vector bundle (\ref{threefold}) with
$p=0$. The non-trivial fibrations $X_p$ with $p>0$ will be recovered
later on by using framing techniques.

A convenient way to describe $\complex^3$ as a toric manifold is
to view it as a $\torus^3$-fibration over $\real^3$. This is achieved
by changing from complex coordinates $z_i$ on $\complex^3$ to polar
coordinates $z_i=|z_i|~\e^{\ii\theta_i}$ along each of the three
complex directions, giving the symplectomorphism
\begin{equation}
\complex^3~\stackrel{\approx}{\longrightarrow}~
\real_{\geq0}^3\times\torus^3 \ , \qquad
\bigl( z_1 \, , \, z_2 \, , \, z_3 \bigr) ~\longmapsto~ \left(
( |z_1|^2 , \theta_1) \, , \, ( |z_2|^2,
\theta_2) \, , \, ( |z_3|^2 , \theta_3 )\right)
\end{equation}
where $\real_{\geq0}^3$ denotes the positive octant of $\real^3$. The
natural $U(1)^3$-action on $\complex^3$ is given by shifts of the
angular coordinates $\theta_i$ in this parametrization. This fibration
is encoded in the toric diagram on the left in Fig.~\ref{C3},
and the corresponding toric graph on the right is obtained by
projecting the three-dimensional diagram onto a plane.
\EPSFIGURE{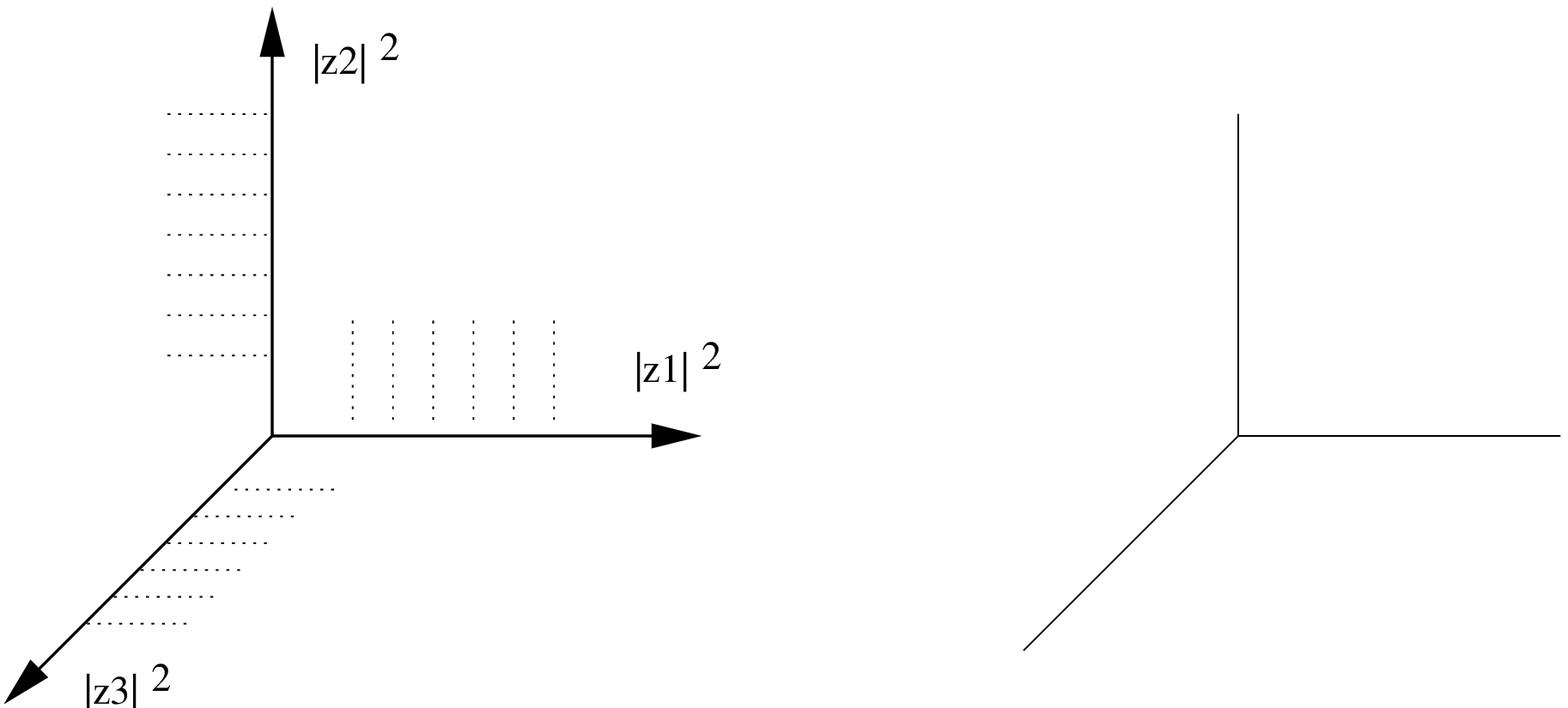,width=4in}{$\complex^3$ as a toric
fibration.\label{C3}} In the diagram a generic point of the
positive octant is associated with a nondegenerate
$\torus^3$ fibre. When one or more coordinates $|z_i|^2$ vanish,
the fibration degenerates. For example, one cycle degenerates to a
single point on each of the planes $|z_i|^2 = 0$, while two cycles
degenerate on each of the lines drawn in the diagram. All three
cycles of the fibration are degenerate at the origin, which is the
fixed point of the natural toric action on $\complex^3$, and the
$\torus^3$ fibre shrinks to a single point.

From this geometry we can construct the threefold
(\ref{threefold}) at $p=0$ by compactifying one
of the new coordinates, say $|z_1|^2$, according to the projection
\begin{equation}
\xymatrix@=10mm{ \complex^3 ~\cong~ \sphere^1 \times \real_{\ge 0} \times
\complex^2 \ar[d]^{~\pi} \\ X_0~\cong~\sphere^1 \times
\big({\real_{\ge 0}}/{\ii\tau \,\zed}\big)  \times \complex^2 }
\end{equation}
where $\tau=- t/2\pi\ii$ is the complex modulus of the dual elliptic
curve. This projection is achieved by imposing the
identifications $|z_1|^2\sim |z_1|^2 + \ii\tau \,n$, $n\in\zed$, and it
is graphically implemented by gluing an edge of the toric diagram to
itself as shown in
Fig.~\ref{C3torus}. \EPSFIGURE{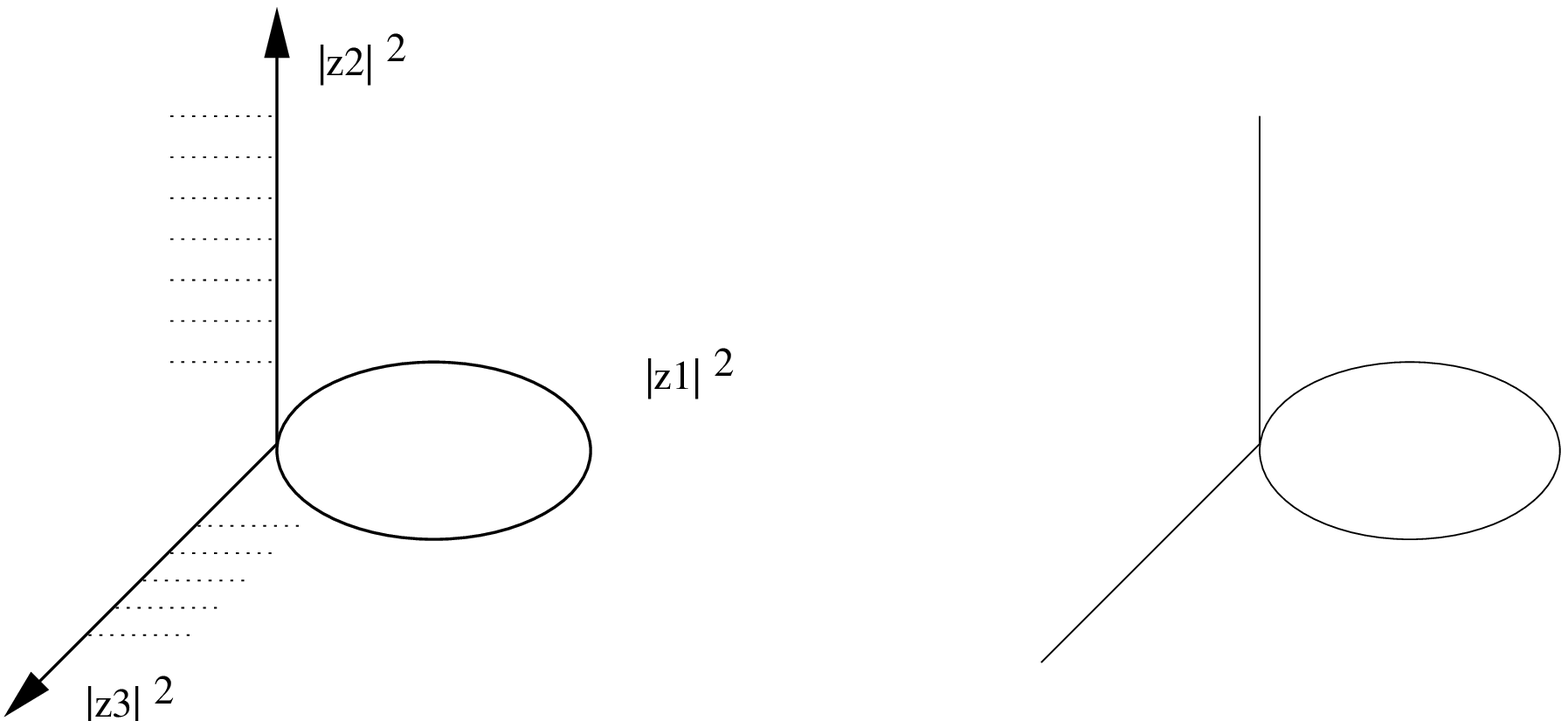,width=4in}{Toric
  construction of the fibration $X_0$.\label{C3torus}} Note that now
the $U(1)$ action given by shifts in $\theta_1$ is no longer
degenerate since the locus $|z_1|^2 = 0$ is identified with
$|z_1|^2 = \tau \, n$ for any $n\in\zed$. While
$X_0=\torus^2\times\complex^2$ contains the algebraic three-torus
$(\complex^\times)^3$ as a dense open subset (where
$\complex^\times:=\complex\setminus\{0\}$ is the punctured complex
plane), and the natural action of $(\complex^\times)^3$ on itself
extends to $X_0$, this toric action is free.\footnote{This can be seen
  explicitly by noting that there is a symplectomorphism
  $\torus^2\times\complex^2\cong(\complex^\times)^2\times\complex$
  obtained by mapping the cylinder into the complex plane.}
As a consequence, the toric graph is no longer trivalent since
the projection $\pi$ eliminates the fixed point of the toric action on
the universal covering space $\complex^3$. Because the fixed
point locus is empty, and hence so is the set of vertices of the toric
graph, the powerful techniques developed in~\cite{Aganagic:2003db} for
the computation of topological string amplitudes on toric Calabi-Yau
manifolds cannot be applied.

\begin{figure}[hbt]
\begin{center}
\epsfxsize=1.6 in\epsfbox{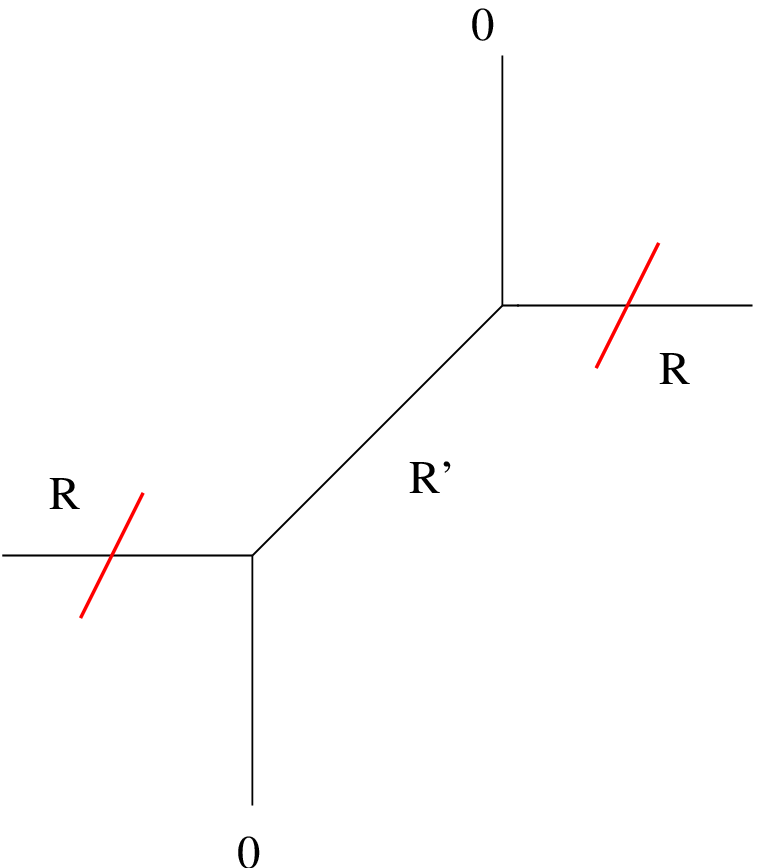}
\end{center}
\caption{Toric graph of the blowup ${\cal X}_0\to X_0$. The dashes
  indicate that the horizontal edges (labelled $R$) are identified. The
  diagonal edge (labelled $R'$) is the new $\PP^1$ at the origin of
  $X_0$.}
\label{toricdiag}\end{figure}

This difficulty is overcome by realizing that this non-toric geometry
can be canonically related to a {\it formal} toric Calabi-Yau
threefold. Let
\beq
{\cal X}_0~\stackrel{\varpi}{\longrightarrow}~X_0
\label{blowup}\eeq
be the blowup of $X_0$ at $z_i=0$. This amounts to excising the origin
of $X_0$ and replacing it with a projective plane $\PP^2$. It
corresponds to a degeneration of the toric graph of $X_0$ obtained by
blowing up an edge representing a projective line
$\mathbb{P}^1\subset\PP^2$ at the origin. The toric
action on $X_0$ can be lifted to ${\cal X}_0$ in such a way that the
natural projection (\ref{blowup}) of the blowup is
$U(1)^3$-equivariant. Composition with the projection $X_0\to\torus^2$
determines an equivariant family of elliptic curves
\beq
{\cal X}_0~\longrightarrow~\torus^2
\label{blowupT2}\eeq
whose fibres over any point $z\neq0$ in $\torus^2$ are ${\cal
  X}_0(z)\cong\complex^2$, while ${\cal
  X}_0(0)\cong\complex^2\amalg\complex^2$. The blowup thus defines a
toric scheme ${\cal X}_0$ whose toric graph is depicted in
Fig.~\ref{toricdiag}.
Since the trivalent graph describing the degeneration locus of the
$\torus^3$-fibration ${\cal X}_0$ is {\it non-planar}, ${\cal X}_0$
is a (regular) ``formal'' toric Calabi-Yau threefold in the
terminology of~\cite{Li:2004uf}.

Although the blowup produces an inequivalent Calabi-Yau space, the
topological string partition function on ${\cal X}_0$ is easily
related to the desired one on the original
background $X_0$ as follows. Let $N^g_{\beta}(X_0)$ denote the genus $g$
Gromov-Witten invariants of the elliptic threefold $X_0$. Since
$H_2(X_0,\zed)=\zed[\torus^2]$, any Calabi-Yau curve class $\beta$ is
of the form $\beta=d\,[\torus^2]$ with $d\in\zed$ and the corresponding
invariant may be denoted $N_{d}^g(X_0)$. On the other hand, the blowup
${\cal X}_0$ has two non-trivial curve classes represented by the two
internal edges in the toric graph of Fig.~\ref{toricdiag}. The non-planar
edge (labelled $R$) is given by the embedding of the torus $\torus^2$
in ${\cal X}_0$ via the family of curves (\ref{blowupT2}),
while the planar edge (labelled $R'$) is the projective line $\PP^1$
from the blowup at the origin. Thus
\beq
H_2({\cal X}_0,\zed)=\zed\big[\torus^2\big]\oplus\zed\big[\PP^1\big] \
,
\label{H2calX0}\eeq
and any curve class $\beta'$ of ${\cal X}_0$ may be expanded as
$\beta'=d\,[\torus^2]+d'\,[\PP^1]$ with $d,d'\in\zed$. The
corresponding formal Gromov-Witten invariants~\cite{Li:2004uf}
$N_{\beta'}^g({\cal X}_0)$ may thus be denoted~$N_{d,d'}^g({\cal
  X}_0)$.

Then one has the equality~\cite{Hu}
\beq
N_{d,0}^g({\cal X}_0)=N_{d}^g(X_0)
\label{GWinvsequal}\eeq
between Gromov-Witten invariants of $X_0$ and of its blowup ${\cal
  X}_0$.\footnote{A more invariant way of writing this identity is
  as $N_{\beta}^g(X_0)=N_{\varpi^!(\beta)}^g({\cal X}_0)$ for all
  $\beta\in H_2(X_0,\zed)$, where $\varpi^!$ is the Gysin pullback on
  homology induced by the natural projection (\ref{blowup}) of the
  blowup.} The topological string amplitude on ${\cal X}_0$ can be
computed by using the degeneration gluing formula
of~\cite{Li:2004uf}. The equality (\ref{GWinvsequal}) then implies
that this partition function agrees with the topological string
amplitude on $X_0$ after blowing down the extra projective line using
(\ref{blowup}) to recover the original elliptic threefold. In
practice this is achieved by regarding the complex K\"ahler class $t'\in
H^{1,1}(\PP^1)$ of the auxilliary $\mathbb{P}^1$, which is the length
of the corresponding edge in the toric graph, as a free parameter
in the theory which can be smoothly continued to $t'=0$ in the
amplitude on ${\cal X}_0$.

To compute the topological string amplitude in the background ${\cal
  X}_0$, we will also need to specify the framings $f$ on the
$U(1)^3$-invariant edges and vertices of the toric graph in
Fig.~\ref{toricdiag}. Since we will set the fields on the external
edges of the graph to zero in order to recover the closed string
geometry of ${\cal X}_0$ (see Sect.~\ref{TopStringAmpl} below for
further discussion), it suffices to specify the framings on the
internal edges. They are determined by the intersection numbers of the
Calabi-Yau space~\cite{Vafa:2004qa}. Each
edge defines a two-cycle in ${\cal X}_0$, which is bounded by two
planes representing the projections of four-cycles in ${\cal X}_0$
onto the toric base. There are two natural four-cycles in ${\cal X}_0$
which are the total transforms $\varpi^!(C^{\phantom{\prime}}_4)$,
$\varpi^!(C_4')$ of the four-cycles in $X_0$ given by the total spaces
of the two holomorphic line sub-bundles of
(\ref{threefold}), with $\varpi^!$ the Gysin pullback on homology
induced by the projection (\ref{blowup}). The two intersection numbers
of these cycles with the embedded $\torus^2$ give the framings of the
two $\PP^1$ edges which are identified in the toric graph of
Fig.~\ref{toricdiag}. Since $p=0$ here, the four-cycles
$C^{\phantom{\prime}}_4\cong C_4'$ are both copies of the total space
of ${\cal O}_{\torus^2}(0)\to\torus^2$. Since the bundle is trivially
fibred over $\torus^2$, one finds the vanishing intersection numbers
\beq
\#\big(\varpi^!(C_4^{\phantom{\prime}})\cap\torus^2\big)=
\#\big(\varpi^!(C_4'\,)\cap\torus^2\bigr)=0 \ .
\label{C4T2ints}\eeq
As a consequence, the internal $R$-edge carries the canonical
framing. The other non-trivial four-cycle in ${\cal X}_0$ is the
exceptional divisor $\PP^2$ over $z_i=0$. The four-cycles
$C_4^{\phantom{\prime}}$ and $C_4'$ are far away from the center of the
blowup, and so
\beq
\#\big(\varpi^!(C_4^{\phantom{\prime}})\cap\PP^1\big)=
\#\big(\varpi^!(C_4'\,)\cap\PP^1\big)~=~0~=~\#\big(\PP^2\cap\torus^2
\big) \ .
\label{C4P1ints}\eeq
By the Calabi-Yau condition, the embedding
$\PP^1\subset\PP^2\hookrightarrow{\cal X}_0$ must be made such that the
normal bundle to $\PP^1$ in ${\cal X}_0$ is of degree $-2$, which
finally identifies the remaining intersection number
\beq
\#\big(\PP^2\cap\PP^1\big)=-2
\label{E4P1ints}\eeq
specifying the canonical framing of the internal $R'$-edge.

The intersection products (\ref{C4T2ints})--(\ref{E4P1ints}) also
uniquely fix the geometry of ${\cal X}_0$ near the two embedded
curves. The normal bundle of the elliptic curve in ${\cal X}_0$ is the
trivial rank~2 holomorphic vector bundle
\beq
{\cal N}_{{\cal X}_0/\torus^2}={\cal O}_{\torus^2}(0)\oplus
{\cal O}_{\torus^2}(0)~\longrightarrow~\torus^2
\label{T2normalbundle}\eeq
with the geometry of the original Calabi-Yau background
$X_0=\complex^2\times\torus^2$, while the normal bundle of the
rational curve is
\beq
{\cal N}_{{\cal X}_0/\PP^1}={\cal O}_{\PP^1}(0)\oplus
{\cal O}_{\PP^1}(-2)~\longrightarrow~\PP^1
\label{P1normalbundle}\eeq
with the geometry of the ALE-type space $\complex\times A_1$. The geometry
of ${\cal X}_0$ itself is given by first forming the disjoint union
$X_0\amalg(\complex\times A_1)$ of these local neighbourhoods. Each of the
two vertices in the toric graph of Fig.~\ref{toricdiag} represents a
local $\complex^3$ patch of the geometry. Using the gluing morphisms
constructed explicitly in~\cite{Li:2004uf}, we glue the normal bundles
in the two $\complex^3$ patches together along their common, but
oppositely oriented, $\torus^2$ and $\PP^1$ two-cycles with K\"ahler
moduli $t$ and $t'$, respectively. The two-cycles intersect each other
precisely when their associated edges share a common vertex, and the
transition functions between the charts corresponding to
(\ref{T2normalbundle}) and (\ref{P1normalbundle}) are respectively
$(z^{\phantom{1}}_1,z^{\phantom{1}}_2,z^{\phantom{1}}_3)\mapsto
(z_1^{-1},z^{\phantom{1}}_2,z^{\phantom{1}}_3)$ and
$(z^{\phantom{1}}_1,z^{\phantom{1}}_2,z^{\phantom{1}}_3)\mapsto
(z_1^{-1},z^{\phantom{1}}_2,z_1^2\,z^{\phantom{1}}_3)$ (with the
corresponding curves defined in these coordinates by $z_2=z_3=0$). This
completely specifies the scheme ${\cal X}_0$. However, it is not clear
how to obtain a description of such a formal toric geometry as a
K\"ahler quotient $\complex^5/\!/\,U(1)\times U(1)$ with respect to a
moment map for the torus action, i.e. as the Higgs branch of a gauged
two-dimensional, $\mathcal{N}=(2,2)$ supersymmetric linear sigma-model
with $U(1)\times U(1)$ gauge symmetry.\footnote{The local elliptic
  Calabi-Yau geometry also arises in the geometric engineering of ${\cal N}=2$
  $U(N)$ gauge theories in five dimensions by a brane web consisting
  of a single NS5 brane and $N$ D5 branes wrapped on a
  circle~\cite{Hollowood:2003cv}. There the blowup (\ref{blowup})
  emerges from turning on a supersymmetry breaking adjoint mass term
  which has the effect of resolving the intersection of the D5 and NS5
  branes.}

\subsection{The Topological String Amplitude\label{TopStringAmpl}}

The perturbative topological string amplitude on the blowup
$\mathcal{X}_0$ can be computed with the topological vertex techniques
of~\cite{Aganagic:2003db,Li:2004uf} which respect the torus
symmetries of the space. Recall that the topological
vertex rules amount to assigning to each trivalent vertex of the toric
graph (representing a $\complex^3$ patch of the geometry) the amplitude
$C_{{\hat R}_{1}{\hat R}_{2}{\hat R}_{3}}(q)$, with $q:=\e^{-g_s}$,
which depends on three $SU(\infty)$ representations $ \hat R_{a}$ as
well as on the orientations and framings of the edges. Its explicit
form is given by
\begin{equation} C_{{\hat R}_{1}
{\hat R}_{2}{\hat R}_{3}}(q) = q^{\frac{1}{2}\,(
\kappa_{{\hat R}_{2}} + \kappa_{{\hat R}_{3}})}\,
\sum_{{\hat Q}_{1} , {\hat Q}_{3} , {\hat Q}} \, N_{{\hat Q}
{\hat Q}_{1}}^{{\hat R}_{1}} \, N_{{\hat Q}
{\hat Q}_{3}}^{{\hat R}_{3}^\top} \, \frac{W_{{\hat R}_{2}^\top
{\hat Q}_{1}} \left( q \right) \, W_{{\hat R}_{2}
{\hat Q}_{3}} \left( q \right)}{ W_{{\hat R}_{2}\bullet} \left( q
\right) } \ ,
\label{topvertexdef}\end{equation}
where $N_{\hat R_{1}\hat R_{2}}^{\hat R_{3}}$ are the Littlewood-Richardson
tensor multiplicity coefficients, $W_{\hat R_{1}\hat R_{2}}(q)$ is the
$SU(\infty)$ Chern-Simons invariant of the Hopf link in $\sphere^3$, and
  $\hat R=\bullet$ denotes the trivial representation with empty Young
  tableau. The topological string amplitude on any (formal) toric
  Calabi-Yau threefold can be computed according to a few simple
  gluing rules.

Given the toric graph associated to the toric geometry, one glues
together two vertices through two edges carrying a representation
${\hat R}$ and its transpose ${\hat R}^\top$ with the Schwinger propagator
$\left( - 1\right)^{|{\hat R}|}~ \e^{ - t \,|{\hat R}|} $, where $t$ is the
K\"ahler parameter of the $\mathbb{P}^1$ two-cycle associated
to the gluing edge~\cite{Aganagic:2003db,Li:2004uf}. This
procedure corresponds to gluing curves with holes along their
boundaries with the opposite orientation as described in
Section~\ref{FTG} above. Since the vertex itself
represents an open string amplitude in $\complex^3$, with the
representations $\hat R_a$ labelling boundary holonomies, the edges come
with a framing ambiguity. Due to the boundary conditions imposed on
each edge, the topological string amplitude depends on a choice of
integers. Geometrically, this corresponds to a choice of a particular
compactification of the non-compact lagrangian submanifolds
$\complex\times\sphere^1\subset\complex^3$, wrapped
by topological A-model D-branes, that determine boundary conditions in
the construction of the vertex
amplitude~\cite{Aganagic:2003db,Aganagic:2002wv}. If we label an edge
of the vertex as $v = (p,q)\in\zed^2$ (corresponding to the
degeneration of the plane-projected $(-q,p)$ cycle of the $\torus^3$
fibration) and the location of the lagrangian submanifold wrapped by
the D-brane with the framing vector $f\in\zed^2$, then the condition
that the lagrangian submanifold is a compact $\sphere^3$ cycle can be
written as the symplectic product~\cite{Aganagic:2003db}
\begin{equation}
f \wedge v =1 \ .
\end{equation}
The framing ambiguity corresponds to the shift $f \rightarrow f -
n \, v$ for any integer $n$. The effect of a change of framing by $n$
units on one edge of a vertex labelled by a representation ${\hat R}$ is
to multiply the vertex amplitude (\ref{topvertexdef}) itself by the
factor $(-1)^{n \, |{\hat R}|}\, q^{ n \, \kappa_{{\hat R}} / 2}$. We will
momentarily only work with the amplitude in the canonical framing in
which the expression above for the topological vertex is derived.

The representations at the ends of unglued edges represent
D-brane degrees of freedom corresponding to asymptotic boundary
conditions at infinity. As explained
in~\cite{Aganagic:2004js,Aganagic:2005dh}, the local Calabi-Yau
geometry given by the sum of two line bundles over a genus $g$
Riemann surface requires precisely $|2g - 2|$ closed string moduli
coming from infinity. In the genus one case, no D-branes are needed to
enforce boundary conditions and so the free edges are labelled by the
trivial representation $\hat R=\bullet$. In other words, we are
building a purely closed string amplitude. The absence of fibre
D-branes is directly related~\cite{Aganagic:2004js} to the fact that
there are no omega-points in the Gross-Taylor string expansion on the
torus. This fact is used implicitly in Sect.~\ref{TQFT} below, and
also in Sect.~\ref{Hurwitz} where we will compare the two-dimensional
string perturbation series with that of topological string theory.

With all of this in mind, the topological string amplitude for the
blowup of the trivial fibration ${\cal X}_0$ is given by
\begin{equation} \label{TopAmp1}
Z_{{\cal X}_0} \left( t \, ,\, t'\,\right)= \sum_{{\hat R} , {\hat R}'}\,
(-1)^{|{\hat R}|+|{\hat R}'|} \,Q^{|{\hat R}|} \, Q'^{\,|{\hat R}'|} ~
C_{\bullet{\hat R}'^\top{\hat R} } (q) \, C_{\bullet{\hat R}'{\hat R}^\top} (q) \ ,
\end{equation}
where $Q := \e^{-t}$ and $Q' := \e^{-t'}$ with $t'$ the auxilliary
K\"ahler parameter. The required topological vertex in (\ref{TopAmp1})
can be represented as
\begin{equation}
C_{\bullet{\hat R}_{1}{\hat R}_{2}} = (-1)^{|{\hat R}_{2}|}~
s_{{\hat R}_{2}^\top} \big( q^{-i+{1}/{2}}\big)\,
s_{{\hat R}_{1}} \big( q^{n_i({{\hat R}_{2}^\top}) - i +
{1}/{2}} \big) \ ,
\end{equation}
where $s_{{\hat R}} (x_i)$ are the Schur
functions whose definition and relevant properties can be found in
Appendix~A. We can thus write (\ref{TopAmp1}) as
\begin{eqnarray} \label{TopAmp2}
Z_{{\cal X}_0} \left( t \, ,\, t'\, \right) &=& \sum_{{\hat R}}\,
(-1)^{|{\hat R}|} \, Q^{|{\hat R}|}\, s_{{\hat R}} \big( q^{- i +
{1}/{2}} \big) \,s_{{\hat R}^\top} \big( q^{- i +{1}/{2}}
\big)\nonumber\\ && \times~ \sum_{{\hat R}'}\, s_{{\hat R}'} \big( - Q' \,
q^{n_i({{\hat R}}) - i + {1}/{2}}\big)\, s_{{\hat R}'^\top}
\big( q^{n_i({{\hat R}^\top}) - i + {1}/{2}} \big) \nonumber\\[4pt] &=&
\sum_{{\hat R}}\, (-1)^{|{\hat R}|} \, Q^{|{\hat R}|} \,s_{{\hat R}}
\big( q^{- i +{1}/{2}} \big)\, s_{{\hat R}^\top} \big( q^{- i +
{1}/{2}} \big) \nonumber\\ && \times~\prod_{i,j\ge 1}\, \left( 1-Q' \,
q^{n_i({{\hat R}^\top}) - i + n_j({{\hat R}}) - j +1 } \right) \ .
\end{eqnarray}

Consider now the identity~\cite{Hollowood:2003cv}
\begin{equation}
\sum_{i,j \ge 1}\, q^{h_{\hat R}(i,j)} = \frac{q}{(q-1)^2} + \sum_{(i,j) \in
{\hat R}}\, \left( q^{h_{\hat R}(i,j)} +  q^{-h_{\hat R}(i,j)} \right) \ .
\label{hookid}\end{equation}
In this formula a given pair of positive integers
$(i,j)$ specifies the location of a box in the Young tableau for the
representation $\hat R$, and $h_{\hat R}(i,j)=n_i({{\hat R}^\top}) - i +
n_j({{\hat R}}) - j +1$ is the hook length of the box $(i,j)$. One
then has the identity
\begin{eqnarray}
&& \prod_{i,j \ge 1}\,\left( 1-Q' \, q^{n_i({{\hat R}^\top}) - i +
n_j({{\hat R}}) - j +1 } \right) \cr && \qquad~
=~ \prod_{k=0}^{\infty}\, \left(1 - Q' \, q^{k+1} \right)^{k+1} ~
\prod_{(i,j)\in {\hat R}}\, \left( 1 - Q' \, q^{h_{\hat R}(i,j)} \right) \,
\left( 1 - Q' \, q^{-h_{\hat R}(i,j)} \right)
\label{hookidfollows}\end{eqnarray}
which follows from (\ref{hookid}) by taking the logarithm of both
sides of eq.~(\ref{hookidfollows}) and expanding. Finally, we can
simplify the expression for the topological string amplitude
(\ref{TopAmp2}) further by using the identity~\cite{MacDonald}
\begin{equation}
s_{{\hat R}} \big( q^{ - i + {1}/{2} } \big) \,
s_{{\hat R}^\top} \big( q^{ - i + {1}/{2} } \big) =
\prod_{(i,j) \in{\hat R}}\,\frac{q^{h_{\hat R}(i,j)}}{ \big( 1 - q^{h_{\hat R}(i,j)}
  \big)^2} \ .
\end{equation}

Combining these identities together we end up with
\bea
Z_{{\cal X}_0}\left( t \, ,\, t'\,\right) &=&
\prod_{k=0}^{\infty}\, \left(1 -
Q' \, q^{k+1} \right)^{k+1} \nonumber\\ && \times
~ \sum_{{\hat R}}\, \left( Q \, Q'\,
\right)^{|{\hat R}|} ~\prod_{(i,j) \in {\hat R}}\, \frac{\big( 1 - Q'
\, q^{h_{\hat R}(i,j)} \big) \, \big( 1 - Q'^{\,-1} \, q^{h_{\hat
    R}(i,j)} \big)}{\big( 1 - q^{h_{\hat R}(i,j)} \big)^2} \ .
\label{TopAmp2a}\eea
The first product in (\ref{TopAmp2a}) is an irrelevant normalization
factor which we will drop in the following. As explained in
Section~\ref{FTG} above (see eq.~(\ref{GWinvsequal})), we can now obtain
the desired partition function of topological string theory on the trivial
fibration $X_0=\torus^2 \times \complex^2$ by using the blowdown
projection (\ref{blowup}). This is accomplished by setting the
K\"ahler parameter $t'$ of ${\cal X}_0$ to zero (equivalently $Q'=1$),
and we arrive finally at
\begin{equation} \label{TopAmp3}
Z_{X_0}(t)=\lim_{t'\to0}\, Z_{{\cal X}_0}\left(t\,,\, t'\,\right) 
= \sum_{{\hat R}} \, Q^{|{\hat R}|} =
\sum_{{\hat R}}\, \e^{-t \, |{\hat R}|} \ ,
\end{equation}
which is the anticipated result from~\cite{Vafa:2004qa}. However, the
free energy corresponding to the partition function
(\ref{TopAmp3}) has no genus expansion. We will see this explicitly in
the next section where we will find that the only non-vanishing
geometric invariants of $X_0$ are those at genus
$g=1$. In Section~\ref{ChernSimons} we will encounter this feature in
an alternative way. Thus to recover two-dimensional Yang-Mills theory,
and its corresponding non-trivial genus expansion, we must consider
topological strings propagating in a non-trivial background geometry.

It is straightforward to implement in the amplitude (\ref{TopAmp3})
the effect of a non-trivial $p>0$ fibration (\ref{threefold}) over the
torus $\torus^2$. As explained above, the framing corresponds to a
particular compactification of a lagrangian submanifold.
Thus a change of framing is reflected in a modification of the target
space geometry. More precisely, a change of the canonical framing by
$n=p$ units alters the intersection numbers of the cycles of the
underlying Calabi-Yau threefold in (\ref{C4T2ints}) to 
\beq
\#\big(\varpi^!(C_4^{\phantom{\prime}})\cap\torus^2\big)=-
\#\big(\varpi^!(C_4'\,)\cap\torus^2\big)=-p \ ,
\label{C4T2intsnew}\eeq
and the corresponding normal bundle to the embedding
$\torus^2\hookrightarrow{\cal X}_p$ is precisely the non-trivial
fibration (\ref{threefold}) over the two-dimensional
torus~\cite{Vafa:2004qa}. The effect of a change of framing by $p$
units in one of the gluing edges labelled by a representation $\hat R$ is
to multiply the vertex amplitude (\ref{topvertexdef}) by a factor
$(-1)^{p \, |{\hat R}|} \, q^{ p \, \kappa_{{\hat R}} / 2}$. Note that we change
the framing of only the non-planar edge, since otherwise the
framed geometry of the planar edge would not survive the blowdown
projection (\ref{blowup}). The effect of this procedure on the final
string amplitude is to modify eq.~(\ref{TopAmp3}) to the expression
\begin{equation}
Z_{X_p}(t)=\sum_{{\hat R}}\, (-1)^{p \,
|{\hat R}|} \,q^{p \, \kappa_{{\hat R}} /2} \, Q^{|{\hat R}|} =\sum_{{\hat R}} \,
\e^{-t \, |{\hat R}|}~ \e^{- g_s \, p \,\kappa_{{\hat R} }/2 }
\label{toptorus}\end{equation}
where we have absorbed the parity factor $(-1)^{p \,
|{\hat R}|}$ into the shift $t \rightarrow t
+  \pi\ii p$ of the K\"ahler modulus, analogously to what was done
in~\cite{Caporaso:2005fp} for the local $\PP^1$ geometry (this is
equivalent to shifting the $\theta$-angle by $\pi\,p$ units).

Confronting eqs.~(\ref{Z2dtorus}) and (\ref{Z2dtorus2}) with
(\ref{toptorus}), we find full consistency (up to overall
normalization) with the conjectured
topological string interpretation of the large $N$ partition function
of $U(N)$ Yang-Mills theory~\cite{Vafa:2004qa}. The basic chiral and
antichiral blocks are obtained from the holomorphic topological string
amplitude (\ref{toptorus}), while a sum over the $U(1)$ charges
is required by the non-perturbative completion provided by
the D-brane partition function of Section~\ref{DPartFn}. The peculiar
dependence on the $U(1)$ charge $\ell$ has been interpreted
in~\cite{Vafa:2004qa,Aganagic:2004js} as coming from a source of
Ramond-Ramond two-form flux through the base Riemann surface which is
wrapped by D2 branes. It would be interesting to obtain a more direct
open string understanding of this phenomenon, as suggested by the
analysis of the amplitude in terms of open branched covering maps
given in Sect.~\ref{YMLargeN}.

\subsection{Topological Quantum Field Theory on the Elliptic
  Curve\label{TQFT}}

For completeness,\footnote{This subsection is not essential to the
  rest of the paper and may be skipped by the uninterested reader.}   
we will now briefly describe another computation of
the topological string amplitude which also proceeds largely in the
spirit of the formal toric geometry techniques employed
above. In~\cite{Bryan:2004iq} a practical algorithm was developed to
build the amplitudes of a two-dimensional topological
quantum field theory, very much akin to the gluing rules of ordinary
two-dimensional Yang-Mills theory, corresponding to the local
Gromov-Witten theory of curves embedded in a Calabi-Yau manifold. The
main idea is to define a partition function that generates
Gromov-Witten residue invariants of the local threefold via
equivariant integration. Up to normalization, this partition function
can be used to define a topological quantum field theory. The cutting
and pasting of base Riemann surfaces corresponds to the cutting and
pasting of the corresponding local Calabi-Yau threefolds, as either
adding or cancelling of D-brane degrees of freedom associated to the
boundaries. The operations of gluing manifolds according to the formal
toric constructions in the topological vertex satisfy all the axioms
of a two-dimensional topological quantum field theory.

Consider the geometric tensor category of 2-cobordisms, whose objects
are disjoint unions of oriented circles and whose morphisms are given
by oriented cobordisms together with a pair of complex line bundles
$(\mathcal{L}_1,\mathcal{L}_2)$ each trivialized over the boundary of
the cobordism. For a connected cobordism, one can label the
isomorphism classes of $(\mathcal{L}_1,\mathcal{L}_2)$ by their {\it
  levels} which are the relative Euler numbers $({\rm
  deg}\,\mathcal{L}_1,{\rm deg}\,\mathcal{L}_2)$, and under
concatenation using the bundle trivializations these levels add. The
partition function in question is then a functor from this category
into the tensor category of $SU(\infty)$ representations. It
associates to the 2-cobordisms, regarded as the building blocks of the
base curve, the algebra of correlators of a topological quantum field
theory~\cite{Bryan:2004iq}. One can
consider the set of generators for the relevant 2-cobordisms and build
through the appropriate gluing rules the most general generating
function of Gromov-Witten invariants for any local threefold of this
type. The fundamental amplitudes are the ``caps'' $C^{(-1,0)}$,
$C^{(0,-1)}$ with the topology of a disc and the ``pants'' $P^{(1,0)}$,
$P^{(0,1)}$ with the topology of a trinion, where the superscripts
represent the levels of the 2-cobordisms. By gluing together
these building blocks one can reconstruct the field theory
amplitudes on a general local threefold over a genus $g$ curve
$\Sigma$ of the form $\mathcal{L}_1 \oplus\mathcal{L}_2 \rightarrow
\Sigma$, where the local Calabi-Yau condition is enforced by taking $\deg
\mathcal{L}_1 = p + 2g - 2$ and $\deg \mathcal{L}_2 = -p$ (one
then has $\mathcal{L}_1\cong\mathcal{O}_\Sigma(p+2g-2)$ and
$\mathcal{L}_2\cong\mathcal{O}_\Sigma(-p)$ up to twisting by degree~1
holomorphic line bundles).

\begin{figure}[hbt]
\begin{center}
\epsfxsize=2 in\epsfbox{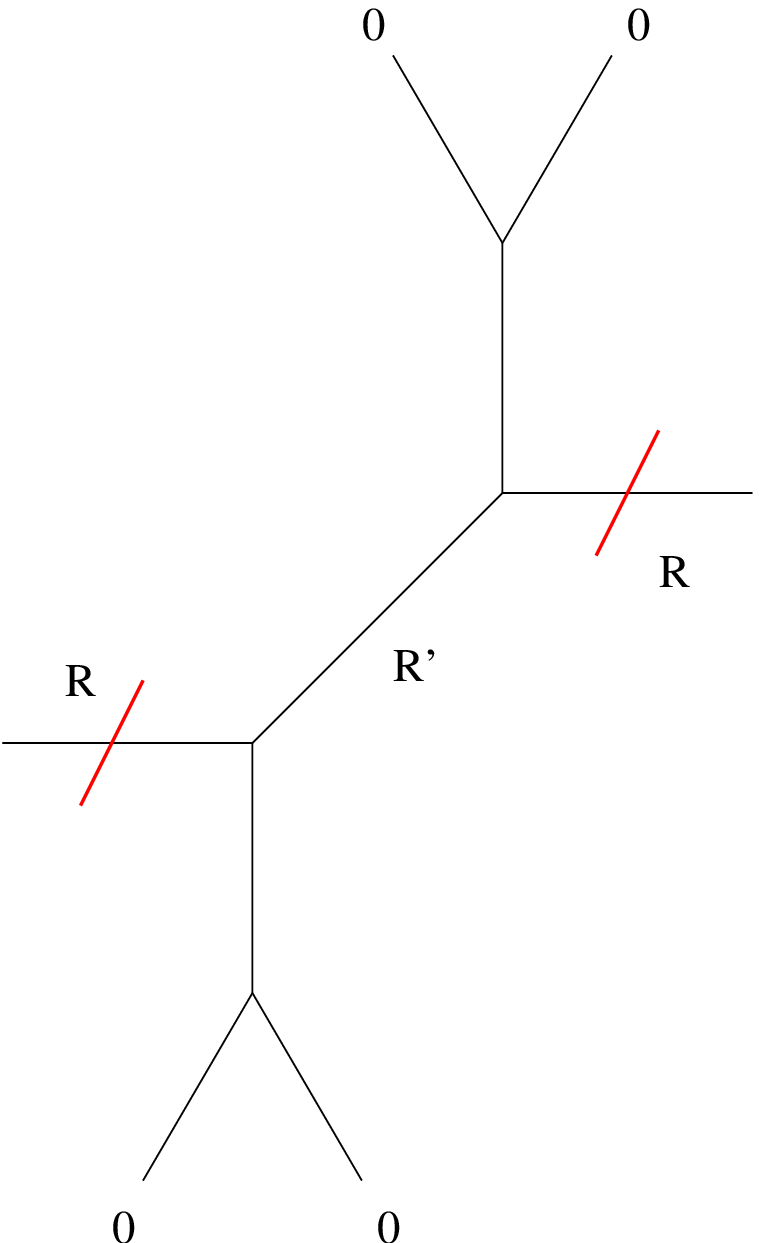}
\end{center}
\caption{The formal toric geometry arising in the topological quantum
  field theory.} \label{TQFTdiag}
\end{figure}

In the case of elliptic curves with trivial fibration $X_0$ one can
build the field theory amplitude by gluing together two pants of the
opposite kind, and to get the overall bundle degree right one
``closes'' the remaining holes with the appropriate caps to cancel the
overall Chern class. In the framework of the topological vertex, we
can interpret the pant as the A-model vertex amplitude
(\ref{topvertexdef}) itself and the cap as the vertex $C_{\hat
  R\bullet\bullet}(q)$ with trivial representations at the ends of two
edges, i.e. with a single stack of D-branes in $\complex^3$. Thus the
topological field theory approach, which is directly related to
two-dimensional Yang-Mills theory~\cite{Aganagic:2004js}, suggests the
formal toric geometry whose toric graph is depicted in
Fig.~\ref{TQFTdiag}.
Using topological vertex techniques, it is straightforward to show
that the topological string amplitude on this geometry coincides with
the amplitude computed in Section~\ref{TopStringAmpl} above, provided
that the two additional internal edges have the same K\"ahler length
and the edge labelled $R'$ is shrunk to zero size as before. The
details of this calculation can be found in Appendix~B.

This new geometry defines a non-singular, regular formal toric
Calabi-Yau scheme $\hat{\mathcal{X}}_0$ which can be obtained as the
blowup of $X_0$ at three points. When the two additional K\"ahler classes
coincide, Gromov-Witten invariants of the threefold
$\hat{\mathcal{X}}_0$ coincide with those of the one-point blowup
${\cal X}_0$ considered above. This is reminiscent of the relations
construed in the construction of~\cite{Caporaso:2005fp}, whereby a
toric variety was built in which the extra K\"ahler moduli
effectively took into account the geometrical effects of fibre
D-branes in the original local $\PP^1$ background.

\section{Invariant Theory of the Local Elliptic
  Threefold\label{InvtTheory}}

The goal of this section is to compute and compare various symplectic
geometry invariants of the threefold $X_p$. We will consider the
Gromov-Witten invariants, which count classes of holomorphic maps into
$X_p$, and their relation with the Hurwitz numbers that count
inequivalent branched coverings of $\torus^2$. The precise connection
that we can make in this case between Gromov-Witten theory and Hurwitz
theory owes to the facts that (a) the only non-trivial two-homology
class of $X_p$ is the base elliptic curve $\torus^2$ itself; (b) the
space $X_p$ carries an action of non-trivial torus symmetries as
described in Sect.~\ref{TopStringElliptic}; and (c) $X_p$ has
properties similar to those of a compact Calabi-Yau threefold. This
yields a formal combinatorial solution for the Gromov-Witten theory of
$X_p$. We will also determine the Gopakumar-Vafa invariants of $X_p$,
which have a more physical origin in the counting of BPS states
arising in a Type~IIA compactification of $X_p$ and are therefore
naturally integer-valued. Finally, we show how the two-dimensional
gauge theory leads to a remarkably simple solution for the
Donaldson-Thomas theory of $X_p$, and compare it with the computation
of the Gopakumar-Vafa invariants.

\subsection{Computing Gromov-Witten Invariants\label{GWCalcs}}

When the topologically twisted A-type sigma-model with target space
$X_p$ is coupled to topological gravity, the topological string free energy
${F}^{X_p}_g(t)$ is given by a sum over genus $g$ worldsheet instanton
sectors. The string path integral localizes onto the space of stable
holomorphic maps from connected genus $g$ curves to the target
Calabi-Yau space $X_p$ lying in two-homology classes $\beta\in
H_2(X_p,\zed)$. It is therefore given by an integration over the moduli
space $\overline{\mathcal{M}}_{g}(X_p , \beta)$ of these maps which
generalizes the (stable compactification of
the) moduli space $\mathcal{M}_{g}$ of genus $g$ algebraic
curves. More precisely, there is a natural forgetful map
$\overline{\mathcal{M}}_{g}(X_p ,\beta)\to\overline{\mathcal{M}}_{g}$,
and an element of $\overline{\mathcal{M}}_{g} (X_p ,\beta)$ is given
by a point $\Sigma\in\overline{\mathcal{M}}_{g}$ together with its
imbedding $\psi : \Sigma\rightarrow X_p$ in the target space
$X_p$. The pushforward of the imbedding map fixes the homology class
$ \beta:=\psi_* [\Sigma] \in H_2 \left( X_p \, , \zed\right)$. Since
$H_2(X_p,\zed)=\zed[\torus^2]$, as in Sect.~\ref{FTG} we can take
$\beta = d \,[ \torus^2]$ with $d\in\zed$. Since we consider only
orientation-preserving holomorphic string maps $\psi$ in the A-model,
we can further restrict to positive degrees $d$.\footnote{The degree~$0$
  geometric invariants can all be shown to
  vanish, reflecting the triviality of the B-model theory of
  degenerate closed worldsheet instantons taking $\Sigma$ to a fixed
  point in $X_p$. Likewise, all genus~$0$ contributions vanish, which
  in the gauge theory framework of Sect.~\ref{YMLargeN} owes to the
  fact that there are no coverings of a torus by a sphere.}
It follows that the free energy of topological string theory on the
elliptic threefold has a perturbative expansion of the form
\begin{equation} \label{GWexpansion}
F_{X_p}(g_s,t)=\sum_{g=1}^\infty\,g_s^{2g-2}~F^{X_p}_g (t) = 
\sum_{g=1}^\infty\,g_s^{2g-2}~\sum_{d=1}^\infty\, N^g_d(X_p)~ Q^{d} \ ,
\end{equation}
where the coefficients $N^g_d(X_p)\in \mathbb{Q}$ are the
Gromov-Witten invariants of $X_p$ which count the number of
holomorphically imbedded curves in $X_p$ of genus $g$ and degree
$d$.

The formal definition and properties of the rational numbers
$N_d^g(X_p)$ will be given in Sect.~\ref{Hurwitz} below. Here we
shall consider explicitly the powerful identification made above
between the topological string amplitude on $X_p$ and the chiral
partition of two-dimensional Yang-Mills theory on the base torus. We
will illustrate how the modularity properties of the
gauge theory can be exploited as an efficient tool to extract
Gromov-Witten invariants of the local threefold.\footnote{The
  modularity properties of B-model topological string amplitudes
  have been studied recently in~\cite{Aganagic:2006wq} from a more
  general perspective.} By matching the two
genus expansions (\ref{chiralFE}) at $\ell=0$ and (\ref{GWexpansion}),
and using the identifications (\ref{tHooftdef}) and (\ref{Kahlerpar})
between the gauge coupling and the string moduli (here we set
$\theta=0$), one has the relation
\beq
F_g^{X_p}(t)=\left(\mbox{$\frac p2$}\right)^{2g-2}~F_g(\lambda) \ .
\label{FgXpFgrel}\eeq
The chiral free energy $F_g(\lambda)$ of Yang-Mills theory on the
torus $\torus^2$ was studied in detail
in~\cite{Griguolo:2004jp,Rudd:1994ta,Douglas1}--\cite{Dijkgraaf1}.

As mentioned in Sect.~\ref{YMLargeN}, the genus $g$ contribution
$F_g(\lambda)$ is a quasi-modular form of weight $6g - 6$ under the
action of the modular group $PSL \left( 2 , \zed \right)$ acting on the
dual Teichm\"uller parameter $\tau=-t/2\pi\ii$. This implies that they can
be expressed as polynomials over $\rat$ in the holomorphic Eisenstein
series
\begin{equation} \label{Eisenstein}
E_k (\tau) = 1 - \frac{2 k}{B_k}\, \sum_{n=1}^{\infty}\, 
\frac{n^{k-1}\,Q^n}{1-Q^n}
\end{equation}
for $k=2,4,6$, where $B_k \in \mathbb{Q}$ is the $k$-th Bernoulli
number. The modular forms $E_4 (\tau)$ and $E_6 (\tau)$ can be expressed
in terms of $E_2 (\tau)$ and its derivatives as
\begin{eqnarray}
E_4 (\tau) &=& E_2 (\tau)^2 + 12  E_2' (\tau) \ , \nonumber\\[4pt]
E_6(\tau)&=& E_2 (\tau)^3 + 18 E_2 (\tau) \, E_2' (\tau) + 36E_2''
(\tau) \ .
\end{eqnarray}
This leads to compact forms for the free energies given by
\begin{eqnarray}
F_1 (\lambda) &=& \mbox{$\frac{\lambda}{48}$}\,\left(N^2 -1\right) -
\log\eta (\tau) \ , \nonumber\\[4pt] F_g
(\lambda) &=& \frac{1}{(2 g -2 )! \, \rho_g}\,
\sum_{k=0}^{3 g -3}~ \sum_{\stackrel{\scriptstyle l,m \in
\nat_0}{\scriptstyle 2 l + 3 m = 3 h - 3 - k} }\, s^{k l}_g ~E_2
(\tau)^k \, E_2'(\tau)^l \, E_2'' (\tau)^m
\label{freeencompact}\end{eqnarray}
with $g \ge 2$ and $\rho_g \, , s_g^{k l} \in \nat$, where
$\eta(\tau)$ is the Dedekind function. Note that the genus one
contribution coincides with the $p=0$ topological string amplitude
(\ref{TopAmp3}) (up to normalization by the ground state
energy corresponding to the empty Young tableau). Explicit formulas
for $F_g(\lambda)$ up to genus $g=8$ can be found
in~\cite{Rudd:1994ta}, where the relevant numerical coefficients
$\rho_g$ and $s_g^{k l}$ were computed explicitly.\footnote{Our
  conventions for the genus~$g$ free energy differ from those
  of~\cite{Rudd:1994ta} by a factor $(-1)^g$.}

As an explicit example, let us analyse in some detail the genus $2$
and $3$ free energy contributions, which are given respectively by
\beq
F_2(\lambda)=-\mbox{$\frac{1}{2! \, 90}$}\,
\big(E_2 (\tau)\, E_2' (\tau) + E_2'' (\tau)
\big) \ , \qquad F_3(\lambda)=-\mbox{$\frac{1}{4! \, 54}$}\,
\big( 7E_2'(\tau)^3 + 3E_2''(\tau)^2  \big) \ .
\eeq
We can substitute the explicit form (\ref{Eisenstein}) for the basic
Eisenstein series $E_2 (\tau)$ and expand the free energy up to the
desired order in $Q$ as
\begin{eqnarray} \label{GWF2}
F_2(\lambda) &=&- 4\,Q^2 - 32\,Q^3 - 120\,Q^4 -
320\,Q^5 - 720\,Q^6 - 1344\,Q^7 - 2480\,Q^8 - 3840\,Q^9 \cr && -\,
6360\,Q^{10} - 8800\,Q^{11} -
  13664\,Q^{12} - 17472\,Q^{13} - 25760\,Q^{14} - 31680\,Q^{15} \cr &&
  -\,44640\,Q^{16} - 52224\,Q^{17}
   - 73332\,Q^{18} -
  82080\,Q^{19} - 111440\,Q^{20} - 125440\,Q^{21} \cr && -\,
  164208\,Q^{22} - 178112\,Q^{23}  - 239040\,Q^{24} - 249600\,Q^{25} -
  323960\,Q^{26} \cr && -\, 348480\,Q^{27} - 439488\,Q^{28} -
  454720\,Q^{29} - 591840\,Q^{30} + 
  {O}\left(Q^{31}\right)
\end{eqnarray}
and
\begin{eqnarray} \label{GWF3}
F_3(\lambda)&=&\mbox{$\frac{4}{3}$}\,Q^2 +
\mbox{$\frac{320}{3}$}\,Q^3 + 1632\,Q^4 + \mbox{$\frac{36608}{3}$}
\,Q^5 + 60368\,Q^6
+ 227712\,Q^7 \cr && +\, \mbox{$\frac{2137856}{3}$}\,Q^8 +
  1918464\,Q^9 + 4676136\,Q^{10} + \mbox{$\frac{30846400}{3}$}
\,Q^{11} + \mbox{$\frac{64214912}{3}$}\,Q^{12} \cr && +\, 41108736\,Q^{13}
  +\mbox{$\frac{230312288}{3}$}\,Q^{14} + 133843072
\,Q^{15} + 230823936\,Q^{16} \cr && +\, 374105600\,Q^{17} + 607542300\,Q^{18} +
  930011328\,Q^{19} + \mbox{$\frac{4317593024}{3}$}
\,Q^{20} \cr && +\,\mbox{$\frac{6322398208}{3}$}
\,Q^{21} + 3135480816\,Q^{22} +
  \mbox{$\frac{13262093696}{3}$}\,Q^{23} + 6380391424\,Q^{24}
\cr && +\,8712698880\,Q^{25} + \mbox{$\frac{36693015512}{3}$}\,Q^{26} +
  16302480768\,Q^{27} + 22343770368\,Q^{28} \cr && +\,
\mbox{$\frac{87248171776}{3}$}\,Q^{29} + 39178738272\,Q^{30} +
  {O}\left(Q^{31}\right) \ .
\end{eqnarray}
From these formulas one can use the relations (\ref{GWexpansion}) and
(\ref{FgXpFgrel}) to read off directly the Gromov-Witten invariants
$N_d^g(X_p)\in\rat$ from the coefficient of $Q^d$ in the expansion
of~$F_g(\lambda)$.

\subsection{Combinatorial Solution via Hurwitz Theory\label{Hurwitz}}

We will now give a formal solution to the counting problem above by
relating the Gromov-Witten invariants of the threefold $X_p$ to the
simple Hurwitz numbers of the base torus. This relationship is
immediately implied by the expansion (\ref{hurwitz}). However, before
spelling this out, it is instructive to understand why such a
simplification of Gromov-Witten theory arises directly from the
point of view of topological string theory, as this provides further
insight into the conjectured nonperturbative completion to
two-dimensional Yang-Mills theory. This is straightforward to do by
appealing directly to the definition of the rational numbers
$N_d^g(X_p)$~\cite{Bryan:2004iq} upon which the topological quantum
field theory of Sect.~\ref{TQFT} is based.

Generally, the Gromov-Witten invariants of a nonsingular, projective
complex algebraic variety $X$ of complex dimension three are defined by
integrals of tautological cohomology classes over the moduli spaces
$\overline{\mathcal{M}}_{g}\left( X ,\beta \right)$ against their
virtual fundamental classes. By the Riemann-Roch theorem, the virtual
dimensions are given by
\beq
\dim\big[~\overline{\mathcal{M}}_{g}
\left( X ,\beta \right)\big]^{\rm vir}=\int_\beta\,c_1(TX) \ .
\label{virtualdim}\eeq
If this dimension is positive, then the Gromov-Witten invariants
depend on a choice of cohomology classes of $X$. Such is the case
when $X$ is the local neighbourhood of a compact Riemann surface
$\Sigma$ of genus $g\neq1$, for which $H_2(X,\zed)=\zed[\Sigma]$ and
the local Calabi-Yau condition is $c_1(TX)=2g-2$. In this case, there
is a natural action of the torus $T=U(1)\times U(1)$ on
$X$ given by scalings of the two line bundles over $\Sigma$. This
toric action lifts to the moduli space
$\overline{\mathcal{M}}_{g}\left( X ,d\right)$. The Gromov-Witten
invariants are then {\it defined} by using the usual
virtual localization formula as residue integrals
over the fixed point locus $\overline{\mathcal{M}}_{g}\left( X
  ,d\right)^T$. This involves integrating the
equivariant Euler class $e_T^{g,d}(X)$ of the virtual normal bundle of
the embedding $\overline{\mathcal{M}}_{g}\left( X
  ,d\right)^T\hookrightarrow
\overline{\mathcal{M}}_{g}\left( X ,d\right)$. The associated
topological string theory is defined in terms of an {\it equivariant}
topological sigma-model with target space $X$.

A stable map to $X$ which is invariant under the toric action factors
through the zero section of the fibration $X\to\Sigma$ and hence there
is an isomorphism
\beq
\overline{\mathcal{M}}_{g}\left( X,d\right)^T\cong
\overline{\mathcal{M}}_{g}\left( \Sigma ,d\right) \ .
\label{modXSigmaiso}\eeq
This implies that $T$-equivariant topological string theory on $X$ reduces
to a topological string theory on the two-dimensional target space
$\Sigma$. The corresponding Gromov-Witten invariants are then related
to the Hurwitz numbers counting covering maps to $\Sigma$. However,
because of the finite-dimensional integration over $e_T^{g,d}(X)$, one
obtains in this way a $q$-deformation of the standard Hurwitz
theory~\cite{Bryan:2004iq,Caporaso:2005fp,Caporaso:2006gk}. Recall
that it is precisely this equivariance which localizes the ${\cal
  N}=4$ D-brane gauge theory in four dimensions to a two-dimensional
gauge theory.

The elliptic threefold $X=X_p$ is special in this regard, as then
$c_1(TX_p)=0$ and the virtual dimension (\ref{virtualdim})
vanishes. In this case the fixed points (\ref{modXSigmaiso}) are
isolated and the residue Gromov-Witten invariants are simply given by
the degrees of the corresponding virtual moduli spaces as
\bea
N_d^g(X_p)&:=&\int_{[~\overline{\mathcal{M}}_{g} (X_p,
d)\,]^{\mathrm{vir}}}~1\nonumber\\[4pt]
&=&\int_{[~\overline{\mathcal{M}}_{g} (X_p,
d)^T\,]^{\mathrm{vir}}}~\frac1{e_T^{g,d}(X_p)}=
\int_{[~\overline{\mathcal{M}}_{g} (\torus^2,d)\,]^{\mathrm{vir}}}~
\frac1{e_T^{g,d}(X_p)} \ ,
\label{defGW}\eea
where the moduli space $\mathcal{M}_{g} (\torus^2,d)$ is precisely the
Hurwitz space of simple branched covers of the underlying elliptic
curve. Note that the isomorphism (\ref{modXSigmaiso}) is crucial for
this correspondence between Gromov-Witten invariants and Hurwitz
numbers, i.e. it is a feature of the {\it equivariant} topological
string theory on $X_p$ defined with respect to the natural torus
symmetries. This is the definition that is provided by the topological
vertex constructions of
Sect.~\ref{TopStringElliptic}~\cite{Li:2004uf}. The invariants
(\ref{defGW}) are rational-valued because of the orbifold nature of
the Deligne-Mumford moduli spaces involved. The equalities
(\ref{modXSigmaiso}) and (\ref{defGW}) show directly that this
topological string theory coincides with that defined by the standard
two-dimensional sigma-model with target space
$\torus^2$~\cite{Bershadsky:1993cx}, with the contributions of
$T$-equivariant Euler characters in (\ref{defGW}) coinciding with the
usual orbifold Euler characters of the analytically compactified
Hurwitz spaces~\cite{Cordes:1994sd}.\footnote{The analogous formulas
  in the case of a curve $\Sigma$ of genus
  $g\neq1$~\cite{Bryan:2004iq} make precise the observation
  of~\cite{deHaro} that the Euler characters of configuration spaces
  of Riemann surfaces (whose orbifold singularity blowups yield
  Hilbert schemes of points in $\Sigma$) continue to appear in the
  large $N$ expansion of $q$-deformed Yang-Mills theory on
  $\Sigma$. In this case one should introduce an additional Euler
  class of the tangent bundle to the moduli space in the moduli space
  integral in order to account for the $|2g-2|$ fibre D-branes arising
from the large $N$ expansion of the gauge theory.}

The simple Hurwitz numbers $H_{g,d}\in\rat$ of the torus may be
extracted from the explicit form for the chiral free energy
(\ref{chiralFE2}) of Yang-Mills theory on $\torus^2$ as derived from
eq.~(\ref{Z2dtorus}). They are given by the
formula~\cite{Griguolo:2004jp}
\begin{equation}
H_{g,d} = \sum_{k=1}^d \,\frac{(-1)^k}{k}~ \sum_{
\stackrel{\scriptstyle \mbf d \in \nat^k}{\scriptstyle
\sum_l\,d_l=d}}~ \sum_{ \stackrel{\scriptstyle \mbf g \in
\nat^k}{\scriptstyle \sum_l\, g_l=g}}\, H^\bullet_{g_1 , d_1} \,\cdots
\,H^\bullet_{g_k , d_k} \ ,
\end{equation}
where $H^\bullet_{g , d}$ are the {\it disconnected} simple Hurwitz
numbers which count reducible coverings of the torus and can be
expressed through the combinatorial formula
\begin{eqnarray} \label{hurwitznumbers}
H^\bullet_{g , d}&=&\sum_{k=1}^d~\sum_{\stackrel{\scriptstyle
\mbf d\in\nat^k\,,\,\sum_l\,d_l=d}{\scriptstyle d_1\geq
d_2\geq\cdots \geq
d_k}}\,\left[~\sum_{l=1}^k\,\frac{(d_l+k-l)\,(d_l+k-l-1)}2\right.
\nonumber\\ && \times~\Biggl.~\prod_{l'\neq l}\,
\left(\frac{d_l-d_{l'}+l'-l-2}{d_l-d_{l'}+l'-l}\right)\Biggr]^{2g-2}
\ .
\end{eqnarray}
By comparing the free energy (\ref{chiralFE2}) in the trivial charge
sector $\ell=0$ with the expansion (\ref{GWexpansion}) and the
relation (\ref{FgXpFgrel}), we arrive at our desired closed formula
for the Gromov-Witten invariants in the form
\begin{equation} \label{GWvsH}
N_d^g  \left( X_p \right) = \frac{1}{( 2  g -2)!}\, \left(
\frac{p}{2} \right)^{2g-2}~H_{g , d} \ .
\end{equation}

We can understand better the geometrical meaning of this result by
looking more closely at the framing dependence of
(\ref{GWvsH}). For $p=0$, only the genus one invariants are non-zero
and they count unramified coverings of the torus of degree $d$, or
equivalently sublattices of $\zed\oplus\tau\,\zed$ of index $d$ which
after taking into account automorphisms is the number
\beq
N_d^1(X_0)=H_{1,d}=\frac1d\,\sum_{k|d}\,k \ .
\label{Nd1X0}\eeq
This is consistent with the free energy computed in (\ref{TopAmp3})
and in (\ref{FgXpFgrel},\ref{freeencompact}), and it has also been
independently derived directly in Gromov-Witten theory~\cite{Pand1}
by using basic constructions related to the virtual fundamental
class. This structure arises because for $p=0$ the
elliptic curve is {\it isolated} in $X_0$, and (\ref{Nd1X0})
represents the expected contributions~\cite{Bershadsky:1993cx} of
isolated genus one curves and their multi-wrappings to the free energy
$F^{X_0}_1(t)$. From eq.~(\ref{TopAmp3}) it follows that
there is no contribution to $F^{X_0}_g(t)$ for $g>1$, in accordance
with~\cite{Bershadsky:1993cx}, due to the absence of non-trivial
two-cycles of higher genus in the background geometry and the fact
that there are no bubbling contributions of genus one curves to higher
genus curves.

In contrast, for $p>0$ the Gromov-Witten invariants (\ref{GWvsH}) are
generically non-vanishing for {\it all} genera. In this case the
elliptic curve is not isolated but belongs to a continuous family
$\psi_z:\torus^2(z)\to X_p$, $z\in\complex^2$ of holomorphic maps
determined by sections of the non-trivial fibration
(\ref{threefold}). Then the contribution to $F^{X_p}_g(t)$ is simply
the total number (modulo automorphisms and an irrelevant overall
rescaling by $p$) of genus~$g$ branched covers of the torus. By
composing covering maps with the holomorphic maps $\psi_z$, one
obtains a family of holomorphic maps from genus~$g$ surfaces to $X_p$
whose parameter space is
$\overline{\mathcal{M}}_{g}(\torus^2)\times\complex^2$, where
$\overline{\mathcal{M}}_{g}(\torus^2)=
\coprod_{d\geq1}\,\overline{\mathcal{M}}_{g}(\torus^2,d)$ can be
described~\cite{Cordes:1994sd} as the base space of an
infinite-dimensional bundle whose total space
parametrizes genus~$g$ holomorphic maps to $\torus^2$. By taking the
Euler class of $\complex^2$ to be~$1$, the free energy
$F^{X_p}_g(t)$ may alternatively be described~\cite{Cordes:1994sd} as
the (orbifold) Euler character of a bundle over the moduli space of
parameters $\overline{\mathcal{M}}_{g}(\torus^2)\times\complex^2$ for
the family of genus~$g$ curves. Because the virtual dimension of the
moduli space of maps $\overline{\mathcal{M}}_{g}(\torus^2,d)$ is zero
in the present case, this is again in agreement with
general expectations~\cite{Bershadsky:1993cx}. There are
no other contributions due to the isomorphism (\ref{modXSigmaiso}). A
more physical picture of these features of the topological
string expansion will be described in Sects.~\ref{GVInt}
and~\ref{DTTheory} below in a
manner akin to the construction of the black hole partition function
in Sect.~\ref{DPartFn}.

\subsection{Gopakumar-Vafa Integrality\label{GVInt}}

Within the framework of Type~IIA superstring theory, the rational
invariants $N^g_d(X_p) $ can be expressed in terms of the {\it
integer} Gopakumar-Vafa invariants $ n^g_d(X_p)
$~\cite{Gopakumar:1998ii,Gopakumar:1998jq} which count
four-dimensional BPS states of wrapped D2 branes in a Type~IIA
compactification on $X_p$. The all genus topological string free
energy (\ref{GWexpansion}) encodes the contributions of these states
through the expansion
\begin{eqnarray} \label{GVexpansion}
{F}_{X_p}(g_s , t) &=& \sum_{r=1}^{\infty} ~\sum_{d=1}^\infty\,
n^r_d(X_p)~ \sum_{k=1}^{\infty}
\,\frac{1}{k}\, \left( 2 \sin \mbox{$\frac{k \, g_s}{2}$}
 \right)^{2r -2}~Q^{k \,d} \ ,
\end{eqnarray}
where $n^r_d(X_p)$ is a twisted supersymmetric index in four
dimensions which receives contributions only from BPS states of $d$ D2
branes which wrap the embedded torus $\torus^2$ in $X_p$. The quantum
number $r$ of these particles labels their left spin representation
\beq
I_r:=(\,\underline{1}\oplus\underline{2}\oplus\underline{1}\,)^{\otimes
  r}
\label{Irspindef}\eeq
of the rotation group $SO(4)\cong SU(2)\times SU(2)$, where
$\underline{m}$ denotes the irreducible $m$-dimensional representation
of $SU(2)$. The integer $k$ in (\ref{GVexpansion}) is the number of D0
branes used to form the given D0--D2 BPS bound state.

A precise mathematical definition of these invariants is not yet
known and the integrality of the numbers $ n^g_d(X_p)$,
although obvious from their Type~IIA definition, has no rigorous proof
directly in the context of the topological string theory on $X_p$. Analogously
to~\cite{Caporaso:2005fp}, we will provide another non-trivial check
of the conjecture of~\cite{Vafa:2004qa,Aganagic:2004js} by showing
that the local elliptic threefolds $X_p$ satisfy the integrality
conjecture. By assuming that the partition function of the chiral sector of
Yang-Mills theory on the torus is equivalent to the partition function
of topological string theory on $X_p$, we will compute the
Gromov-Witten invariants using the technique explained in
Sect.~\ref{GWCalcs} above and extract from them the Gopakumar-Vafa
invariants by using an inversion formula. The displayed integrality of
the closed BPS invariants obtained in this manner then provides
further strong evidence that two-dimensional Yang-Mills theory indeed
does provide a non-perturbative definition of the topological string.

Gopakumar-Vafa invariants are related to Gromov-Witten invariants by
equating the instanton sum (\ref{GWexpansion}) to the index
(\ref{GVexpansion}). This relationship was explicitly inverted
in~\cite{Bryan:2000ve}, and we will now briefly review the
derivation. The idea is simply to match the coefficients of both
series regarded as functions of $Q$ and $g_s$, and thereby determine
$n_d^r(X_p)$ recursively in terms of $N_d^g(X_p)$. For this, we write
$k= \frac{n}{d}\in\nat$ in (\ref{GVexpansion}) for fixed $n$ and match
the coefficients of the $Q^{n}$ term in each series. This gives the
relation
\begin{equation} \label{GWtoGV1}
\sum_{r=1}^\infty\, N^r_{n }(X_p) \, n^{3 - 2r} ~ x^{2r -2} =
\sum_{r=1}^\infty~\sum_{d|n}\,  n^r_{d } (X_p) \, d ~ \left( 2 \sin
\mbox{$\frac{x}{2d}$} \right)^{2r -2}
\end{equation}
where $x = n \, g_s$. We now apply the M\"obius inversion
formula. If $g$ is a function of positive integers and $f(n)=
\sum_{d|n} \,g(d)$ then $g(d) = \sum_{k|d}\, \mu \big(
\frac{d}{k}\big)\, f(k)$, where $\mu:\nat\to\{0,\pm\,1\}$ is the
M\"obius function defined by $\mu(1)=1$, $\mu(n)=0$ if $n$ has a
square divisor and $\mu (n)=(-1)^s$ if $n$ can be factorized into a
product of $s$ distinct primes. Applying this inversion formula to
(\ref{GWtoGV1}) and defining $y = \sin\frac{x}{2 d}$ gives
\begin{equation}
\sum_{r=1}^\infty\, n^r_{d } (X_p)~ y^{2r-2} = \sum_{r=1}^\infty~
\sum_{k|d}\, \mu\big( \mbox{$\frac{d}{k}$} \big)\,
\left(\mbox{$\frac{d}{k}$}
\right)^{2r-3} N^r_{k } (X_p)~ \left( 2 \arcsin \mbox{$\frac{y}{2}$}
\right)^{2r-2} \ .
\label{GWGVMobius}\end{equation}

The final step consists of trading the sum over $k$ for one over
${d}/{k}$, expanding the inverse sine function in a power series,
and equating the coefficients of $y^{2r -2}$ on both sides of
(\ref{GWGVMobius}). In this way we arrive at
\begin{equation} \label{inversionformula}
n^r_{d } (X_p) = \sum_{g=1}^{r}\, \alpha^{r}{}_{ g}~
\sum_{k|d}\, \mu( k)\,k^{2g-3} ~N_{{d}/{k} }^{g} (X_p)
\end{equation}
where the rational number $\alpha^r{}_{g}$ is the coefficient of
$u^{r-g}$ in the power series expansion of the function
$\big[{\arcsin(\frac{\sqrt{u}}{2})}/{(\frac{\sqrt{u}}{2})}\big]^{2g-2}$,
which may be determined explicitly through the recursion relations
\begin{eqnarray}
\alpha^r{}_r &=& 1  \ , \nonumber \\[4pt]
\alpha^r{}_{g}&=&\frac1{r-g}\,\sum_{s=1}^{r-g}\,\frac{\big(g\,(2s+1)-r-s
\big)\,(2s)!}{2^{4s}\,(2s+1)\,(s!)^2}~\alpha^{r-s}{}_g \ , \quad g<r \ .
\end{eqnarray}
The inversion formula (\ref{inversionformula}) can now be applied
to the Gromov-Witten invariants computed from the quasi-modular
expansions (\ref{GWF2}) and (\ref{GWF3}) to obtain the
Gopakumar-Vafa invariants. For spin $r=2$ one finds the BPS invariants
up to degree $d=30$ as provided in the table
\begin{eqnarray}
\begin{array}{|c|c|c|c|c|c|}
\hline 0 &-4\,p^2&-32\,p^2&-112\,p^2&-320\,p^2&-644\,p^2\\\hline
-1344\,p^2&-2240\,p^2&-3744\,p^2&-5700\,p^2&-8800\,p^2&-11888\,p^2
\\\hline -17472\,p^2&-23044\,p^2&-30560\,p^2&-39680\,p^2&-52224\,p^2&
  -63684\,p^2
   \\\hline -82080\,p^2&-98160\,p^2&-121184\,p^2&-146564\,p^2&
  -178112\,p^2&-204992\,p^2 \\\hline
  -248000\,p^2&-288964\,p^2&-336960\,p^2&-387184\,p^2&-454720\,p^2&
-508100\,p^2 \\\hline
\end{array}
\end{eqnarray}
where the degree increases from left to right and from top to
bottom. For spin $r=3$ one likewise has the table
\begin{small}
\begin{eqnarray}
\begin{array}{|c|c|c|}
\hline 0 & \frac{-p^2}{3} + \frac{4\,p^4}{3}&\frac{-8\,p^2}{3} +
\frac{320\,p^4}{3}  \\ \hline \frac{-28\,p^2}{3} + \frac{4864\,p^4}{3} &
  \frac{-80\,p^2}{3} + \frac{36608\,p^4}{3}&\frac{-161\,p^2}{3} +
  \frac{178436\,p^4}{3}  \\\hline -112\,p^2 +
  227712\,p^4 &
  \frac{-560\,p^2}{3} + \frac{2098688\,p^4}{3}&-312\,p^2 +
  1915584\,p^4 \\\hline -475\,p^2 + 4578348\,p^4&
  \frac{-2200\,p^2}{3} + \frac{30846400\,p^4}{3}&\frac{-2972\,p^2}{3}
  + \frac{62634752\,p^4}{3} \\\hline -1456\,p^2 +
  41108736\,p^4&
  \frac{-5761\,p^2}{3} + \frac{224845828\,p^4}{3}&\frac{-7640\,p^2}{3} +
  \frac{400500800\,p^4}{3}  \\ \hline
  \frac{-9920\,p^2}{3} + \frac{675368960\,p^4}{3} & -4352\,p^2 +
  374105600\,p^4 &-5307\,p^2 + 590587692\,p^4 \\\hline
  -6840\,p^2 + 930011328\,p^4 & -8180\,p^2 + 1401585920\,p^4&
  \frac{-30296\,p^2}{3} + \frac{6303843776\,p^4}{3}\\ \hline
  \frac{-36641\,p^2}{3} + \frac{9159665924\,p^4}{3} & \frac{-44528\,p^2}{3} +
  \frac{13262093696\,p^4}{3}&\frac{-51248\,p^2}{3} +
  \frac{18570790400\,p^4}{3} \\\hline
   \frac{-62000\,p^2}{3} +\frac{26133520640\,p^4}{3} &
  \frac{-72241\,p^2}{3} + \frac{35706397060\,p^4}{3}&- 28080\,p^2 +
  16250682240\,p^4  \\ \hline
  \frac{-96796\,p^2}{3} + \frac{65187144448\,p^4}{3} &
  \frac{-113680\,p^2}{3} +
  \frac{87248171776\,p^4}{3}&\frac{-127025\,p^2}{3} +
  \frac{113930816900\,p^4}{3} \\ \hline
\end{array} \ .
\end{eqnarray}
\end{small}

These tables exhibit the remarkable feature that each invariant
$n^r_d (X_p)$ is an integer for every $p\in \zed$. The full set of
Gromov-Witten and Gopakumar-Vafa invariants up to genus $g=6$, spin
$r=6$ and degree $d=15$ can be found in Appendix~C, with the same
integrality properties. It is also possible to obtain a formal
combinatorial solution to the BPS state counting problem in terms of
Hurwitz numbers by using (\ref{GWvsH}) to write
\begin{equation} \label{GVvsH}
n^r_{d } (X_p) = \sum_{g=1}^{r}\, \frac{\alpha^r{}_{g }}{( 2g
  -2)!}\,\left(\frac{p}{2} \right)^{2g-2}~\sum_{k|d}\, \mu( k)
\,k^{2 g -3} ~H_{g,d/k} \ .
\end{equation}
Given this rather explicit formula, it should be feasible to construct
a combinatorial proof of the integrality of the BPS invariants of
$X_p$, but we have not been able to do so.

The physical and geometrical significances of this result can be
understood by again closely analysing the framing dependence of
(\ref{GVvsH}). Consider first the case $p=0$. Then only the $g=1$ term
survives in (\ref{GVvsH}), and since $\alpha^r{}_1=\delta_{r,1}$ the
only non-zero invariant is
\beq
n^1_d(X_0)=\sum_{k|d}\, \frac{\mu(
  k)}k~H_{1,d/k}=\frac1d\,\sum_{k|d}\,\mu(k)~\sum_{n|d/k}\,n=1
\label{n1dX0}\eeq
where the last equality can be derived by applying M\"obius inversion
to the function $f(n)=\sigma_1(n):=\sum_{d|n}\,d$. This result means
that $d$ D2 branes in representation $I_1$, containing two spin~$0$
particles and one spin~$\frac12$ particle, form a single bound state for all
$d$ and there are no stable bound states of D2 branes in the higher
spin representations $I_r$, $r>1$. Geometrically, it arises as a
consequence of the fact that, for D2 branes wrapping an isolated
elliptic curve, only curves $\Sigma$ whose genera $g$ coincide with
the arithmetic genus $r=1$ realize that class.

Quantum mechanically, the configuration of D2 branes fluctuates
over the D-brane moduli space ${\cal M}_{\rm D}(X_0,d)$. We can
algebraically deform the set of $d$ D2 branes wrapping the isolated
$\torus^2$ into a single large D2 brane wrapped around the cycle $d$
times, as such a deformation defines a point in the same moduli
space. One component of this moduli space then comes from the $U(1)$
flux of the bound D0 branes, which is the moduli space
${\cal M}_{1,0}(1,0)$ of flat $U(1)$ line bundles over $\torus^2$. From
eq.~(\ref{modspinstNq}) it follows that this moduli space is the
dual torus $\tilde\torus^2$, which is just the jacobian torus of the
elliptic curve. The D-brane moduli space in this case is thus ${\cal
  M}_{\rm D}(X_0,d)=\overline{\cal
  M}_1(X_0,d)^T\times\tilde\torus^2$. The supersymmetric quantum
ground states correspond to cohomology classes in
$H^\sharp({\cal M}_{\rm D}(X_0,d),\complex)\cong
H^\sharp(\,\overline{\cal M}_1(X_0,d)^T,\complex)\otimes
H^\sharp(\tilde\torus^2,\complex)$. There is an isomorphism
$H^2(\tilde\torus^2,\complex)=\omega_{\torus^2}\cdot
H^0(\tilde\torus^2,\complex)$. The K\"ahler class induces a Lefschetz
operator which defines a representation of the group $SU(2)$ on the cohomology
ring $H^\sharp(\tilde\torus^2,\complex)$ through the Lefschetz
decomposition
\beq
\underline{2}=H^0\big(\tilde\torus^2\,,\,\complex\big)\oplus
H^2\big(\tilde\torus^2\,,\,\complex\big)\cong \complex\oplus
t\,\complex \ , \qquad
\underline{1}\oplus\underline{1}=H^1\big(\tilde\torus^2\,,\,
\complex\big)\cong\complex^2 \ .
\label{cohringT2}\eeq
It follows that the full Hilbert space of ground states is given by
its decomposition
\beq
H^\sharp\big({\cal M}_{\rm D}(X_0,d)\,,\,\complex\big)=
H^\sharp\big(\,\overline{\cal
  M}_1(X_0,d)^T\,,\,\complex\big)\otimes I_1
\eeq
as an $SU(2)$ representation.

On the other hand, for $p>0$ the Gopakumar-Vafa invariants
(\ref{GVvsH}) are generically non-vanishing for all arithmetic genera
$r$. As the elliptic curve is no longer isolated inside the threefold
$X_p$, by composing with covering maps of $\torus^2$ there are
contributions to the number of BPS states from {\it all} allowed
genera $g$ of the D2 branes with $1\leq g\leq r$ and all degrees $d'$
dividing their number $d$ for which the M\"obius function
$\mu\big(\frac d{d'}\big)$ is non-vanishing. The sum over Hurwitz
numbers $H_{g,d'}$ at fixed $g$ is generically expected from adding up
all torus invariant bound states. The quantum states contributing
at each genus $g$ correspond to the jacobian torus $\tilde\torus^{2g}$
(the moduli space of flat $U(1)$ bundles on a genus $g$ curve), whose 
complex cohomology ring admits a Lefschetz decomposition
$H^\sharp(\tilde\torus^{2g},\complex)\cong I_g$ as a representation of
$SU(2)$. The spin $r$ invariants (\ref{GVvsH}) can thus be thought of
as the virtual number of genus~$r$ jacobians contained in the D-brane
moduli space ${\cal M}_{\rm D}(X_p,d)$. This moduli space should be
fibred over the geometric moduli spaces $\coprod_{g\geq1}\,\overline{\cal
  M}_g(X_p,d)^T$ in (\ref{modXSigmaiso}) with fibre
$\tilde\torus^{2g}$ over each imbedded curve $(\Sigma,\psi)\in\overline{\cal
  M}_g(X_p,d)^T$. The integers (\ref{GVvsH}) then give the
coefficients in the decomposition of the fibrewise representation of
$SU(2)$ on $H^\sharp({\cal M}_{\rm D}(X_p,d),\complex)$ in the basis
given by the cohomologies of the jacobian tori.

However, the full D-brane moduli space at $p>0$ at present is not
known. To specify a D2 brane wrapping a holomorphic curve $\Sigma$ one
needs to fix a holomorphic line bundle, as above, or more generally a
semi-stable coherent sheaf over $\Sigma$. In~\cite{Hosono:2001gf} it
was proposed to take ${\cal M}_{\rm D}(X_p,d)$ as the normalized
moduli space of degree $k$ semi-stable sheaves on $X_p$ (with $k$
corresponding to the D0 brane charges in (\ref{GVexpansion})) of
pure dimension~$1$ with Hilbert polynomial $d\,k+1$, with the natural
morphism onto the moduli space of support curves of sheaves of degree $d$ in
$X_p$. In~\cite{Katz:2006gn} it was suggested that the BPS invariants
(\ref{GVvsH}) could be defined in terms of Donaldson-Thomas invariants
of this moduli space, at least in some special
circumstances. In~\cite{Dijkgraaf:2006um} a physical relation between
the Gopakumar-Vafa and Donaldson-Thomas invariants was proposed in
terms of the counting of black hole microstates in four and five
dimensions. Understanding
the connection between our formula (\ref{GVvsH}) and Donaldson-Thomas
theory could help shed light on the precise geometrical definition of
Gopakumar-Vafa invariants for generic threefolds, and we briefly
explore this connection in Sect.~\ref{DTTheory} below.

Note that for $d=1$ and $r>1$, all invariants (\ref{GVvsH}) vanish,
while again $n_d^1(X_p)=1$ for all $p\geq0$ and all $d\geq1$. This
owes to the geometrical fact that a simple branched covering of the
torus by a curve of genus $g>1$ requires at least two sheets (so that
$\overline{\cal M}_g(X_p,1)^T=\emptyset$ for all $g>1$). Physically,
this means that there are no higher spin bound states of a single
wrapped D2 brane. On the other hand, bound states of multiply wrapped
D2 branes on a non-isolated $\torus^2$ appear to exist for all
arithmetic genera $r$. 

\subsection{Donaldson-Thomas Theory\label{DTTheory}}

The topological A-model on the local Calabi-Yau threefold $X_p$
localizes onto a sum over worldsheet instantons that can be described
as curves of particular genus embedded in the target space. However,
there is another point of view that naturally arises in the counting
of complex curves. Each curve can be characterized by a set of
equations on the manifold for which the zero locus defines an ideal
sheaf on $X_p$. The two points of view are complementary. While the
first one is natural in the context of worldsheet instanton
contributions leading to the Deligne-Mumford moduli space of
holomorphic curves in $X_p$ and Gromov-Witten theory, the second one
is more apt to the counting of D-branes in a dual theory leading to
the moduli space of ideal sheaves on $X_p$ and Donaldson-Thomas
theory. This dual theory should provide the proper construction of the
BPS invariants (\ref{GVvsH}) as the multiplicities of $SU(2)$
Lefschetz representations on the cohomology of the moduli space of
sheaves on $X_p$. Since $X_p$ is a local Calabi-Yau threefold fibred
over $\torus^2$, this cohomology should reduce to that of an
appropriate moduli space of rank $d$ bundles on $\torus^2$, consistent
with the expectation that the dual theory is realized as a
six-dimensional topological $U(1)$ gauge theory on
$X_p$~\cite{Okounkov:2003sp,Iqbal:2003ds}. We will verify these
expectations presently.

Each ideal sheaf $\mathcal{I}$ on $X_p$ defines a subscheme $Y\subset
X_p$ through the short exact sequence of sheaves given by
\begin{equation}
0 ~\longrightarrow~ \mathcal{I}~ \longrightarrow~ \mathcal{O}_{X_p}~
\longrightarrow ~\mathcal{O}_Y ~\longrightarrow~ 0
\end{equation}
where $\mathcal{O}_{X_p}$ (resp.~$\mathcal{O}_Y$) is the structure
sheaf of holomorphic functions on $X_p$ (resp.~$Y$). When $Y$ contains
the support curve of the sheaf, it is characterized
by its homology class $\beta= \left[ Y\right]=d\,[\torus^2] \in H_2
(X_p , \zed)$ and by the holomorphic Euler characteristic $\chi
(\mathcal{O}_Y) = m$. The degree of $Y$ may be characterized as the
intersection number $\#(Y\cap(X_p)_z)=d$ over a generic base point
$z\in\torus^2$. The set of isomorphism classes is the
moduli space of ideal sheaves $I_m (X_p ,d)$ which is isomorphic to
the Hilbert scheme of curves of the given topology in $X_p$. Counting
elements of $I_m (X_p ,d)$ corresponds physically to counting BPS
bound states of a single D6 brane wrapping $X_p$ and $d$ D2 branes
wrapped on the holomorphic two-cycles $[Y]$ with $|m|$ D0 branes.

Generally, an application of the Grothendieck-Riemann-Roch theorem
shows that the virtual dimensions of these moduli spaces are again
given by the formula (\ref{virtualdim}). The Donaldson-Thomas
invariants of $X_p$ are defined as the degrees of the zero-dimensional
virtual fundamental classes of $I_m (X_p,d)$ through
\begin{equation} \label{defDT}
D^m_{d}(X_p) = \int_{[I_m (X_p ,d)]^{\mathrm{vir}}}~ 1 \ .
\end{equation}
Analogously to (\ref{defGW}), the toric action on $X_p$ canonically
lifts to the moduli space of ideal sheaves and the integration is
defined by virtual localization onto the isolated $T$-fixed point set
$I_m (X_p,d)^T\subset I_m(X_p ,d)$ as an equivariant
residue~\cite{MNOP,OP}. Compared with (\ref{defGW}), the Euler
characteristic of the sheaf now plays the role of the genus of the
curve. Any support curve $Y\subset X$ associated to an element of
$I_m(X_p,d)^T$ is preserved by the torus action and hence is supported
on the base $\torus^2$. The genus of $Y$ may be defined as the integer
$g(Y)=1-\chi({\cal O}_Y)$ and can be negative. The invariants
(\ref{defDT}) are integer-valued since no orbifold is taken in the
definition of the moduli space.

These invariants can be encoded in the Donaldson-Thomas partition
function
\begin{equation}
\mathcal{Z}^{\rm DT}= 1+\sum_{d=1}^\infty\,Q^d~{\cal Z}_d^{\rm DT}=
1+\sum_{d=1}^\infty\, Q^d~\sum_{m=-\infty}^\infty\,
D^m_d(X_p)~(-q)^m \ .
\label{DTpartfn}\end{equation}
The leading degree $d=0$ term is ${\cal Z}_0^{\rm DT}=1$ in this case
since $\chi(X_p)=c_1(TX_p)=0$~\cite{MNOP,OP}. It was conjectured
in~\cite{MNOP}, and proven in~\cite{OP} for local Calabi-Yau
threefolds fibred over a Riemann surface, that the Donaldson-Thomas partition
function (\ref{DTpartfn}) coincides with the perturbative topological
string partition function $Z_{X_p}=\e^{F_{X_p}}$ after the analytic
continuation $g_s\to-\ii g_s$ in the all genus expansion
(\ref{GWexpansion}).\footnote{We will encounter similar (but
  non-analytic) continuations of coupling constants in the next
  section.} This continuation is a result of the
relation~\cite{Iqbal:2003ds} between the string coupling $g_s$ and the
$\theta$-angle of the noncommutative topologically twisted $U(1)$
gauge theory on $X_p$ given by $q=-\e^{\ii\theta}$. The noncommutative
deformation of the present geometry can be understood in terms of the
formal toric blowup ${\cal X}_p\to X_p$ constructed in 
Sect.~\ref{TopStringElliptic}. Torsion-free
sheaves on $X_p$ can be lifted to ${\cal X}_p$ where they may be
identified with line bundles. Most of the ensuing results in the
rest of this section should therefore be understood as being derived on
${\cal X}_p$ {\it after} the blowdown projection is taken, as described in
Sect.~\ref{FTG}. Indeed, one has a completely analogous equality as
(\ref{GWinvsequal}) for the corresponding Donaldson-Thomas
invariants~\cite{HuLiDT}.

The main impetus into this identification is the
observation~\cite{Okounkov:2003sp,Iqbal:2003ds} that the gluing rules
of the topological vertex used to construct the topological string
amplitude in Sect.~\ref{TopStringAmpl} naturally correspond to torus
invariant ideal sheaves in the threefold $X_p$. In particular, the
$SU(\infty)$ representations $\hat R$ contributing to the chiral gauge
theory partition function (\ref{toptorus}) on $\torus^2$ are in
bijective correspondence through their Young diagrams with the sets
$\{(i,j)\in\hat R~|~z_2^i\,z_3^j\notin I_{\hat R}\}$, where $I_{\hat
  R}$ is a one-dimensional monomial ideal in the coordinate ring
$\complex[z_1^{\pm1},z^{\phantom{1}}_2,z^{\phantom{1}}_3]$ generated
by $z_2^{n_i(\hat R)}\,z_3^{i-1}$, $i=1,\dots,N-1$. The homogeneous ideal
$I_{\hat R}$ is the restriction of an ideal sheaf ${\cal I}_{\hat R}\in
I_m(X_p,d)^T$, with $d=|\hat R|$, to the intersection of $\complex^3$
patches of the formal toric geometry of $X_p$ constructed in
Sect.~\ref{TopStringElliptic}. This naturally ties the two-dimensional
gauge theory to the melting crystal picture of~\cite{Okounkov:2003sp}.

By matching the partition functions (\ref{DTpartfn}) and
(\ref{toptorus}) we can write down a formal combinatorial solution for
the Donaldson-Thomas theory of $X_p$. After a rearrangement of the
expansion (\ref{toptorus}) into a sum over Young tableaux $\hat R$
with fixed numbers of boxes $|\hat R|=d$ and fixed quadratic Casimir
invariants $\kappa_{\hat R}$, one finds
\beq
D_d^m(X_p)=(-1)^m~\#\big\{\hat R~\big|~\hat R\vdash d ~ , ~
\mbox{$\frac12$}\,p\,\kappa_{\hat R}=m\big\} \ .
\label{DdmXpnumYoung}\eeq
In particular, the Donaldson-Thomas invariants vanish whenever
$m\notin p\,\zed$.

Once again, let us study the framing dependence of
(\ref{DdmXpnumYoung}). For $p=0$ the Donaldson-Thomas invariants
vanish for all $m\neq0$, and only bound states {\it without} D0
branes exist. For $m=0$ one has
\beq
D_d^0(X_0)=\Pi(d)
\label{Dd0X0Pid}\eeq
where $\Pi(d)$ is the number of proper unordered partitions of the
degree $d$ into positive integers. For instance, the first few
numbers of D6--D2 instanton bound states are $D_1^0(X_0)=1$,
$D_2^0(X_0)=2$, $D_3^0(X_0)=3$, $D_4^0(X_0)=5$, $D_5^0(X_0)=7$, and so
on. The instanton number (\ref{Dd0X0Pid}) grows asymptotically
according to the Hardy-Ramanujan formula as
$D_d^0(X_0)=\e^{\pi\,\sqrt{2d/3}}/4\,\sqrt3\,d$ for $d\to\infty$. To
understand this result geometrically, we associate as above to
any partition $\hat R\vdash d$ a monomial ideal $I_{\hat R}$ in the
fibre coordinates $(z_1,z_2)\in\complex^2\subset X_0$. It defines a
subscheme $Y_{\hat R}\subset X_0$ supported on $\torus^2$ at
$z_1=z_2=0$ of degree $d$ and Euler characteristic $\chi({\cal
  O}_{Y_{\hat R}})=0$ due to the triviality of the normal bundle
(\ref{T2normalbundle}). These are the only such
subschemes, and each of these ideals defines an isolated point of the
moduli space $I_0(X_0,d)^T$. Eq.~(\ref{Dd0X0Pid}) now follows since
there are $\Pi(d)$ such points. On the other hand the moduli space
$I_1(X_0,1)^T$, for example, consists of ideal sheaves of support
$Y=\torus^2\cup\{z\}$ with $z\in X_0\setminus\torus^2$. It
follows~\cite{Katz:2004js} that $I_1(X_0,1)^T$ is the equivariant
blowup of $X_0$ along the elliptic curve $\torus^2$, and so
$D_1^1(X_0)=-\chi(I_1(X_0,1)^T)=-\chi(X_0)=0$.

Now let us turn to the cases $p>0$. For a Young tableau $\hat R$
containing $d$ boxes, the quadratic Casimir invariant can be written
as~\cite{MacDonald}
\beq
\kappa_{\hat R}=d+\sum_{i=1}^{N-1}\,n_i(\hat R)\,\big(n_i(\hat R)-2i
\big)=2\,\sum_{(i,j)\in\hat R}\,(j-i) \ .
\label{kappacontent}\eeq
The integer $j-i$ is called the {\it content} of the box $(i,j)\in\hat
R$. For a given fixed degree $d$, the only non-vanishing
Donaldson-Thomas invariants (\ref{DdmXpnumYoung}) are those for which
the Euler characteristic $m$ is determined by the total content of a
Young diagram $\hat R\vdash d$ to be
\beq
m=m_{\hat R}(p):=p\,\sum_{(i,j)\in\hat R}\,(j-i) \ .
\label{mhatR}\eeq
There are $\Pi(d)$ distinct Young tableaux $\hat R$ and hence $\Pi(d)$
distinct integers (\ref{mhatR}), since (\ref{kappacontent}) is
an invariant of the corresponding inequivalent $SU(\infty)$
representations. The combinatorial relation (\ref{mhatR}) selects the
allowed D0 brane Ramond-Ramond charges in a given number $d$ of D2
branes for which stable bound states form, and for each such charge
there is a unique (virtual) equivariant D6--D0 brane bound state
giving the degree
\beq
D_d^{m_{\hat R}(p)}(X_p)=(-1)^{m_{\hat R}(p)} \ .
\label{DdmhatRXp}\eeq

Note that if $\hat R$ is a partition of $d$, then so is its transpose
$\hat R^\top$ and the bijection $\hat R\to\hat R^\top$ sends
$(i,j)\to(j,i)$ for each box $(i,j)\in\hat R$ in (\ref{mhatR}). This
is equivalent to the reflection symmetry $m_{\hat R}(p)\to m_{\hat
  R^\top}(p)=-m_{\hat R}(p)$, proving that the Donaldson-Thomas partition
function (\ref{DTpartfn}) is invariant under the transformation
$q\to1/q$ exchanging D0 branes with antibranes as conjectured
in~\cite{MNOP}. Since the normal bundle to the non-planar edge in the
formal toric geometry of $X_p$ is
$\mathcal{O}_{\torus^2}(p)\oplus\mathcal{O}_{\torus^2}(-p)$, the
formula (\ref{mhatR}) coincides with the general formula for the Euler
characteristic $\chi({\cal O}_{Y_{\hat R}})$ derived in~\cite{MNOP}
using elementary calculations in toric geometry. Alternatively, this
formula can be derived by calculating the six-dimensional instanton
contributions to the deformed $U(1)$ gauge theory on
$X_p$~\cite{Iqbal:2003ds} (with the one-loop fluctuation determinants
on the blowup ${\cal X}_p$ given by the products in
eq.~(\ref{TopAmp2a})), demonstrating the equivalence between the
two-dimensional and six-dimensional gauge theory descriptions of the
topological string. The first few allowed charges are provided in the
table
\begin{eqnarray}
\begin{array}{|c|c|}
\hline d& m_{\hat R}(p)  \\
\hline 1 & 0\\
\hline 2 & \pm\,p \\
\hline 3 & 0 ~,~ \pm\,3p \\
\hline 4 & 0 ~,~ \pm\,2p ~,~ \pm\,6p \\
\hline 5 & 0 ~,~ \pm\,2p ~,~ \pm\,5p ~,~\pm\,10p
 \\
\hline
\end{array} \ .
\label{allowedmtable}\end{eqnarray}

Compared to the Gromov-Witten and Gopakumar-Vafa invariants, we see
that the counting of torus invariant bound states of D-branes in
Donaldson-Thomas theory is provided by very simple formulas. The
combinatorics, as well as the non-trivial framing of the geometry,
come into play only in the enumeration of the total number of bound states
at a given degree $d$ and in the determination of the contributing
configurations of D0 branes using the formula (\ref{mhatR}). The total number
of bound states again exhibits an exponential behaviour in its growth
at fixed large $d$. We can
obtain a formal relationship between all of the invariants, as well as
giving an implicit definition of Gopakumar-Vafa invariants in terms of
the moduli spaces of ideal sheaves, by appealing to the Gopakumar-Vafa
expansion (\ref{GVexpansion}) of the topological string free
energy. By equating the exponentiation of this
expansion with (\ref{DTpartfn}) we may express the Donaldson-Thomas
partition function in terms of Gopakumar-Vafa invariants
as~\cite{Katz:2004js}
\beq \label{ZDTexpandinGV}
\mathcal{Z}^{\rm DT}= \frac1{\eta(\tau)}~\prod_{d=2}^\infty~
\prod_{r=2}^{\infty}~
\prod_{k=0}^{2r-2}\, \left( 1 -q^{r-1-k}\,Q^d
\right)^{(-1)^{k+r}\,{{2r-2}\choose k}\, n_d^r(X_p) } \ ,
\eeq
where we have used the fact that the genus zero BPS
invariants $n_d^0(X_p)$ all vanish along with
$n_1^r(X_p)=\delta_{r,1}$ and $n_d^1(X_p)=1$ for all $d$. As expected,
the genus one contribution represented by the Dedekind function in
(\ref{ZDTexpandinGV}) factors out separately in the partition
function. Using the Donaldson-Thomas invariants computed above,
eq.~(\ref{ZDTexpandinGV}) gives a recursive procedure for computing 
the involved Gopakumar-Vafa invariants of Sect.~\ref{GVInt} above.
This has the virtue of formally making various aspects of the BPS
invariants more natural within the context of Donaldson-Thomas
theory.

\section{Chern-Simons Theory on Torus Bundles\label{ChernSimons}}

In this section we will take a somewhat different approach to
analysing topological string dynamics on the local elliptic
threefold $X_p$. We will explore the relation between the
two-dimensional Yang-Mills theory, describing the four-dimensional
instanton counting explained in Sect.~\ref{DPartFn}, and Chern-Simons
gauge theory. While a complete description of the ${\cal N}=4$ gauge
theory in four dimensions is presently out of reach, the
three-dimensional gauge theory can be solved exactly and used to
elucidate the connection between the two-dimensional gauge theory and
its higher dimensional origins. This relation has been
suggested in~\cite{Aganagic:2004js}. The basic idea is
that the four-dimensional physics is captured, in this context, by a
three-dimensional Chern-Simons theory that lives on the boundary
of the non-compact four-cycle $C_4={\cal O}_{\torus^2}(-p)\to\torus^2$
in $X_p$.

To understand the geometry of this three-manifold, it is convenient to
introduce a hermitean metric on the fibres of the holomorphic line
bundle. Then the four-manifold $C_4$ can be identified with the total
space of the unit disk bundle $\disk(\mathcal{L}^{\otimes -
  p})\to\torus^2$, where $\mathcal{L}\to\torus^2$ is the canonical
hermitean (monopole) line bundle over the torus. The bundle
$\disk(\mathcal{L}^{\otimes -p})$ contains those fibre vectors of
$\mathcal{L}^{\otimes -p}$ with norm $\leq1$, and its boundary is  the
unit circle bundle
\beq
\torus^3_p:=\partial C_4 = \sphere\big(\mathcal{L}^{\otimes -p}\big)
\label{toruspdef}\eeq
over $\torus^2$.  This gives the three-manifold $\partial C_4$ the
structure of a Seifert manifold (the total space of a circle bundle
over a Riemann surface). For instance, for the trivial fibration one
has $\disk({\cal L}^0) \cong \mathbb{T}^2 \times \mathbb{C} \cong
\mathbb{T}^2 \times \sphere^1 \times \mathbb{R}_{\geq0} $ and  $
\partial\disk(\mathcal{L}^0) =:\sphere (\mathcal{L}^0) \cong
\torus^2\times\sphere^1=\mathbb{T}^3$.

The emergence of the $U(N)$ Chern-Simons theory can be understood
heuristically as follows. An instanton excitation on the four-manifold
$C_4$ is described by the topological density $\Tr(F\wedge
F)$. Locally, on each contractible $\complex^2$ patch of $C_4$, this
closed 4-form is exact and can be written as the exterior derivative of the
Chern-Simons 3-form $\Tr(A\wedge \mathrm{d} A + \frac{2}{3}\, A \wedge A
\wedge A)$. This formally means that the four-dimensional instanton
counting problem should reduce to the evaluation of the Chern-Simons
partition function
\begin{equation}
\mathcal{Z}_{U(N)}^{\rm CS}\big(\torus_p^3 \,,\, k\big) = \int \DD A ~\exp
\left[\frac{\ii k}{4 \pi }\, \int_{\torus_p^3}\, \Tr\left(A \wedge
    \mathrm{d} A + \mbox{$\frac{2}{3}$}\, A \wedge A \wedge A \right)
\right]
\label{CSpartfn}\end{equation}
with $k\in\nat_0$.\footnote{Negative levels $k\in\zed$ can be obtained by
  reversing the orientation of the three-manifold, a transformation
  under which the Chern-Simons action in (\ref{CSpartfn}) is odd.}
This rather crude correspondence carries with it certain subtleties
which we will explain in detail later on. Note that one should set the
$\theta$-angle in (\ref{4dobservables}) to zero in order to work out
the equivalence with Chern-Simons gauge theory, because there is no
$\theta$-angle in three dimensions.

According to~\cite{Beasley:2005vf} (see also~\cite{Blau:2006gh}), the
partition function (\ref{CSpartfn}) naturally localizes onto a
two-dimensional gauge theory on the base torus of the Seifert
fibration $\torus_p^3\to\torus^2$. The purpose of this section is to
give an explicit realization of this localization in terms of the
previously studied Yang-Mills gauge theory on $\torus^2$, and to match
it to the problem of BPS state counting described in
Sect.~\ref{SDuality}. This relates the closed topological string
theory on $X_p$ to a dual open topological string theory on the
cotangent bundle $T^*\torus_p^3$.

\subsection{Chern-Simons Gauge Theory on a Mapping Torus}

We will now explicitly compute the Chern-Simons partition function
(\ref{CSpartfn}), as a first step in connecting four-dimensional and
two-dimensional physics with gauge theory in three dimensions. Later
on this connection will be used to analyse the D-brane partition
function of Sect.~\ref{DPartFn}. We begin by considering Chern-Simons
theory with gauge group $SU(N)$ on the trivial Seifert fibration over
$\torus^2$ which is the three-torus
$\torus^3_{p=0}\cong\torus^3$. Generally, the Chern-Simons path
integral on any three-manifold of the form $\Sigma \times\sphere^1$,
with $\Sigma$ a compact Riemann surface, can be readily computed in the
hamiltonian formalism~\cite{Witten:1988hf}. The idea is to construct
the finite-dimensional Hilbert space $\mathcal{H}^k(\Sigma)$ of the
topological gauge theory on each time slice $\Sigma$ and then study
the propagation of physical states
in the time direction $\mathbb{I}=[0,1]$ with the vanishing Chern-Simons
hamiltonian. The circle $\sphere^1$ is built by identifying the
endpoints of the unit interval $\mathbb{I}$, i.e. by identifying the
initial and final states. This is implemented by taking a trace over
the quantum Hilbert space, giving finally the partition function
\begin{equation}
\mathcal{Z}^{\rm CS}\big(\Sigma \times\sphere^1\,,\,k\big) =
\Tr_{\mathcal{H}^k (\Sigma) }(\id) = \dim \mathcal{H}^k (\Sigma) \ .
\label{ZCSdimHgen}\end{equation}
In particular, the Hilbert space $\mathcal{H}^{k}_{SU(N)}(\torus^2)$
is the space of integrable representations of $SU(N)$ at level
$k$~\cite{Witten:1988hf,Periwal:1993yu}. A complete orthonormal basis
for this space is thus built through $SU(N)$ representations $\hat{R}$
with at most $k$ boxes in their associated Young tableaux. We will
denote by $ | \hat{R} \rangle$ a generic vector of this basis. This
Hilbert space has dimension~\cite{Periwal:1993yu}
\begin{equation}
\dim \mathcal{H}^{k}_{SU(N)} \left( \torus^2 \right) = {{N-1+k}\choose
k} \ .
\end{equation}

Next we consider $SU(N)$ Chern-Simons theory on the
Seifert manifold $\torus^3_p$ for $p>0$. For this, it is convenient to
have an alternative description of the three-manifold in question
which exploits the uniqueness theorem for Seifert fibrations. The
idea is that in the time-slicing construction above we have turned the
(trivial) $\sphere^1$-bundle over $\torus^2$ ``on its side'' and
effectively treated it as a $\torus^2$ fibration over
$\sphere^1$. This can also be accomplished for the generic circle
bundles $\torus_p^3\to\torus^2$ with $p>0$ by taking a closer look at its
underlying Seifert structure. Excise a disk $\disk^2$ from the base
$\torus^2$ of the fibration and denote the resulting surface by
$E$. Then the circle bundle $M_p:=E\times\sphere^1$ has boundary
$\partial M_p=\partial E\times\sphere^1\cong\sphere^1\times\sphere^1$
which is another two-torus. Each non-contractible circle in this torus
is isotopic to a unique ``linear'' $\sphere^1$ which lifts to a line
$y=-p\,x$ of slope $-p$ in the universal cover
$\real^2\to\sphere^1\times\sphere^1$. In $M_p$, attach a solid torus
$\disk^2\times\sphere^1$ along its $\torus^2$ boundary to the torus
$\partial M_p$ by the diffeomorphism which takes a meridian circle
$\partial\disk^2\times\{y\}$ of $\partial\disk^2\times\sphere^1$  to a
circle of slope $-p$ in $\partial M_p$. The slope $-p$ uniquely
specifies the resulting three-manifold. The $\sphere^1$-fibring of $M_p$
extends to a Seifert fibring of this three-manifold via the standard
circle fibration of each attached solid torus. By the uniqueness
of Seifert fibrations, this three-manifold coincides with the original
$\torus_p^3\cong\sphere({\cal L}^{\otimes-p})$.

We will match this Seifert structure to that of a particular torus
bundle. Generally, a mapping torus is defined for any complex curve
$\Sigma$ and a diffeomorphism $\beta : \Sigma \rightarrow \Sigma$ as
\begin{equation}
\Sigma\times_\beta\sphere^1:=
\Sigma \times [0,1] ~/~ \big( x \,,\, 0 \big) \sim 
\big( \beta(x) \,,\, 1 \big) \ .
\label{Sigmabeta}\end{equation}
In the case where $\Sigma$ is a two-torus, every diffeomorphism
$\beta:\torus^2\to\torus^2$ is isotopic to a uniquely determined
large diffeomorphism ${\sf K}\in SL(2,\zed)$ which is
essentially the map $\beta_*$ on
$H_1(\torus^2,\zed)\cong\zed^2$ induced by $\beta$. The
modular group $SL(2, \zed)$ is generated by the elements
\begin{equation}\label{SLgenerators}
{\sf S} = \begin{pmatrix} 0~ & -1 \\ 1~ & 0  \end{pmatrix}
\ , \qquad {\sf T} = \begin{pmatrix} 1~ &~ 1 \\ 0~ &~ 1
 \end{pmatrix}
\end{equation}
which satisfy the relations ${\sf S}^2 = ({\sf S}\,{\sf T})^3 =
\id$. The torus bundle $\torus^2\times_{\sf K}\sphere^1\to\sphere^1$
is {\it irreducible} (i.e. every two-sphere in $\torus^2\times_{\sf
  K}\sphere^1$ is contractible), exactly like our Seifert manifold
$\torus_p^3$, since its universal cover is $\real^3$.

Consider now a Seifert fibring of $\torus^2\times_{\sf K}\sphere^1$
determined by the modular transformation matrix ${\sf K}={\sf
  K}^{(p)}$, where
\beq
{\sf K}^{(p)}:={\sf S}\,{\sf T}^{-p}\,{\sf S}=-\begin{pmatrix}
1~&~0\\ p~&~1\end{pmatrix} \ .
\label{sfKpdef}\eeq
Then the generic torus fibre of the $\torus^2$-bundle $\torus^2\times_{{\sf
    K}^{(p)}}\sphere^1$ is isotopic to a vertical surface (a union of
$\sphere^1$ fibres), and so its complement in $\torus^2\times_{{\sf
    K}^{(p)}}\sphere^1$ is $\torus^2\times\interval$ which has only
the trivial product Seifert fibring up to isomorphism. Regarding
$\torus^2\times\interval$ as a trivial circle bundle, the linear
transformation (\ref{sfKpdef}) preserves the slope in annuli whose
boundaries have the same slope in both ends $\torus^2\times\{0\}$ and
$\torus^2\times\{1\}$. This is the same Seifert fibring as for our
original three-manifold and so there is an isomorphism
\beq
\torus^3_p\cong\torus^2\times_{{\sf K}^{(p)}}\sphere^1 \ .
\label{Tp3torusbun}\eeq
It is interesting to use this point of view to consider the
distinction from the trivial fibration with $p=0$. In that case ${\sf
  K}^{(0)}=\id$ (in $PSL(2,\zed)$) and the generic torus fibre is a
{\it horizontal} surface (transverse to the $\sphere^1$ fibres),
because $\torus_0^3$ is exactly the same three-manifold
$\torus^2\times\sphere^1$ in both of its fibre bundle descriptions
which are given by the canonical projections onto the first and second
factors.

We can apply this geometric construction to the evaluation of the
Chern-Simons partition function on $\torus^3_p$. Generally, the
identifications used to construct the mapping torus (\ref{Sigmabeta})
imply that the initial and final states are identified by the
diffeomorphism $\beta$ in the quantum gauge theory. This modifies the
result (\ref{ZCSdimHgen}) of the functional integration to
\begin{equation}
\mathcal{Z}^{\rm CS}\big(\Sigma\times_{\beta}\sphere^1\,,\,k\big) =
\Tr_{\mathcal{H}^k (\Sigma)}\big(O_{\beta}\big)
\end{equation}
where $O_{\beta}$ is the operator acting on $\mathcal{H}^k (\Sigma)$,
induced by the gluing morphism $\beta$, which arises from the
representation of the mapping class group of $\Sigma$ on the physical
Hilbert space. Applying this formula to the modular
transformation (\ref{sfKpdef}) we find
\bea
\mathcal{Z}_{SU(N)}^{\rm CS} \big(\torus_p^3 \,,\, k\big) =
\Tr_{\mathcal{H}^{k}_{SU(N)}(\torus^2)}\big(O_{{\sf K}^{(p)}}\big)&=&
\Tr_{\mathcal{H}^{k}_{SU(N)}(\torus^2)}\big(O_{{\sf T}^{-p}}\big)
\nonumber \\[4pt] &=&
\sum_{\hat{R}\in\mathcal{H}^{k}_{SU(N)}(\torus^2)}\,
\langle \hat{R} \, | \, O_{{\sf T}^{-p}} \, | \, \hat{R} \rangle \ .
\label{ZCShatRsum}\eea
In this formula we see the origin of the mass deformation
$p\,\Tr(\Phi^2)$ introduced in (\ref{qYMaction}), which was the only
term in the partition function accounting for the
non-triviality of the fibration. In the Chern-Simons formulation the
insertion of this operator corresponds to the presence of the gluing
operator ${\sf T}^{-p}$. The three-dimensional gauge theory
thereby naturally explains the occurence of the quadratic
superpotential which was argued in~\cite{Vafa:2004qa} to arise in the
four-dimensional topologically twisted gauge theory on $C_4$ as a
result of integrating out a holomorphic section of the line bundle
${\cal O}_{\torus^2}(p)$ (having $p$ zeroes). This
fact has also been noticed using surgery techniques
in~\cite{deHaro1,Blau:2006gh,Aganagic:2002wv,Jeffrey:1992tk,Marino:2002fk},
and it agrees with the topological string calculation of
Sect.~\ref{TopStringAmpl} wherein the amplitudes on non-trivial
fibrations were constructed by insertions of the framing operator
$q^{p\,C_2(\hat R)/2}$ (which as we show below corresponds to ${\sf
  T}^{-p}$) in the path integral for the trivial bundle.

The lift $O_{{\sf T}}$ to $\mathcal{H}^{k}_{SU(N)}(
\torus^2)$ of the modular transformation $\sf T$ in
(\ref{SLgenerators}) is a diagonal operator in the Verlinde basis of
integrable $SU(N)$ representations which can be expressed
as~\cite{Jeffrey:1992tk}
\begin{equation} \label{Tintrep}
\langle \hat Q\, | \, O_{{\sf T}} \, | \,\hat R\rangle = 
\delta_{\hat Q,\hat R} ~\exp \left[
\frac{ \pi\ii}{k+N} \, \Tr\big({\mbf n}(\hat R) + \mbf\rho\big)^2 - 
\frac{\pi\ii }{N} \, \Tr\big(\mbf\rho^2\big)\right] \ ,
\end{equation}
where $\mbf\rho$ is the Weyl vector (the half sum of the positive
roots of $SU(N)$) and
\begin{equation}
\Tr\big(\mbf n(\hat R) + \mbf\rho\big)^2= 
C_2 \big(\hat R\big) + \Tr\big( \mbf\rho^2\big)
\end{equation}
is given in terms of the natural inner product on the Lie algebra
$\mathfrak{su}(N)$. We can read off the norm of the Weyl vector (the
Freudenthal-de~Vriess formula) by expressing (\ref{Tintrep}) in terms
of conformal invariants of the $SU(N)$ Wess-Zumino-Witten model at
level $k$ as~\cite{Witten:1988hf,Jeffrey:1992tk}
\begin{equation}
\langle \hat Q\, | \, O_{{\sf T}} \, | \,\hat R\rangle = 
\delta_{\hat Q,\hat R} ~\e^{
 2 \pi \ii (\Delta_{\hat R} - {c}/{24} )} \ ,
\end{equation}
where
\beq
\Delta_{\hat R}=\frac{C_2 (\hat R)}{2 \left( k+N \right)} \ , \qquad
c= \frac{k\, \dim SU(N)}{k+N} = \frac{k \, \left(N^2 -1\right)}{k+N}
\eeq
are respectively the conformal weight of a primary field in the
representation $\hat R$ and the central charge of the conformal field
theory at level $k$. Since the contribution from the
central charge is independent of the $SU(N)$ representation, it can be
regarded as an overall normalization and the partition function
(\ref{ZCShatRsum}) of $SU(N)$ Chern-Simons theory on the mapping torus
$\torus_p^3$ finally reads
\begin{equation} \label{CSpartfun1}
\mathcal{Z}_{SU(N)}^{\rm CS}\big(\torus_p^3 \,,\, k\big) = 
\sum_{\hat{R}\in\mathcal{H}^{k}_{SU(N)}(\torus^2)}\, \exp \left[-
\frac{\pi \ii p}{k+N} \, C_2 (\hat R) + \frac{\pi \ii k\, p }{N+k}\,
\frac{N \, \left(N^2 -1\right)}{12}\right] \ .
\end{equation}

It remains to extend these calculations to gauge group
$U(N)$, which is the one relevant for the counting of black hole
microstates through the D-brane partition function. To start, as
obtained in~\cite{Douglas:1994ex} using the free fermion
representation of the Chern-Simons gauge theory, we have
\begin{eqnarray} 
\dim \mathcal{H}^{k}_{U(N)}\left(\torus^2\right) &=& {{N+k}\choose
k}= {{N-1+k}\choose k}\, \frac{N \, (N+k)}{N^2} \nonumber \\[4pt]
&=& \frac{1}{N^2} \, \left[ \dim
\mathcal{H}^{k}_{SU(N)}\left(\torus^2\right) \right]
 \, \left[ \dim
   \mathcal{H}^{N\,(k+N)}_{U(1)}\left(\torus^2\right)\right] \ .
\label{dimH}\end{eqnarray}
This dimension formula reflects the decomposition $U(N) = SU(N)\times U(1) /
\zed_N$, where $N \, (k+N)$ is the effective $U(1)$ Chern-Simons
coupling. As before, for $p=0$ the partition function computes
the dimension of the Hilbert space of physical states. The corresponding
free energy is the logarithm of (\ref{dimH}), as computed
in~\cite{Periwal:1993yu}, and it has a large $N$ expansion beginning
at order $N$. This fact was stressed in~\cite{Douglas:1994ex} as
evidence that Chern-Simons theory on $\torus^3$ does {\it not} admit a
closed string theory interpretation. Indeed, we will later on show
that a closed string theory interpretation of Chern-Simons
gauge theory on a torus bundle over the circle, if it exists,
is provided by two-dimensional Yang-Mills theory interpreted as a
topological string theory as in the previous sections. In the case of
the three-torus with $p=0$ all higher genus geometric invariants of
the corresponding local Calabi-Yau background $X_0$ vanish and there
is no closed string genus expansion, in complete agreement with the
observations of~\cite{Douglas:1994ex}.

A complete basis for the irreducible, integrable representations of
the unitary group $U(N)$ at level $k$ is given by Young tableaux
$R$ whose rows are labelled by a set of integers $\mbf n(R)=\{
n_i(R) \}$ obeying the constraints
\begin{equation} \label{UNconfig}
- \mbox{$\frac{k+N-1}{2}$} \le n_N(R) < n_{N-1}(R) < \cdots
<n_1(R) \le \mbox{$\frac{k+N-1}{2}$} \ .
\end{equation}
As was done in Sect.~\ref{LargeNExp}, we will rewrite the $U(N)$
representations $R$ in terms of $SU(N)$ representations $\hat R$
according to the decomposition $U(N) = SU(N)\times U(1) /
\zed_N$. First we perform the change of weight variables
\begin{equation}
\hat{n}_i(\hat R) = n_i(R) - n_N(R) + i -N \ , \quad i = 1,\dots,N-1
\label{UNweightchange}\end{equation}
to get weights which describe $SU(N)$ representations $\hat R$ at
level $k$, i.e. which obey the constraints
\begin{equation}
0 \le \hat{n}_{N-1}(\hat R) \le \hat{n}_{N-2}(\hat R) \le \dots \le
\hat{n}_1(\hat R) \le k \ .
\end{equation}
As in Sect.~\ref{LargeNExp}, the $U(1)$ charge $m$ of a representation
$R$ has the general form
\beq
m= \sum_{i=1}^{N-1}\, \hat n_i(\hat R) + N\,  r \ .
\eeq
The range of the integer $r$ can be determined from (\ref{dimH}) by
the dimension constraint
\begin{eqnarray}
{{N+k}\choose k}= \sum_{- \frac{k+N-1}{2} \le n_N <
\cdots < n_1 \le \frac{N+k-1}{2} }~1 = \sum_m ~\sum_{0
\leq \hat n_{N-1} \leq \cdots \leq \hat n_1 \leq k } ~1 \ ,
\end{eqnarray}
which fixes $r= n_N(R)-\frac{k }{2}$ with $-k -\frac{N-1}{2}\leq r
\leq -\frac{N-1}{2}$ (here we tacitly assume that $k$ is even
and $N$ is odd).

Finally, using eqs.~(\ref{C2RC2hatR}) and~(\ref{kappahatRdef}) we can
compute explicitly the $U(N)$ Chern-Simons partition function in the
form (\ref{CSpartfun1}) as
\begin{equation}
\mathcal{Z}_{U(N)}^{\rm CS}\big(\torus_p^3 \,,\, k\big) =
\sum_{(\hat R,m)\in\mathcal{H}^{k}_{U(N)}(\torus^2)}\, \exp
\left[ -\frac{\pi \ii p}{k+N} \, \left( C_2 (\hat R) +
\frac{m^2}{N} \right) \right]
\label{ZUNCShatR}\end{equation}
up to an overall normalization factor. The argument of the exponential
in (\ref{ZUNCShatR}) can be written as
\bea
C_2 (\hat{R}) + \frac{m^2}{N} &=& \sum_{i=1}^{N-1}\, \hat{n}_i(\hat R) \,
\big( \hat{n}_i(\hat R) + N+1 - 2i \big) - \frac{1}{N}\, \biggl(\,
\sum_{i=1}^{N-1}\, \hat{n}_i(\hat R) \biggr)^2 \nonumber\\ && +\,
\frac{1}{N}\,\biggl(\,
\sum_{i=1}^{N-1}\, \hat{n}_i(\hat R) + r \, N \biggr)^2 \ .
\eea
Returning to the original $U(N)$ weight variables using
eq.~(\ref{UNweightchange}) we then have
\begin{equation}
C_2( \hat{R}) + \frac{m^2}{N} = \sum_{i=1}^N\, \left(
n_i(R) - \frac{N+k-1}{2} \right)^2 + \frac{N}{12}\, \left( N^2
- 1 \right) \ ,
\end{equation}
and so one finds that up to an irrelevant overall
normalization factor the amplitude (\ref{ZUNCShatR}) reads
\begin{eqnarray}
\mathcal{Z}_{U(N)}^{\rm CS}\big(\torus_p^3 \,,\, k\big) &=&
\sum_{- \frac{k+N-1}{2} \le
n_N < \cdots < n_1 \le \frac{N+k-1}{2} }\, \exp \left[- \frac{\pi \ii
p}{k + N}\, \sum_{i=1}^N\, \left( n_i - \frac{N+k-1}{2} \right)^2
\right]\nonumber \\[4pt]
 &=& \frac{1}{N!}\, \sum_{\stackrel{\scriptstyle n_1,\dots,n_N= 0}
{\scriptstyle n_i \neq n_j }}^{N + k -1}\, \exp \left[-
 \frac{\pi \ii p}{k+N}\, \sum_{i=1}^N\, n_i^2
 \right] \ .
\label{ZCSwithcasimir}\end{eqnarray}

\subsection{Flat Connections\label{FlatConns}}

Thus far in this section we have derived the explicit form of the
Chern-Simons partition function on the pertinent mapping torus. In
order to build the bridge to four-dimensional instanton physics we
will need to understand how the expression (\ref{ZCSwithcasimir})
arises from a localization formula, as we did for the two-dimensional
Yang-Mills theory in Sect.~\ref{YMNLF}. We will now provide a
classification of the gauge field contributions to such a localization
and use it to formally demonstrate the equivalence between
Chern-Simons theory on $\torus^3_p$ and Yang-Mills theory on
$\torus^2$. For this construction it is most convenient to return to
the original description of the three-manifold $\torus_p^3$ as a Seifert
fibration. In Sect.~\ref{CSNLF} below we will demonstrate how the
expression (\ref{ZCSwithcasimir}) explicitly fits into this
framework.

The partition function of Chern-Simons gauge theory on
$\mathbb{T}_p^3$, and in general on any Seifert manifold,
localizes onto the critical points of the Chern-Simons action
functional~\cite{Beasley:2005vf,Marino:2002fk,Rozansky:1994wv}. The
path integral (\ref{CSpartfn}) is thus given by a sum over all flat
$U(N)$ gauge connections on $\mathbb{T}_p^3$. Flat
connections on a manifold have a nice geometrical
characterization. The vanishing of the curvature implies that the
holonomy of the connection along a closed path depends only on the
homotopy class of the path. We can thus characterize the flat
connection by giving its holonomies along a basis of non-contractible
one-cycles of $\mathbb{T}_p^3$. This implies that, modulo gauge
transformations, flat $U(N)$ gauge bundles over $\mathbb{T}_p^3$ are in
one-to-one correspondence with homomorphisms of the fundamental group
$\pi_1 (\torus^3_p)$ into the gauge group $U(N)$, i.e. $N$-dimensional
unitary representations of $\pi_1 (\torus^3_p)$.

The group $\pi_1 (\torus^3_p)$ has a
presentation~\cite{Beasley:2005vf,Furuta} consisting of three
generators $a$, $b$ and $h$ subject to the relations
\beq
a\,b=h^{-p}\,b\,a \ , \qquad a\,h=h\,a \ , \qquad b\,h=h\,b \ .
\label{pi1rels}\eeq
The elements $a$ and $b$ arise from the two non-contractible one-cycles
of the two-torus which are the generators of the abelian group $\pi_1
(\mathbb{T}^2) = \mathbb{Z}^2$. The central generator $h$
characterizes the winding of flat connections along the generic
$\sphere^1$ fiber over $\torus^2$. Alternatively, we may characterize
the fundamental group $\pi_1(\torus^3_p)$ as a central extension of
$\pi_1(\torus^2)$ through the exact sequence of groups given by
\begin{equation}
1 ~\longrightarrow~ \big\langle\, h\,\big\rangle~ \longrightarrow~ \pi_1
\big(\mathbb{T}_p^3\big)~\longrightarrow~ \pi_1\big(\mathbb{T}^2\big)~
\longrightarrow~ 1 \ .
\end{equation}
In particular, $\pi_1(\torus^3_{p=-1})$ is the universal central
extension of $\pi_1(\torus^2)$.

Before proceeding to the explicit classification of representations of
the fundamental group, it is useful to first understand the geometric
meaning of the classifying integers that will arise. If
${\cal E}\to\torus_p^3$ is any $U(N)$ gauge bundle, then its first
Chern class $c_1({\cal E})$ is an element of the second cohomology group
$H^2(\torus^3_p,\zed)$. Understanding the structure of this group will
thereby produce the allowed magnetic charges of the flat gauge field
configurations on $\torus^3_p$, and also aid in detailing the
relation between these flat connections and solutions of the
Yang-Mills equations on the base $\torus^2$. For this, we use the fact
that $\torus_p^3\to\torus^2$ is a circle bundle to write the
Thom-Gysin exact sequence~\cite{Furuta}
\bea
0&~\longrightarrow~&H^1\big(\torus^2\,,\,\zed\big)~\longrightarrow~
H^1\big(\torus_p^3\,,\,\zed\big)~\longrightarrow~H^0\big(\torus^2\,,\,
\zed\big)~\stackrel{\delta}{\longrightarrow}~
H^2\big(\torus^2\,,\,\zed\big)~\longrightarrow \nonumber\\
&~\longrightarrow~&H^2\big(\torus_p^3\,,\,\zed\big)~\longrightarrow~
H^1\big(\torus^2\,,\,\zed\big)~\longrightarrow~0 \ .
\label{GysinThomseq}\eea
The maps $H^n(\torus^2,\zed)\to H^n(\torus_p^3,\zed)$ are the
pullbacks $\pi^*$ on cohomology induced by the bundle projection
$\pi:\torus_p^3\to\torus^2$. Recalling that $H^2(\torus^2,\zed)\cong
H^1(\torus^2,U(1))$ classifies hermitean line bundles over $\torus^2$,
the map $\delta$ is defined by taking the generator of
$H^0(\torus^2,\zed)\cong\zed$ to the class
$[\mathcal{L}^{\otimes-p}]=-p\,[{\cal L}]$ in
$H^2(\torus^2,\zed)=\zed[{\cal L}]$.

When $p>0$, the map $\delta$ is injective and its cokernel is torsion
of order $p$. In particular, from (\ref{GysinThomseq}) it follows that
\beq
H^2\big(\torus_p^3\,,\,\zed\big)=H^1\big(\torus^2\,,\,\zed\big)\oplus
H^2\big(\torus^2\,,\,\zed\big)/\big\langle[{\cal L}^{\otimes-p}]
\big\rangle \ ,
\label{H2H1calL}\eeq
and since $H^1(\torus^2,\zed)\cong\zed^2$ we arrive finally
at\footnote{This isomorphism can also be deduced from the presentation
  of the fundamental group $\pi_1(\torus_p^3)$ in (\ref{pi1rels}). By
  Poincar\'e duality, the cohomology group $H^2(\torus_p^3,\zed)\cong
  H_1(\torus^3_p,\zed)$ can be computed as the abelianization of
  $\pi_1(\torus_p^3)$. This formally sets $h^p=1$ in
  (\ref{pi1rels}), illustrating the fact that the torsion charges
  arise from liftings on $\torus^2$ along the $\sphere^1$ fibres of
  $\torus^3_p$. However, this construction does not exhibit the
  explicit mapping between the two-dimensional and three-dimensional
  gauge theories.}
\beq
H^2\big(\torus_p^3\,,\,\zed\big)\cong\left\{\begin{array}{ccc}
\zed^2\oplus\zed_p & \ , & \quad p>0 \ , \\
\zed^3 & \ , & \quad p=0 \ . \end{array} \right. 
\label{H2torusp3}\eeq
The torsion subgroup $\zed_p$ of the cohomology group (\ref{H2torusp3})
for $p>0$ induces $p\,N$-torsion in the relevant group
$H^1(\torus_p^3,U(N))$ classifying principal $U(N)$ bundles over $\torus_p^3$
through the map $\det:H^1(\torus_p^3,U(N))\to H^2(\torus_p^3,\zed)$
which assigns to any unitary vector bundle its corresponding
determinant line bundle. We will find below that all {\it flat} $U(N)$
gauge bundles over $\torus^3_p$ carry torsion magnetic charges $m$ in the
$\zed_{p\,N}$ subgroup of this cohomology group. Moreover, from
(\ref{H2H1calL}) it follows that all such torsion bundles over
$\torus^3_p$ are pullbacks of bundles over $\torus^2$ with ordinary
(integer) Chern classes. Below we will construct this map explicitly
by establishing a one-to-one correspondence between flat connections
on $\torus_p^3$ and instantons on $\torus^2$. For $p=0$, any flat
bundle over $\torus^3$ has vanishing Chern class and arises from a
flat bundle over $\torus^2$.

We now turn to the classification of the unitary representations of
the fundamental group $\pi_1(\torus_p^3)$. Since
$\pi_1(C_4)\cong\pi_1(\partial C_4)$, any homomorphism
\beq
\gamma\,:\,\pi_1\big(\torus_p^3\big)~\longrightarrow~ U\big(N\big)
\eeq
also determines an asymptotic flat connection corresponding to a
finite action $U(N)$ instanton on the four-cycle $C_4$. Consider first
the case $p=0$. Since $\pi_1(\torus^3)\cong\zed^3$ is an abelian
group, any $U(N)$ representation is of the form
$\gamma=\bigoplus_{i=1}^r\,\gamma_i^{\oplus N_i}$ with $\gamma_i$ a
one-dimensional irreducible representation of $\zed^3$ of multiplicity
$N_i$ in $U(N)$ with $N=\sum_{i=1}^r\,N_i$. In this case the flat
connections are simply labelled by partitions $\{N_i\}$ of the gauge
group rank $N$ with vanishing Chern class.

The situation is more interesting for $p>0$. Since $h$ is a central
generator of $\pi_1(\torus_p^3)$, $\gamma(h)$ must lie in the
centralizer subgroup $U_{\gamma(h)}$ of the element $\gamma(h)\in
U(N)$. Consider first the case where $U_{\gamma(h)}=U(N)$. Then
$\gamma(h)$ lies in the center $U(1)$ of the $U(N)$ gauge group. Taking
the determinant of both sides of the first relation in (\ref{pi1rels})
shows that $\gamma(h)^{-p\,N}=\id_N$, and hence $\gamma(h)=\e^{-2\pi\ii
  m/p\,N}~\id_N$ for some integer $m$ with $0\leq m<p\,N$. The
integer $m$ is simply the torsion charge of the flat connection
described by $\gamma$ as inferred from (\ref{H2torusp3}). The element
$\gamma(h)^{-p}$ lives in the center $\zed_N$ of the $SU(N)$ subgroup
of $U(N)$, and the algebraic relation
\beq
\gamma(a)\,\gamma(b)=\e^{2\pi\ii m/N}~\gamma(b)\,\gamma(a)
\label{Weylalgrho}\eeq
defines the Heisenberg-Weyl group for $SU(N)$. It possesses a unique
irreducible representation $\gamma_w$ of dimension $w={\rm gcd}(m,N)$,
and $\gamma=\gamma_w^{\oplus N/w}$.

In the generic case the centralizer
subgroup $U_{\gamma(h)}$ is of the form (\ref{centralsubgp}) and we can
repeat the above construction in each $U(N_i)$ block. Thus
$\gamma(h)=\bigoplus_{i=1}^r\,\e^{-2\pi\ii m_i/p\,N_i}~\id_{N_i}$ with
Chern classes $m_i$ obeying $0\leq m_i<p\,N_i$, and $\gamma(h)^{-p}$
lives in the center of the symmetry breaking pattern
$\zed_{N_1}\times\cdots\times\zed_{N_r}\subset
SU(N_1)\times\cdots\times SU(N_r)$. The decomposition of $\gamma$ into
irreducible representations is given by
\beq
\gamma=\bigoplus_{i=1}^r\,\gamma_{w_i}^{\oplus N_i/w_i}
\label{redrepWeyl}\eeq
with $w_i={\rm gcd}(m_i,N_i)$. Thus a generic flat connection
on a $U(N)$ gauge bundle over $\torus_p^3$ is uniquely characterized
by a partition $N=\sum_{i=1}^r\,N_i$ and a set of topological charges
$\mbf m=\{m_i\}$ with $0\leq m_i<p\,N_i$ for each $i=1,\dots,r$.

Except for the finite ranges of the Chern numbers $m_i$, this
classification is identical to the classification of Yang-Mills
connections on $\torus^2$ that we found in Sects.~\ref{YMNLF}
and~\ref{SDuality}. This owes to the algebraic fact that both
classification schemes are based on central extensions ($\Gamma_\real$
in two dimensions and $\pi_1(\torus_p^3)$ in three dimensions) of the
fundamental group $\pi_1(\torus^2)$ of the underlying
two-torus. Heuristically, a flat connection in three dimensions can be
lifted from a (not necessarily flat) Yang-Mills connection in two
dimensions by assigning to it a fixed holonomy along the fibre
generator $h$. We can make this correspondence more precise as
follows~\cite{Furuta}.

Let $A^{\phantom{(}}_0=A_0^{(1)}$ be the standard one-monopole connection on the
canonical line bundle $\mathcal{L}\to\mathbb{T}^2$. Its curvature
$F_{A_0}=\dd A_0=2\pi\,\omega_{\torus^2}$ is proportional to the unit volume
form on $\torus^2$. The monopole connection $A_0^{(-p)}=-p\,A_0$ of
magnetic charge $-p$ on ${\mathcal L}^{\otimes -p}$ has curvature
$F_{A_0^{(-p)}}=-2\pi\,p\,\omega_{\torus^2}$. Let $\tilde m$ be an integer
with $0 \le\tilde m < N$. Regarding the monopole connection as a fixed
fiducial background gauge field, a constant curvature instanton on
$\torus^2$ corresponds to a principal $U(N)$ bundle ${\cal
  P}_N\to\torus^2$ with first Chern class $\tilde m$ and a Yang-Mills
connection $A$ of curvature $F^{\phantom{(}}_A= \frac{\tilde
  m}{N}\,F_{A_0^{(-p)}}~\id_N=-\frac{2\pi\,m}N\,\omega_{\torus^2}~\id_N$ where
$m:=p\,\tilde m$. Such two-dimensional $U(N)$ gauge connections are in
a one-to-one correspondence with flat $SU(N)$ gauge connections
$\underline{A}$ on $\torus^3_p$ with fixed $p\,N$-torsion holonomy
\beq
\exp\biggl(\ii\oint_{\sphere^1}\,\underline{A}\,\biggr)=
\e^{2 \pi \ii {m}/p\,{N}}~\id_N
\label{torholonomy}\eeq
in the center $\zed_N$ of $SU(N)$ along the $\sphere^1$ fibres over
$\torus^2$.

The three-dimensional flat connection can be explicitly constructed in
terms of the two-dimensional constant curvature connection in the
following way. The pull-back of the bundle projection $\pi :
\torus_p^3\rightarrow \mathbb{T}^2 $ naturally lifts the
two-dimensional data $({\cal P}_N,A)$ to three-dimensional data
$(\,\underline{\cal P}_{\,N}\,,\,\underline{A}\,)$. Recall from
eqs.~(\ref{GysinThomseq}) and~(\ref{H2H1calL}) that the pull-back
$\pi^*({\cal L})$ of the line bundle ${\cal L}\to\torus^2$ is
$p$-torsion. It follows that $\pi^*(\mathcal{L}^{\otimes - p
})\cong\torus_p^3 \times \mathbb{C} $ is the trivial line bundle
over $\torus_p^3$ with connection $\lambda = \frac{1}{N}\,
\pi^*\big(A^{(-p)}_0\big)$ (to which we have the freedom to add any
trivial connection). We can now construct the bundle $\underline{\cal
  P}_{\,N} = \pi^*({\cal P}_N)\otimes\pi^*(\mathcal{L}^{\otimes-m})$
whose curvature vanishes by construction. Moreover, the structure
group of $\underline{\cal P}_{\,N}$ is $SU(N)$, because the
determinant line bundle $\det\underline{\cal P}_{\,N}= \det
\pi^*({\cal P}_N) \otimes \pi^*(\mathcal{L}^{\otimes-m\,N})$ is
endowed with a vanishing connection and the holonomy along the fibre
is in the center of $SU(N)$ by construction.

To better understand this latter point, let $\kappa$ be a connection
on the principal $U(1)$ bundle $\torus_p^3\to\torus^2$ whose curvature
two-form is the pull-back of the Euler form given by 
\beq
\dd\kappa=-p\,\pi^*(\omega_{\torus^2}) \ .
\label{dkappaomega}\eeq
The one-form $\kappa$ defines a contact structure on the Seifert
manifold $\torus_p^3$~\cite{Beasley:2005vf,Blau:2006gh}. Then the
pull-back of the constant curvature $F_A$ on $\torus^2$ is given by
\beq
\pi^*(F_A)=-\mbox{$\frac{2\pi\,m}N$}\,\pi^*(\omega_{\torus^2})~\id_N=
\mbox{$\frac{2\pi\,m}{p\,N}$}~\dd\kappa~\id_N \ ,
\label{piFA}\eeq
and thus the pull-back of the irreducible instanton connection may be
taken to be
\beq
\underline{A}=\pi^*(A)=\mbox{$\frac{2\pi\,m}{p\,N}$}\,\kappa~\id_N \ .
\label{pullbackA}\eeq
Since the integral of $\kappa$ along any $\sphere^1$ fibre over
$\torus^2$ is given by~\cite{Beasley:2005vf,Blau:2006gh}
$\oint_{\sphere^1}\,\kappa=1$, eq.~(\ref{torholonomy}) follows. We can
also use the contact structure to compute the value of the
Chern-Simons action in (\ref{CSpartfn}) for the given flat
connection. Using the one-loop quantum shift in the Chern-Simons level
$k\to k+N$ by the dual Coxeter number $\breve{c}_{\mathfrak{su}(N)}=N$
of the $SU(N)$ gauge group~\cite{Beasley:2005vf,Blau:2006gh}, along
with the fact that the connection (\ref{pullbackA}) is irreducible,
one finds
\bea
S_{U(N)}^{\rm CS}(\,\underline{A}\,)&=&\frac{\ii(k+N)}{4\pi}\,
\int_{\torus_p^3}\,\Tr(\,\underline{A}\wedge\dd\underline{A}\,)
\nonumber\\[4pt] &=&\frac{\ii(k+N)}{4\pi}\,\left(\frac{2\pi\,m}{p\,N}
\right)^2\,N\,\int_{\torus_p^3}\,\kappa\wedge\dd\kappa\nonumber\\[4pt]
&=&-\frac{\pi\ii(k+N)\,m^2}{p\,N}\,\int_{\torus^2}\,\omega_{\torus^2}
=-\frac{\pi\ii(k+N)\,m^2}{p\,N} \ .
\label{SUNCSflat}\eea

This construction easily generalizes to generic reducible instantons
on $\torus^2$ with centralizer subgroups (\ref{centralsubgp}). In this
background the two-dimensional data $({\cal P}_N,A)$ decompose
according to (\ref{riduco}). The argument outlined above can be
applied to each sub-bundle ${\cal P}_{N_i}$ with its constant
curvature Yang-Mills connection $A_i$ to lift the
two-dimensional reducible connection to a Chern-Simons critical
point. The only essential difference is that now the holonomy along
the circle fiber is no longer generally an element of the center of $SU(N)$
but respects the gauge symmetry breaking pattern, i.e. the holonomy is
given by a block diagonal matrix whose $i$-th entry is of the form
$\e^{2\pi \ii {m_i}/{p\,N_i}}~\id_{N_i}$ where $N_i$ is the rank of
the corresponding subgroup and $m_i$ is the topological charge with $0
\le m_i < p\,N_i$. The value of the Chern-Simons action on a reducible
flat connection is a sum of contributions of the form
(\ref{SUNCSflat}) for each irreducible component.

\subsection{The Nonabelian Localization Formula\label{CSNLF}}

To compare the Chern-Simons partition function (\ref{ZCSwithcasimir})
with the nonabelian localization onto flat connections on
$\torus_p^3$, and hence with the partition function of two-dimensional
Yang-Mills theory, we have to perform a modular transformation. An
efficient way to do this is to resort to the integral representation
introduced in~\cite{Griguolo:2004jp}. This amounts to inserting the
contour integral representation for the Kronecker delta-function given
by
\begin{equation}
\delta_{m,n} = \oint\, \frac{\mathrm{d} z}{2 \pi
\ii z}~ z^{m-n} \ ,
\end{equation}
where $m,n \in \zed$ and the line integral goes around the origin
$z=0$ of the complex plane in the counterclockwise direction. The
effect of this insertion is to trade the constraint on the sums in
(\ref{ZCSwithcasimir}) for an extra integration, and one finds
\begin{eqnarray}
\mathcal{Z}^{\rm CS}_{U(N)}\big(\torus_p^3 \,,\, k\big) &=& \frac{1}{N!}\,
\oint\,\frac{\mathrm{d} z}{2 \pi \ii z^{N+1}}~
\prod_{n=0}^{N+k-1} \, \left( 1 + z ~\e^{-\frac{\pi \ii
p}{k + N} \,n^2}  \right) \nonumber \\[4pt]
&=& \frac{1}{N!}\, \oint\,
\frac{\mathrm{d} z}{2 \pi \ii z^{N+1}}~\exp \left[ -
\sum_{m=1}^{\infty}\, \frac{\left( -z \right)^m}{m}~
\sum_{n=0}^{N+k-1} \,  \e^{-\frac{\pi \ii p \, m}{k + N} \, n^2}\right] \ .
\end{eqnarray}
By using the Gauss sum reciprocity formula~\cite{Jeffrey:1992tk}
\begin{equation}\label{gaussum}
\frac1{\sqrt q}\,\sum_{n=0}^{q-1}\,\e^{-\pi\ii
  n^2\,p/q}=\frac{\e^{-\pi\ii/4}}{\sqrt p}\,\sum_{n=0}^{p-1}\,
\e^{\pi\ii n^2\,q/p} \ ,
\end{equation}
we can write
\begin{equation}\label{int2}
\mathcal{Z}_{U(N)}^{\rm CS}\big(\torus_p^3 \,,\, k\big) = \frac{1}{N!}\,
\oint\,\frac{\mathrm{d} z}{2 \pi \ii z^{N+ 1}}~\exp 
\left[ - \sqrt{\frac{N + k}{\ii p}}~\sum_{m=1}^{\infty}\, \frac{\left( -z
\right)^m}{m^{{3}/{2}}} ~ \sum_{n=0}^{p \, m - 1} \,  
\e^{\pi \ii n^2 \,\frac{k + N}{p \, m} }  \right] \ .
\end{equation}

Let us now compare this expression with the partition function of
two-dimensional Yang-Mills theory on the torus given in
eq.~(\ref{instantonrep}). It can be
written in the same form~\cite{Griguolo:2004jp}
\beq
\label{int} \mathcal{Z}^{\rm YM}  = \frac{1}{N!}\,
\oint\,\frac{\mathrm{d} z}{2 \pi \ii z^{N+ 1}}~
\exp \left[ -\sqrt{  \frac{2 \pi}{g_s \,p}}~
\sum_{m=1}^{\infty}\, \frac{\left( -z \right)^m}{m^{{3}/{2}}}
~\sum_{ s =-\infty}^{\infty} \,(-1)^{(N-1)\,s}~
\e^{ - \frac{2 \pi^2}{g_s \, p}\, \frac{ s
 ^2}{m}}\right] \ ,
\eeq
provided that the coupling constants of the two gauge
theories are identified as\footnote{The two-dimensional and
  three-dimensional gauge theories are also compared
  in~\cite{Blau:2006gh} from the point of view of their
  representations as the two-dimensional BF-theory with action
  (\ref{qYMaction}), where they are shown to agree under the
  identification (\ref{gskrel}) for any local curve $\Sigma$.}
\begin{equation}
g_s =\frac{2 \pi \ii}{k + N} \ .
\label{gskrel}\end{equation}
Observe that while in the partition function (\ref{int}) the
degree $p$ appears only in the combination $g_s \,p $ and can thus
be absorbed in a redefinition of the string coupling $g_s$, the
dependence of the partition function (\ref{int2}) on $p$ is
non-trivial. By looking at the reciprocity formula (\ref{gaussum}) we
see that after the modular transformation the integer $p$ appears in
the combination $p \, m -1$ that limits the allowed topological
charges, in complete agreement with the classification of flat
connections carried out in Sect.~\ref{FlatConns} above. The difference
lies in the fact that while the representations of the central
extension $\pi_1(\torus_p^3)$ depend sensitively on the the degree $p$
of the Seifert fibration, those of the universal central extension
$\Gamma_\real\cong\pi_1(\torus_{p=-1}^3)$ are independent of $p$.

To further clarify this point, let us write the dual integer $s$ in
(\ref{int}) as $s =\hat s + p \,m  \, l $ where $\hat s=0,1,\dots,p\,
m -1$ and $l \in \zed$. One can then rewrite (\ref{int}) in terms of
the Jacobi theta-functions (\ref{Jacobitheta}) as
\begin{eqnarray}
\mathcal{Z}^{\rm YM}  = \frac{1}{N!}\,\nonumber \oint\,
\frac{\mathrm{d} z}{2 \pi \ii z^{N + 1}} && \exp \bigg[
-\sqrt{  \frac{2 \pi}{g_s \,p}}~ \sum_{m=1}^{\infty}\, \frac{\left(
-z \right)^m}{m^{{3}/{2}}}~
\sum_{\hat s =0}^{p \, m - 1} \,  \e^{ - \frac{2 \pi^2}{g_s \, p}\, \frac{\hat s
 ^2}{m}}\\ &&\qquad \times \,
(-1)^{(N-1)\,\hat s}\, \vartheta_3\big(\mbox{$\frac{2\pi\ii
p\,m}{g_s} $}\,,\,
 \mbox{$\frac{2 \pi\ii \hat s }{g_s}+ \frac{p\,m\,(N-1)}{2}$}\big)
  \bigg] \ .
\label{ZYMhats}\end{eqnarray}
This formula demonstrates the connection between the Yang-Mills partition
function on the two-torus and the Chern-Simons partition function
on the mapping torus $\torus_p^3$. The rewriting of the integers $s$
in (\ref{int}), at the level of the Yang-Mills theory, is completely
arbitrary due to the trivial $p$-dependence. It is just the
identification of the Yang-Mills coupling with the (complex)
Chern-Simons coupling that sets the correct periodicity. Note that
substitution of (\ref{gskrel}) with $k\in\zed$ into (\ref{ZYMhats})
leads to a divergence coming from the
theta-functions. This means that the continuation between the two
partition functions is non-analytic, owing to the fact that for $g_s$
imaginary the model becomes periodic and the infinite sums collapse to
an infinite number of copies of the finite sums over integrable
representations of the gauge group. On the Chern-Simons side, the
invariance of the path integral (\ref{CSpartfn}) under large gauge
transformations necessitates that $q=\e^{-g_s}$ be a root of
unity. However, the Reshetikhin-Turaev approach to Chern-Simons theory
also makes sense for $q$ real and consists in replacing the
finite-dimensional Hilbert space of integrable $U(N)$ representations
by the infinite-dimensional Hilbert space of {\it all} $U(N)$
representations that naturally arises in two-dimensional Yang-Mills
theory. The very different natures of the Hilbert spaces for
$q$ a root of unity and $q\in\real$ is the source of the
non-analyticity of the continuation, which can be thought of as a map
between a rational and an irrational conformal field theory in two
dimensions. The divergences arising in this continuation were also
encountered in~\cite{deHaro1}.

When the integration in eq.~(\ref{int2}) is explicitly carried out,
the constraints on the dual integers are implemented by a sum over
partitons of the rank $N$ and one obtains the formula
\begin{eqnarray} \label{ZCSfinal}
\mathcal{Z}_{U(N)}^{\rm CS}\big(\torus_p^3 \,,\, k\big) &=&
\sum_{\stackrel  {\scriptstyle{\mbf \nu } \in
\nat_0^N} {\scriptstyle\sum_a\,a\,\nu_a = N }}\,
\mathcal{Z}^{\rm CS}_{\mbf \nu}
\left( \torus_p^3 \,,\, k \right)  \\[4pt] &:=&
 \sum_{\stackrel{\scriptstyle{\mbf \nu} \in \nat_0^N}{
\scriptstyle\sum_a\,a\,\nu_a = N} }~ \prod_{a=1}^{N}\,
\frac{\left( -1 \right)^{\nu_a} }{\nu_a !}\, \left(
\frac{ k +N}{\ii p\,a^3} \right)^{{\nu_a}/{2}}\,
\left( \, \sum_{m=0}^{p \, a -1}\, \e^{\pi\ii \frac{ k +N}{p}\,
\frac{\pi \, m^2}{a}} \right)^{\nu_a} \ . \nonumber
\end{eqnarray}
This rewriting identifies the values
\beq
S_{\mbf \nu}^{\rm CS}(\mbf m)=-\frac{\pi\ii (k+N)}{p}\,
\sum_{a=1}^N\,\frac1a~\sum_{j=1+\nu_1+\dots+
\nu_{a-1}}^{\nu_1+\dots+\nu_a}\,m_j^2
\label{classCSaction}\eeq
of the semi-classical Chern-Simons actions. The sums in (\ref{ZCSfinal}) go
over the flat $U(N)$ gauge connections on $\torus_p^3$, as found in
Sect.~\ref{FlatConns} above, and the action (\ref{classCSaction})
coincides with (\ref{SUNCSflat}) evaluated on a generically reducible
flat connection. However, the actions (\ref{classCSaction}) are not
generally produced by gauge inequivalent
solutions of the Chern-Simons equations of motion. This situation
naturally parallels that of the Yang-Mills partition function in the
instanton representation as described in Sect.~\ref{SDuality}. This is
of course expected due to the relations unveiled
in~\cite{Beasley:2005vf}. Here the sum over partitions of $N$ nicely
appears, describing the reduction of the flat bundle near the critical
points of Chern-Simons theory on $\torus_p^3$, and the gauge
inequivalent reorganisation into a sum over a flat $U(N)$ gauge
bundles corresponds to the decompositions (\ref{redrepWeyl}) into
irreducible representations of the Heisenberg-Weyl groups.

To understand this point better, we proceed as in the case of the
Yang-Mills theory of Sect.~\ref{SDuality}. The $U(2)$ Chern-Simons
partition function (\ref{ZCSfinal}) can be written explicitly as
\begin{eqnarray}
\mathcal{Z}_{U(2)}^{\rm CS}\big(\torus_p^3\,,\,k\big)  =
\sum_{m_1,m_2=0}^{p-1}\,\frac{ 2 \pi}{ g_s\, p
}~\e^{-\frac{2 \pi^2}{ g_s\, p }\,(m_1^2+m_2^2)}-\sum_{m=0}^{2p-1}\,
\frac{1}{ \sqrt{2}}\,\sqrt{\frac{ 2 \pi}{ g_s\, p
}}~\e^{-\frac{ \pi^2}{ g_s \,p }\,m^2} \ ,
\end{eqnarray}
and again one can collect contributions corresponding to
gauge inequivalent flat connections to get
\begin{eqnarray}
\mathcal{Z}_{U(2)}^{\rm CS}\big(\torus_p^3\,,\,k\big)&=&\nonumber
\sum_{ m_1,m_2=0}^{p-1}\,\left(\,\frac{ 2 \pi}{ g_s\, p }-
\delta_{m_1,m_2}\,\frac{1}{\sqrt{2}}\,\sqrt{\frac{ 2 \pi}{ g_s\, p
}}~\right)~ \e^{
-\frac{ 2 \pi^2}{ g_s\, p }\, ( m_1^2+
m_2^2)} \\ &&-\,\sum_{ m=0}^{p-1}\,\frac{1}{\sqrt{2}}\,
\sqrt{\frac{ 2 \pi}{ g_s \,p}}~\e^{
-\frac{ \pi^2}{ g_s\, p }\, (2m+1)^2} \ .
\label{CSgaugeineq}\end{eqnarray}
In analogy with the case of Yang-Mills theory on $\torus^2$ the same
non-trivial moduli spaces of non-isolated (flat) connections on
$\torus_p^3$, with the singular structure of symmetric product
orbifolds, arise in the Chern-Simons gauge theory. In this case the
moduli spaces of flat connections ${\rm
  Hom}(\pi_1(\torus_p^3),U(N))/U(N)$ coincide with
(\ref{modspinstNq}), as this is also the moduli space of
representations of the Heisenberg-Weyl
algebra~\cite{Paniak:2002fi}. The source of the non-isolated flat
connections is zero modes of the gauge connections on $\torus_p^3$
giving an integral of the Ray-Singer torsions over the moduli space of
flat connections.

Now we can rewrite the $U(2)$ Yang-Mills partition function
(\ref{su22}) in a way which clarifies its relation to the Chern-Simons
partition function (\ref{CSgaugeineq}), similarly to the rewriting
(\ref{ZYMhats}). One finds
\begin{eqnarray}
\mathcal{Z}^{\rm YM}  &=&\nonumber \sum_{\hat m_1,
\hat m_2=0}^{p-1}\,\left((-1)^{\hat m_1+\hat m_2}\,\frac{
2 \pi}{ g_s\, p }\,
    \vartheta_3\big(\mbox{$\frac{2 \pi \ii  \hat m_1 }{g_s}+
\frac{p}{2}    $}\,,\,\mbox{$\frac{2\pi \ii p }{g_s} $} \big)\,
\vartheta_3\big(\mbox{$\frac{2 \pi \ii  \hat m_2 }{g_s}+
\frac{p}{2}    $}\,,\,\mbox{$\frac{2\pi \ii p }{g_s} $} \big)\right.\\
&&\left.\qquad\qquad+\,\delta_{m_1,m_2}\,\frac{1}{ \sqrt{2}}\,
\sqrt{\frac{ 2 \pi}{g_s\, p }}\,\vartheta_3\big(\mbox{$\frac{4\pi
    \ii\hat m_1}{g_s}    $}\,,\,\mbox{$\frac{4
\pi \ii p }{g_s} $} \big)\right)~\e^{-\frac{ 2 \pi^2}{ g_s\, p }\, (\hat
m_1^2+\hat m_2^2)}\nonumber\\ && -\,\sum_{\hat
m=0}^{p-1}\,\frac{1}{\sqrt{2}}\,\sqrt{\frac{ 2 \pi}{ g_s\, p
}}\,\vartheta_3\big(\mbox{$\frac{2 \pi \ii (2
\hat m+1) }{g_s}    $}\,,\,\mbox{$\frac{4\pi \ii p }{g_s} $}
\big)~ \e^{-\frac{ \pi^2}{ g_s\, p } (2\hat m+1)^2} \ .
\label{conn} \end{eqnarray}
A similar expression for the genus zero fibrations has been derived
in~\cite{Caporaso:2005ta}. In eq.~(\ref{conn}), the Yang-Mills
partition function on the two-torus is given by a sum over gauge
inequivalent Chern-Simons flat connections. Each term is
multiplied by a fluctuation factor that results from the mixing of
two kinds of contributions. The first contributions are the
characteristic theta-functions that arise when we compare Yang-Mills
theory and Chern-Simons theory, as was found in~\cite{Caporaso:2005ta}. The
second contributions, as explained at length in Sect.~\ref{SDuality},
come from the one-loop fluctuation determinants around the classical
solutions and they become explicit when we collect together the gauge
inequivalent solutions. The same exercise can in principle be repeated
for any $N$. The general structure is clear but unfortunately we have
not been able to find closed expressions for the fluctuation integrals
over the singular moduli spaces.

Our lack of explicit knowledge of quantities associated to the
singular moduli spaces for generic gauge group rank $N$ also presents
an obstacle to describing the large $N$ geometric
transition between the Chern-Simons gauge theory and closed
topological string theory, as explicitly hinted to by the relations
above. Recall that Chern-Simons gauge theory on a three-manifold $M$
is equivalent at large $N$ to the A-model on the cotangent bundle
$T^*M$. If $M$ can be described as the total space of a
$\torus^2$-fibration over the interval $\interval$, then $T^*M$ has
the structure of a $\torus^2\times\real$-fibration over
$\real^3$~\cite{Aganagic:2002wv} and thus admits a natural description
in terms of toric geometry. In the present case, the description of
$\torus_p^3$ as a torus bundle over the circle gives its cotangent
bundle $T^*\torus_p^3$ the structure of a
$\torus^2\times\real$-fibration over $\sphere^1\times\real^2$ with no
$\torus^2$ cycle degenerations. A formal toric geometry for
$T^*\torus_p^3$ could now be constructed using blowup techniques as in
Sect.~\ref{FTG}, and it would be interesting to pursue this
geometrical construction further.

\acknowledgments

We thank S. de Haro, M. Mari\~no, S.~Ramgoolam, C.~S\"amann, E. Sharpe
and A. Sinkovics for helpful discussions. The work of N.~C. was
supported in part by a fellowship from the Fondazione Della
Riccia. The work of M.~C. and R.~J.~S. was supported in
part by the EU-RTN Network Grant MRTN-CT-2004-005104 and by PPARC
Grant PPA/G/S/2002/00478.

\appendix
\addcontentsline{toc}{section}{Appendices}

\section{Schur Functions}

In this appendix we will summarize some useful identities
for Schur functions. For a more complete treatment we refer
the reader to~\cite{MacDonald}, and to~\cite{Zhou:2003zp} which
contains a brief review of most of the properties commonly used in
topological vertex computations. Symmetric polynomials are
polynomials of $N$ independent variables $x_1,\dots,x_N$ which are
invariant under the natural action of the symmetric group ${S}_N$ by
permutations of the $x_i$. They form a subring $\Lambda_N$ of the ring
of polynomials with integer coefficients given by
\begin{equation} \label{ringsymm}
\Lambda_N = \zed[x_1 , \dots, x_N]^{{S}_N} \ .
\end{equation}
A convenient basis for the ring (\ref{ringsymm}) is provided by the
Schur functions $s_{\hat{ R}} ({x_i})$ which are defined in terms of
$SU(N)$ representations $\hat R$ as
\begin{equation}
s_{\hat{ R}} (x_i) = \frac{\det\big( x_j^{n_i(\hat R) +
N -i}\big)}{\det\big(x_j^{N -i}\big)} \ .
\label{Schurdef}\end{equation}
The invariant polynomial (\ref{Schurdef}) is homogeneous of
degree $|\hat{ R}|$.

The Schur functions satisfy the orthogonality relations
\begin{eqnarray} \label{fortext1}
\sum_{\hat{ R}}\, s_{\hat{ R}} \left( x_i \right) \,
s_{\hat{ R}} \left( {y_i} \right) &=& \prod_{i , j\geq1}\, \left( 1 - x_i
\, y_j \right)^{-1} \ , \nonumber\\[4pt]
\sum_{\hat{ R}}\, s_{\hat{ R}} \left( x_i \right) \,
s_{\hat{ R}^\top} \left( {y_i} \right) &=& \prod_{i , j\geq1}\, \left( 1 +
x_i \, y_j \right) \ .
\end{eqnarray}
These two properties are not independent of each other because the
second identity follows from the first identity after applying the
operation $\Xi$ defined by
\begin{equation}\Xi_{y_i} \left( s_{\hat{ R}} \left( {y_i} \right)
\right) = s_{\hat{ R}^\top} \left( {y_i} \right)
\end{equation}
to the variables ${y_i}$~\cite{MacDonald}. On the right-hand side of
(\ref{fortext1}), $\Xi_{x_i}$ acts as
\begin{equation}
\Xi_{x_i} \big(\,\mbox{$\prod\limits_{i \ge 1}\, (1+ x_i \,t)$}
\big) = \prod_{i \ge1}\, (1 - x_i \,t)^{-1}
\end{equation}
for every $t$. The operation $\Xi$ is an involution of the ring
$\Lambda_N$.

It is convenient to define the skew Schur functions $s_{\hat{ R} /
\hat{ Q}} ({x_i})$ as
\begin{equation}
s_{\hat{ R} / \hat{ Q}} ({x_i}) = \sum_{\hat Q'}\, N_{\hat{ Q}
\hat Q'}^{\hat{ R}} ~ s_{\hat Q'} ({x_i}) \ ,
\end{equation}
where $N_{\hat{ Q} \hat Q'}^{\hat{ R}}$ are the fusion
coefficients defined by
\begin{equation}
s_{\hat{ Q}} ({x_i})\, s_{\hat Q'} ({x_i}) = \sum_{\hat{ R}}\,
N_{\hat{ Q} \hat Q'}^{\hat{ R}} ~ s_{\hat{ R}} ({x_i}) \ .
\end{equation}
It is immediate to see that
\begin{eqnarray} \label{forapp2}
\sum_{\hat{ R}}\, s_{\hat{ R} / \hat{ Q}} \left( {x_i}
\right) \, s_{\hat{ R}} \left( {y_i} \right) &=& \sum_{\hat Q'}
\, s_{\hat Q'} \left( {x_i} \right) \, s_{\hat{ Q}} \left( {y_i}\right)
\, s_{\hat Q'} \left( {y_i} \right) \ , \nonumber\\[4pt]
\sum_{\hat{ R}}\, s_{\hat{ R}^\top / \hat R^{\prime\,\top}} \left(
{x_i} \right) \, s_{\hat{ R}} \left( {y_i} \right) &=& \sum_{\hat Q'}
\, s_{\hat Q^{\prime\,\top}} \left( {x_i} \right) \, s_{\hat R'} \left(
{y_i}\right) \, s_{\hat Q'} \left( {y_i} \right) \ ,
\end{eqnarray}
where the second identity follows from applying the involution
$\Xi_{x_i}$ to the first identity. The orthogonality relations
(\ref{fortext1}) can be generalized to the skew Schur functions as
\begin{eqnarray}\label{forapp1}
\sum_{\hat R'}\, s_{\hat R' / \hat{ R}} \left( {x_i}
\right) \, s_{\hat R' / \hat{ Q}} \left( {y_i} \right) &=& \prod_{i
, j\geq1}\, \left( 1 - x_i \, y_j \right)^{-1}~ \sum_{\hat T}\,
s_{\hat{ Q} / \hat T} \left( {x_i} \right) \, s_{\hat{ R} /
\hat T} \left( {y_i} \right) \ , \nonumber\\[4pt]
\sum_{\hat R'}\, s_{\hat R' / \hat{ R}^\top} \left( {x_i}
\right) \, s_{\hat R'^\top / \hat{ Q}} \left( {y_i} \right) &=&
\prod_{i , j\geq1}\, \left( 1 + x_i \, y_j \right)~ \sum_{\hat T}\,
s_{\hat{ Q}^\top / \hat T} \left( {x_i} \right) \, s_{\hat{ R}
/ \hat T^\top} \left( {y_i} \right) \ .
\end{eqnarray}

\section{The Topological Quantum Field Theory Amplitude}

The topological string amplitude associated with the formal toric
geometry depicted in Fig.~\ref{TQFTdiag} can be computed by means of
the topological vertex gluing rules described in
Sect.~\ref{TopStringAmpl}. It is given by
\begin{eqnarray}
{\cal Z}^{\rm TQFT} = \sum_{\stackrel{\scriptstyle\hat R_{1},\hat
    R_{2}}{\scriptstyle\hat R_{3} ,\hat R}}
&& (-1)^{|\hat{R}| + |\hat{R}_{1}| +
|\hat{R}_{2}| + |\hat{R}_{3}|}\,Q^{|\hat{R}|}\, Q_1^{|\hat{R}_{1}|}\,
Q_2^{|\hat{R}_{2}|}\,Q_3^{|\hat{R}_{3}|}  \cr && \times \,C_{\bullet\bullet
\hat{R}_{2}}(q) \, C_{\hat{R}\hat{R}_{2}^\top\hat{R}_{1}}(q)
\, C_{\hat{R}^\top\hat{R}_{3}^\top\hat{R}_{1}^\top }(q)\,
C_{\hat{R}_{3}\bullet\bullet}(q)
\end{eqnarray}
where $Q_i:=\e^{-t_i}$ with $t_i$ the auxilliary K\"ahler moduli
corresponding to the three blowup lines in the toric graph of
Fig.~\ref{TQFTdiag}. The explicit expression (\ref{topvertexdef}) for
the topological vertex can be written in terms of Schur functions as
\beq
C_{\hat{R}_{1}\hat{R}_{2}\hat{R}_{3}}(q)=
q^{\frac{1}{2}\,( \kappa_{\hat{R}_{2}} +
\kappa_{\hat{R}_{3}})}\, s_{\hat{R}_{2}^\top} \big(
q^{-i+{1}/{2}} \big)\, \sum_{\hat{P}}\, s_{\hat{R}_{1} /
\hat{P}} \big( q^{n_i({\hat{R}_{2}^\top}) -i + {1}/{2}}
\big) \, s_{\hat{R}_{3}^\top / \hat{P}} \big(
q^{n_i({\hat{R}_{2}}) -i +{1}/{2}} \big) \ .
\eeq
We can then write the topological string amplitude as
\begin{eqnarray}
{\cal Z}^{\rm TQFT}&=& \sum_{\stackrel{\scriptstyle\hat R_{1},\hat
    R_{2}}{\scriptstyle\hat R_{3} ,\hat R}}\,
(-1)^{|\hat{R}| + |\hat{R}_{1}| +
|\hat{R}_{2}| + |\hat{R}_{3}|} \, Q^{|\hat{R}|}\,
Q_1^{|\hat{R}_{1}|}\,Q_2^{|\hat{R}_{2}|}\,
Q_3^{|\hat{R}_{3}|}~
 s_{\hat{R}_{2}}  \big( q^{-i+{1}/{2}} \big) \,
 s_{\hat{R}_{3}}  \big( q^{-i+{1}/{2}} \big)
 \cr && \times\,
\left[ q^{ \frac{1}{2}\,( \kappa_{\hat{R}} +
\kappa_{\hat{R}_{2}^\top}) }\, s_{\hat{R}^\top} \big(
q^{-i+{1}/{2}} \big)\, \sum_{\hat S} \, s_{\hat{R}_{1} /
\hat{S}} \big( q^{n_i({\hat{R}^\top}) -i + {1}/{2}} \big)
\, s_{\hat{R}_{2} / \hat{S}} \big( q^{n_i({\hat{R}}) -i +
{1}/{2}} \big)  \right] \nonumber\\[2pt] && \times\,
\left[ q^{ \frac{1}{2}\,( \kappa_{\hat{R}^\top} +
\kappa_{\hat{R}_{3}^\top}) } \,s_{\hat{R}} \big(
q^{-i+{1}/{2}} \big)\, \sum_{\hat T} \, s_{\hat{R}_{1}^\top /
\hat T} \big( q^{n_i({\hat{R}}) -i + {1}/{2}} \big)
\, s_{\hat{R}_{3} / \hat T} \big( q^{n_i({\hat{R}^\top}) -i
+{1}/{2}} \big)  \right] \ . \cr &&
\end{eqnarray}
It is convenient to absorb the factors $q^{
\kappa_{\hat{R}_{2}^\top}/2}$ and $q^{
\kappa_{\hat{R}_{3}^\top}/2}$ by using Proposition~4.1
of~\cite{Zhou:2003zp} which asserts the identity
\begin{equation}
q^{\kappa_{\hat{R}^\top}/2} \,s_{\hat{R}} \big(
q^{-i+{1}/{2}} \big) = s_{\hat{R}} \big( -q^{i-{1}/{2}}
\big) \ .
\end{equation}

Now let us perform the sum over $\hat{R}_{1}$ by using the
formulas (\ref{forapp1}), and expand the sums over $\hat{R}_{2}$
and $\hat{R}_{3}$ by using eqs.~(\ref{forapp2}) to get
\begin{eqnarray}
{\cal Z}^{\rm TQFT} &=&\nonumber  \sum_{\hat{R} , \hat{P} , \hat T} \,
(-1)^{|\hat{R}|}\,Q^{|\hat{R}|}\,
s_{\hat{R}^\top} \big( q^{-i+{1}/{2}} \big)
\, s_{\hat{R}} \big( q^{-i+{1}/{2}} \big) \cr && \qquad\qquad \times\,
 (-1)^{|\hat{P}|}\,Q_1^{|\hat{P}|}~ \prod_{i , j\geq1}\, \left( 1 -
Q_1 \, q^{n_i({\hat{R}^\top}) -i + n_j({\hat{R}})
-j +1} \right) \\ && \times\,
\sum_{\hat Q'}\, s_{\hat Q'} \big( q^{n_i({\hat{R}})
-i+{1}/{2}} \big) \, s_{\hat{P}} \big( Q_2 \,
q^{i-{1}/{2}} \big) \, s_{\hat Q'} \big(  Q_2 \,
q^{i-{1}/{2}} \big) \cr && \times\,
\sum_{\hat{S}}\, s_{\hat{S}} \big( q^{n_i({\hat{R}^\top})
-i+{1}/{2}} \big) \, s_{\hat T} \big( Q_3 \,
q^{i-{1}/{2}} \big) \, s_{\hat{S}} \big( Q_3 \,
q^{i-{1}/{2}} \big) \cr && \times\,
\sum_{\hat{ Q}} \, s_{\hat T^\top / \hat{ Q}} \big( - Q_1 \,
q^{n_i({\hat{R}^\top}) -i + {1}/{2}} \big) \, s_{\hat{P}^\top
/ \hat{ Q}^\top} \big(  q^{n_i({\hat{R}}) -i +{1}/{2}}
\big) \ .
\end{eqnarray}
We can perform the sums over $\hat Q'$ and $\hat{S}$ using
the orthogonality relations (\ref{fortext1}), and expand the sums over
$\hat{P}$ and $\hat T$ by using eqs.~(\ref{forapp2}) again to write
\begin{eqnarray}
{\cal Z}^{\rm TQFT}  &=& \sum_{\hat{R} , \hat{ Q} }
\,(-1)^{|\hat{R}|}\, Q^{|\hat{R}|}\,
 s_{\hat{R}^\top} \big( q^{-i+{1}/{2}} \big)
\, s_{\hat{R}} \big( q^{-i+{1}/{2}} \big) \nonumber \\ && \times\,
\prod_{i , j\geq1}\, \frac{\left( 1 - Q_1 \, q^{n_i({\hat{R}^\top}) -i +
n_j({\hat{R}}) -j+1} \right)}{\left( 1 - Q_2 \, q^{n_i({\hat{R}}) -i +
j} \right) \,
\left( 1 - Q_3 \, q^{n_i({\hat{R}^\top}) -i + j } \right)} \cr &&
\times\, \sum_{\hat{P}} \, s_{\hat{P}^\top} \big( q^{n_i({\hat{R}})
-i+{1}/{2}} \big) \, s_{\hat{ Q}} \big( -  Q_2 \, Q_1 \,
q^{i-{1}/{2}} \big) \, s_{\hat{P}} \big( - Q_1 \, Q_2 \,
q^{i-{1}/{2}} \big) \cr && \times\,
\sum_{\hat T} \, s_{\hat T^\top} \big( - Q_1 \,
q^{n_i({\hat{R}^\top}) -i+{1}/{2}} \big) \, s_{\hat{ Q}^\top}
\big(  Q_3 \, q^{i-{1}/{2}} \big) \, s_{\hat T} \big(
 Q_3 \, q^{i-{1}/{2}} \big) \ .
\end{eqnarray}
Finally, we can repeatedly apply eqs.~(\ref{fortext1}) to obtain
\begin{eqnarray}
{\cal Z}^{\rm TQFT}  &=& \sum_{\hat{R}}\, (-1)^{|\hat{R}|} \, Q^{|\hat{R}|}\,
s_{\hat{R}^\top} \big( q^{-i+{1}/{2}} \big) \, s_{\hat{R}}
\big( q^{-i+{1}/{2}} \big) \nonumber \\ && \times\,
\prod_{i , j\geq1}\, \frac{\left( 1 - Q_1 \, Q_2 \, q^{i +
n_j({\hat{R}}) -j } \right) \,\left( 1 - Q_1 \, Q_3 \, q^{i +
n_j({\hat{R}}) -j} \right)}{\left( 1 - Q_2 \, q^{n_i({\hat{R}}) -i +
j} \right) \,
\left( 1 - Q_3 \, q^{n_i({\hat{R}^\top}) -i + j} \right)}
\nonumber\\[2pt] && \times\, \prod_{i , j\geq1}\,  
\Big( 1 - Q_1 \, Q_2 \, Q_3 \, q^{ i + j -1} \Big) \,
\left( 1 - Q_1 \, q^{n_i({\hat{R}^\top}) -i +
n_j({\hat{R}}) - j + 1} \right) \ .
\end{eqnarray}
This expression reduces up to a normalization factor to the amplitude
(\ref{TopAmp2}) for the equal K\"ahler moduli $t_2=t_3$,
i.e. $Q_2=Q_3$, and when the auxilliary edge shrinks to zero length
$t_1\to0$, i.e. $Q_1\to1$.

\section{Gromov-Witten and Gopakumar-Vafa Invariants}

In this Appendix we will list the Gromov-Witten and the Gopakumar-Vafa
invariants up to genus $6$ and degree $15$, as computed with the
techniques outlined in Sect.~\ref{InvtTheory}.

The following tables collect the Gromov-Witten invariants
$N_d^g(X_p)\in\rat$:

\begin{small}
\begin{flushleft}

\begin{eqnarray} \nonumber
\begin{array}{|c|c|c|c|c|c|c|c|c|}
\hline N_{d}^g(X_p) / p^{2g-2} & d=2 & d=3 & d=4 & d=5 & d=6 & d=7  \\
\hline
g=2 & -4 & -32 & -120 & -320 & -720 & -1344 \\
\hline
g=3 & \frac{4}{3} & \frac{320}{3} & 1632 & \frac{36608}{3} & 60368
& 227712 \\ 
\hline g=4 & - \frac{8}{45}   & -\frac{5824}{45}  & -
\frac{24512}{3}  &
  - \frac{1544960}{9}  & - \frac{5770592}{3}  & - \frac{214381696}{15}  \\
\hline g=5 & \frac{4}{315} & \frac{5248}{63} & \frac{2230912}{105}
& \frac{394608128}{315} & \frac{3334331408}{105} &
\frac{6962181376}{15}\\ 
\hline g=6 & - \frac{8}{14175}  & - \frac{472384}{14175}
 & - \frac{3579712}{105}  & -
\frac{15917295872}{2835}
& - \frac{43419217184}{135}  & - \frac{4850481974144}{525}  \\
\hline
\end{array}
\end{eqnarray}
\end{flushleft}


\begin{eqnarray} \nonumber
\begin{array}{|c|c|c|c|c|}
\hline N_{d}^g(X_p) / p^{2g-2} & d=8 & d=9 & d=10 & d=11  \\
\hline g=2 & -2480 & -3840&-6360&-8800 \\
\hline g=3 & \frac{2137856}{3}&1918464&4676136&\frac{30846400}{3} \\
\hline g=4 & - \frac{715188224}{9}  &-355731968&-
\frac{4044479440}{3}  &-
\frac{40169500480}{9}  \\
\hline g=5 &
\frac{1448560219136}{315}&\frac{1193578001408}{35}&
\frac{21220261061096}{105}&\frac{63082147862912}{63} 
\\ 
\hline g=6 & - \frac{66030144109568}{405}  &-
\frac{630459322936832}{315}  &
  - \frac{1947915582268112}{105}  &- \frac{77922934815152576}{567}
  \\
\hline
\end{array}
\end{eqnarray}


\begin{eqnarray} \nonumber
\begin{array}{|c|c|c|c|c|}
\hline N_{d}^g(X_p) / p^{2g-2} & d=12 & d=13 & d=14 & d=15 \\
\hline g=2 &-13664&-17472&-25760 & -31680 \\
\hline g=3 & \frac{64214912}{3}&41108736&\frac{230312288}{3} & 133843072  \\
\hline g=4 & - \frac{597369157888}{45}  &- \frac{539401977088}{15}
 &-
\frac{814373960768}{9}  & - \frac{637260238720}{3}  \\
\hline g=5 & \frac{192944817460736}{45}&
\frac{243795677644288}{15}&\frac{2502721471395104}{45}&
\frac{18289716970807552}{105} 
\\ 
\hline g=6 &
  - \frac{12018743197554581248}{14175}  &-
  \frac{2358402999142145792}{525}  & 
  - \frac{59331004145048405312}{2835}  & -
  \frac{82517399556049550464}{945}  \\ 
\hline
\end{array}
\end{eqnarray}

\end{small}


\bigskip

The following tables collect the Gopakumar-Vafa invariants
$n_{d}^r(X_p)\in\zed$:

\begin{flushleft}

\begin{small}
\begin{eqnarray} \nonumber
\begin{array}{|c|c|c|c|c|}
\hline
n_{d}^r(X_p) & d=2 & d=3 \\
\hline r=2 &-4\,p^2 & -32\,p^2 \\
\hline r=3 & \frac{-p^2}{3} + \frac{4\,p^4}{3} &
\frac{-8\,p^2}{3} + \frac{320\,p^4}{3}
\\
\hline r=4 & \frac{-2\,p^2}{45} + \frac{2\,p^4}{9} -
\frac{8\,p^6}{45} & \frac{-16\,p^2}{45} + \frac{160\,p^4}{9} -
\frac{5824\,p^6}{45} \\
\hline r=5 &  \frac{-p^2}{140} + \frac{7\,p^4}{180} -
\frac{2\,p^6}{45} + \frac{4\,p^8}{315} & \frac{-2\,p^2}{35} +
\frac{28\,p^4}{9} - \frac{1456\,p^6}{45} + \frac{5248\,p^8}{63} \\
\hline r=6 &
\begin{array}{c}
\frac{-2\,p^2}{1575} + \frac{41\,p^4}{5670} - \frac{13\,p^6}{1350}
\\ + \frac{4\,p^8}{945} - \frac{8\,p^{10}}{14175}
\end{array}
&
\begin{array}{c} \frac{-16\,p^2}{1575}
+ \frac{328\,p^4}{567} - \frac{4732\,p^6}{675} \\ +
\frac{5248\,p^8}{189} - \frac{472384\,p^{10}}{14175}
\end{array}
\\
\hline
\end{array}
\end{eqnarray}


\begin{eqnarray} \nonumber
\begin{array}{|c|c|c|}
\hline
n_{d}^r(X_p) & d=4 & d=5  \\
\hline r=2 & -112\,p^2 & -320\,p^2 \\
\hline r=3 & \frac{-28\,p^2}{3} + \frac{4864\,p^4}{3} & 
\frac{-80\,p^2}{3} + \frac{36608\,p^4}{3} \\
\hline r=4 &   \frac{-56\,p^2}{45} + \frac{2432\,p^4}{9} -
\frac{367424\,p^6}{45} & \frac{-32\,p^2}{9} + \frac{18304\,p^4}{9}
- \frac{1544960\,p^6}{9} \\
\hline r=5 & \frac{-p^2}{5} + \frac{2128\,p^4}{45} -
\frac{91856\,p^6}{45} + \frac{956032\,p^8}{45} & \frac{-4\,p^2}{7}
+ \frac{16016\,p^4}{45} - \frac{386240\,p^6}{9} +
\frac{394608128\,p^8}{315} \\
\hline r=6 &
\begin{array}{c}
\frac{-8\,p^2}{225} + \frac{24928\,p^4}{2835} -
\frac{298532\,p^6}{675} \\ + \frac{956032\,p^8}{135} -
\frac{483257024\,p^{10}}{14175}
\end{array}
&
\begin{array}{c}
 \frac{-32\,p^2}{315} + \frac{187616\,p^4}{2835} -
\frac{251056\,p^6}{27} \\ + \frac{394608128\,p^8}{945} -
\frac{15917295872\,p^{10}}{2835}
\end{array}
\\
\hline
\end{array}
\end{eqnarray}


\begin{eqnarray} \nonumber
\begin{array}{|c|c|c|}
\hline n_{d}^r(X_p) & d=6 & d=7  \\
\hline r=2 & -644\,p^2&-1344\,p^2 \\
\hline r=3 &\frac{-161\,p^2}{3} + \frac{178436\,p^4}{3}  &
-112\,p^2 +227712\,p^4  \\
\hline r=4 & \frac{-322\,p^2}{45} + \frac{89218\,p^4}{9} -
\frac{86370568\,p^6}{45} & \frac{-224\,p^2}{15} + 37952\,p^4 -
\frac{214381696\,p^6}{15} \\
\hline r=5 & \frac{-23\,p^2}{20} + \frac{312263\,p^4}{180} -
\frac{21592642\,p^6}{45} + \frac{1428518108\,p^8}{45} &
\frac{-12\,p^2}{5} + \frac{33208\,p^4}{5} -
\frac{53595424\,p^6}{15} + \frac{6962181376\,p^8}{15} \\
\hline r=6 &
\begin{array}{c}
\frac{-46\,p^2}{225} + \frac{1828969\,p^4}{5670} -
\frac{140352173\,p^6}{1350} \\ + \frac{1428518108\,p^8}{135}
-\frac{4558775786248\,p^{10}}{14175}
\end{array}
&
\begin{array}{c}
\frac{-32\,p^2}{75} + \frac{389008\,p^4}{315} -
\frac{174185128\,p^6}{225} \\ + \frac{6962181376\,p^8}{45} -
\frac{4850481974144\,p^{10}}{525}
\end{array}
 \\
\hline
\end{array}
\end{eqnarray}


\begin{eqnarray} \nonumber
\begin{array}{|c|c|c|}
\hline n_{d}^r(X_p) & d=8 & d=9  \\
\hline r=2 & -2240\,p^2 & -3744\,p^2 \\
\hline r=3 &\frac{-560\,p^2}{3} + \frac{2098688\,p^4}{3} &
-312\,p^2 + 1915584\,p^4 \\
\hline r=4 & \frac{-224\,p^2}{9} + \frac{1049344\,p^4}{9} -
\frac{712835072\,p^6}{9} &
\frac{-208\,p^2}{5} + 319264\,p^4 -\frac{1778502592\,p^6}{5} \\
\hline r=5
&
\begin{array}{c}
   -4\,p^2 + \frac{918176\,p^4}{45} \\ -
\frac{178208768\,p^6}{9} + \frac{206814792704\,p^8}{45}
\end{array}
&
\begin{array}{c}
\frac{-234\,p^2}{35} + \frac{279356\,p^4}{5} \\  -
\frac{444625648\,p^6}{5} + \frac{1193571625088\,p^8}{35}
\end{array}
\\
\hline r=6 &
\begin{array}{c}
 \frac{-32\,p^2}{45} + \frac{10755776\,p^4}{2835} -
 \frac{579178496\,p^6}{135} \\ + \frac{206814792704\,p^8}{135} -
    \frac{462161522828288\,p^{10}}{2835}
\end{array}
&
\begin{array}{c}
\frac{-208\,p^2}{175} + \frac{3272456\,p^4}{315}
   -\frac{1445033356\,p^6}{75} \\ +
\frac{1193571625088\,p^8}{105} -
\frac{3152295581580352\,p^{10}}{1575}
\end{array}
 \\
\hline
\end{array}
\end{eqnarray}


\begin{eqnarray} \nonumber
\begin{array}{|c|c|c|}
\hline n_{d}^r(X_p) & d=10 & d=11  \\
\hline r=2 & -5700\,p^2 & -8800\,p^2  \\
\hline r=3 &-475\,p^2 +
  4578348\,p^4 &\frac{-2200\,p^2}{3} + \frac{30846400\,p^4}{3}  \\
\hline r=4 & \frac{-190\,p^2}{3} + 763058\,p^4 -
\frac{4027998200\,p^6}{3} &
 \frac{-880\,p^2}{9} + \frac{15423200\,p^4}{9} -
 \frac{40169500480\,p^6}{9} \\ 
\hline r=5 &
\begin{array}{c}
\frac{-285\,p^2}{28} + \frac{2670703\,p^4}{20} -
\frac{1006999550\,p^6}{3} \\ + \frac{21203424343468\,p^8}{105}
\end{array}
 &
\begin{array}{c}
\frac{-110\,p^2}{7} + \frac{2699060\,p^4}{9} -
\frac{10042375120\,p^6}{9} \\ + \frac{63082147862912\,p^8}{63}
\end{array}
 \\
\hline r=6 &
\begin{array}{c}
  \frac{-38\,p^2}{21} + \frac{15642689\,p^4}{630} -
  \frac{1309099415\,p^6}{18} 
 \\  + \frac{21203424343468\,p^8}{315} -
\frac{389522748612056\,p^{10}}{21}
\end{array}
&
\begin{array}{c}
\frac{-176\,p^2}{63} + \frac{31617560\,p^4}{567} -
\frac{6527543828\,p^6}{27} \\
 + \frac{63082147862912\,p^8}{189} -
\frac{77922934815152576\,p^{10}}{567}
\end{array}
 \\
\hline
\end{array}
\end{eqnarray}


\begin{eqnarray} \nonumber
\begin{array}{|c|c|c|}
\hline n_{d}^r(X_p) & d=12 & d=13  \\
\hline r=2 & -11888\,p^2 & -17472\,p^2  \\
\hline r=3 & \frac{-2972\,p^2}{3} + \frac{62634752\,p^4}{3} &
-1456\,p^2+41108736\,p^4   \\
\hline r=4 & \frac{-5944\,p^2}{45} + \frac{31317376\,p^4}{9} -
\frac{594509989696\,p^6}{45} & \frac{-2912\,p^2}{15} + 6851456\,p^4 -
\frac{539401977088\,p^6}{15} \\ 
\hline r=5 &
\begin{array}{c}
  \frac{-743\,p^2}{35} + \frac{27402704\,p^4}{45} -
  \frac{148627497424\,p^6}{45} \\ + 
  \frac{1349318703070592\,p^8}{315}
\end{array}
 &
\begin{array}{c}
\frac{-156\,p^2}{5} + \frac{5995024\,p^4}{5} -
\frac{134850494272\,p^6}{15} \\ + \frac{243795677644288\,p^8}{15}
\end{array}
 \\
\hline r=6 &
\begin{array}{c}
 \frac{-5944\,p^2}{1575} + \frac{321003104\,p^4}{2835} -
 \frac{483039366628\,p^6}{675} \\ 
  + \frac{1349318703070592\,p^8}{945} -
  \frac{12016399468490766016\,p^{10}}{14175} 
\end{array}
&
\begin{array}{c}
\frac{-416\,p^2}{75} + \frac{70227424\,p^4}{315} -
\frac{438264106384\,p^6}{225} \\ + \frac{243795677644288\,p^8}{45}
- \frac{2358402999142145792\,p^{10}}{525}
\end{array}
 \\
\hline
\end{array}
\end{eqnarray}


\begin{eqnarray} \nonumber
\begin{array}{|c|c|c|}
\hline n_{d}^r(X_p) & d=14 & d=15  \\
\hline r=2 & -23044\,p^2 & -30560\,p^2
  \\
\hline r=3 &  \frac{-5761\,p^2}{3} + \frac{224845828\,p^4}{3} &
\frac{-7640\,p^2}{3} +
 \frac{400500800\,p^4}{3}
  \\
\hline r=4 & \frac{-11522\,p^2}{45} + \frac{112422914\,p^4}{9} -
\frac{4051289026568\,p^6}{45} &
 \frac{-3056\,p^2}{9} + \frac{200250400\,p^4}{9} -
 \frac{1911401650880\,p^6}{9} \\ 
\hline r=5 &
\begin{array}{c}
\frac{-823\,p^2}{20} + \frac{393480199\,p^4}{180} -
\frac{1012822256642\,p^6}{45} \\ +
\frac{2500047993276124\,p^8}{45}
\end{array}
 &
\begin{array}{c}
\frac{-382\,p^2}{7} + \frac{35043820\,p^4}{9} -
\frac{477850412720\,p^6}{9} \\ + \frac{10973657170889344\,p^8}{63}
\end{array}
 \\
\hline r=6 &
\begin{array}{c}
 \frac{-1646\,p^2}{225} + \frac{2304669737\,p^4}{5670} -
 \frac{6583344668173\,p^6}{1350} 
 \\  + \frac{2500047993276124\,p^8}{135} -
 \frac{296587967662108631048\,p^{10}}{14175}
\end{array}
&
\begin{array}{c}
\frac{-3056\,p^2}{315} + \frac{58644760\,p^4}{81} -
\frac{310602768268\,p^6}{27} \\ +
\frac{10973657170889344\,p^8}{189} -
\frac{35364555026212714688\,p^{10}}{405}
\end{array}
 \\
\hline
\end{array}
\end{eqnarray}

\end{small}

\end{flushleft}

\end{document}